\documentclass[journal, twocolumn, 10pt]{IEEEtran}
\usepackage[inkscapelatex=false]{svg}
\usepackage{subfig}
\usepackage[T1]{fontenc}
\usepackage{lipsum}
\usepackage{blindtext}
\usepackage{multicol}
\usepackage{graphicx}
\usepackage{cite}
\usepackage{amssymb,amsmath}
\usepackage{algorithm}
\usepackage{algpseudocode}
\usepackage{flushend}
\usepackage{multirow}% http://ctan.org/pkg/multirow
    \usepackage{lgrind}
\usepackage{hhline}% http://ctan.org/pkg/hhline
\usepackage{tcolorbox}
\usepackage{graphicx}
\usepackage{cite}
\usepackage{amssymb,amsmath}
\usepackage{algorithm}
\usepackage{color,soul,colortbl}
\usepackage{cleveref}
\usepackage{booktabs}
\usepackage{xcolor}
\usepackage{graphicx}
\usepackage[export]{adjustbox}
\usepackage{caption}
\usepackage{multirow}% http://ctan.org/pkg/multirow
\usepackage{hhline}% http://ctan.org/pkg/hhline
\usepackage{psfrag}
\usepackage{widetext}
\usepackage{flushend}
\usepackage{cuted}
\usepackage{float}
\usepackage{lipsum}

\newcounter{tempEquationCounter}
\newcounter{thisEquationNumber}

\usepackage[compact]{titlesec}         % you need this package
\titlespacing{\section}{0pt}{0pt}{0pt} % this reduces space between (sub)sections to 0pt, for example
\AtBeginDocument{%                     % this will reduce spaces between parts (above and below) of texts within a (sub)section to 0pt, for example - like between an 'eqnarray' and text
  \setlength\abovedisplayskip{0pt}
  \setlength\belowdisplayskip{0pt}}

\usepackage{amsthm}

\theoremstyle{plain}

\theoremstyle{plain}

\theoremstyle{plain}

\providecommand{\lemmaname}{Lemma}
\providecommand{\propositionname}{Proposition}
\providecommand{\theoremname}{Theorem}
\providecommand{\lemmaname}{Lemma}
\providecommand{\propositionname}{Proposition}
\providecommand{\theoremname}{Theorem}
\makeatother
\providecommand{\lemmaname}{Lemma}
\providecommand{\propositionname}{Proposition}
\providecommand{\theoremname}{Theorem}

\definecolor{G}{gray}{0.9}
\definecolor{LC}{rgb}{0.88,1,1}

\usepackage[nolist]{acronym}
\begin{document}
\title{{\LARGE{The Triple-C Paradigm: 
Cooperative, Complementary, and Competitive Modes for TBS–HAPS–LEO Integration}}}

\author{{ Eros Kuikel, Sidrah Javed,~\IEEEmembership{Member, IEEE}, Baha Eddine Youcef Belmekki, ~\IEEEmembership{Senior Member, IEEE}, Yunfei Chen,~\IEEEmembership{Senior Member, IEEE}, Ning Wang,~\IEEEmembership{Senior Member, IEEE}, Mohamed-Slim Alouini, ~\IEEEmembership{Fellow, IEEE} } 
 \thanks{ E. Kuikel and M.S. Alouini are with the CEMSE Division, King Abdullah University of Science and Technology (KAUST), Thuwal, Makkah Province, 23955-6900 Saudi Arabia. E-mail:\{eros.kuikel, slim.alouini\}@kaust.edu.sa.  S. Javed and N. Wang are with the School of Electrical, Electronic and Mechanical Engineering, University of Bristol, BS8 1QU, England. E-mail: \{sidrah.javed, n.wang\}@bristol.ac.uk. BEY Belmekki is with the Microwaves \& Antenna Engineering Group, Heriot-Watt University, Edinburgh EH14 4AS, Scotland.  E-mail: B.Belmekki@hw.ac.uk,  and Y. Chen is with the Department of Engineering, University of Durham, DH1 3LE, England. E-mail: yunfei.chen @durham.ac.uk. \\
Copyright (c) 2025 IEEE. Personal use of this material is permitted. However, permission to use this material for any other purposes must be obtained from the IEEE by sending a request to pubs-permissions@ieee.org.}}

 \maketitle
 \begin{acronym}
 \acro{UA}{user association}
  \acro{NIB}{network-in-a-box}
  \acro{UAV}{unmanned aerial vehicle}
  \acro{TN}{terrestrial network}
    \acro{NTN}{non-terrestrial network}
  \acro{SIC}{successive interference cancellation}
    \acro{HAPS}{high-altitude platform station}
    \acro{NOMA}{non-orthogonal multiple access}
   \acro{LAPS}{low-altitude platform station}
  \acro{SCA}{successive convex approximation}
  \acro{GU}{ground user}
\acro{6G}{sixth generation}
  \acro{RATs}{radio access technologies}
  \acro{RAT}{radio access technology}
\acro{RZF}{regularized zero-forcing}
\acro{SWaP}{size, weight, and power}
\acro{BER}{bit error rate}
\acro{AWGN}{additive white Gaussian noise}
\acro{CDF}{cumulative distribution function}
\acro{CSI}{channel state information}
\acro{FDR}{full-duplex relaying}
\acro{HDR}{half-duplex relaying}
\acro{IC}{interference channel}
\acro{IGS}{improper Gaussian signaling}
\acro{MHDF}{multi-hop decode-and-forward}
\acro{SIMO}{single-input multiple-output}
\acro{MIMO}{multiple-input multiple-output}
\acro{MISO}{multiple-input single-output}
\acro{MRC}{maximum ratio combining}
\acro{PDF}{probability density function}
\acro{PGS}{proper Gaussian signaling}
\acro{RSI}{residual self-interference}
\acro{RV}{random vector}
\acro{r.v.}{random variable}
\acro{HWD}{hardware distortion}
\acro{cHWD}[HWD]{Hardware distortion}
\acro{AS}{asymmetric signaling}
\acro{GS}{geometric shaping}
\acro{PS}{probabilistic shaping}
\acro{HS}{hybrid shaping}
\acro{SISO}{single-input single-output}
\acro{QAM}{quadrature amplitude modulation}
\acro{PAM}{pulse amplitude modulation}
\acro{PSK}{phase shift keying}
\acro{DoF}{degrees of freedom}
\acro{ML}{maximum likelihood}
\acro{MAP}{maximum a posterior}
\acro{SNR}{signal-to-noise ratio}
\acro{SCP}{successive convex programming}
\acro{RF}{radio frequency}
\acro{CEMSE}{Computer, Electrical, and Mathematical Sciences and Engineering}
\acro{KAUST}{King Abdullah University of Science and Technology}
\acro{DM}{distribution matching}
\acro{CapEx}{capital expenditure}
\acro{OpEx}{operational expenditure}
\acro{SatCom}{satellite communication}
\acro{RIS}{reconfigurable intelligent surface}
\acro{OGS}{optical ground station}
\acro{CoMP}{coordinated multi-point}
\acro{HARQ}{hybrid automatic repeat request}
\acro{SAN}{satellite access network}
\acro{ARPU}{average revenue per user}

\end{acronym}
\vspace{-10cm}
\begin{abstract}
The growing demands of ubiquitous and resilient global coverage have pushed existing networks to their operational limits, making it increasingly difficult to meet all requirements on their own. Integrating \emph{Terrestrial Base Stations (TBS), High Altitude Platform Stations (HAPS)} and \emph{Low-Earth-Orbit (LEO)} satellites is envisioned as a promising solution, yet the coordination across these heterogeneous platforms remains an open challenge. This paper proposes a novel unifying \emph{Triple-C framework: Cooperation, Complementarity, and Competition}, that systematically defines the TBS–HAPS–LEO interaction to deliver seamless resilient and scalable connectivity. For each C, we detail the architectural methodology, required pre-requisites, and measurable deliverables that govern when and how the three layers should collaborate, complement each other, or contend. We further identify the enabling technologies across physical, logical, and cognitive layers to operationalize the proposed 3C paradigm. A rich portfolio of use cases and targeted applications demonstrates how this technological leap will make such integration both feasible and impactful.  
Comprehensive performance analysis and emulation results quantify the trade-offs of such integrated networks. In addition, we examine the economical, environmental, safety, privacy, standardization, and regulatory implications that shape the real-world implementation of the proposed framework. eventually, we provide the gap analysis, outline key technical/non-technical challenges, and a road-map of future research directions needed to unlock the full potential of  Cooperation, Complementarity, and Competition operations in TBS–HAPS–LEO integrated networks. 
\end{abstract}
%\acresetall
\begin{IEEEkeywords}
Non-terrestrial network, satellites, unmanned aerial vehicles, heterogeneous network, integrated hybrid aerial-space-terrestrial architecture, multi-tier connectivity, and 6G. 
\end{IEEEkeywords}

\section{Introduction} 
%%%%%%%%%%%%%%%%%%%%%%%%%%%%%  Why global connectivity? Why standalone TN/NTN are insufficient? Why integrate?
The pursuit of global connectivity represents the core design principle of sixth-generation (6G) communication networks, promising connectivity to everyone, everything, and everywhere. Yet, achieving this ``Internet of Everywhere'' relies on the interplay of \acp{TN} and \acp{NTN}, as envisioned by leading standardization bodies and research alliances (e.g., ITU-R, 3GPP, and the 6G Flagship) \cite{itu2023framework,rajatheva2020white}. Traditional TNs, though foundational to mobile broadband, remain constrained by limited coverage, costly expansion and scalability, restricted deployability, and operational rigidity, leaving billions without reliable connectivity. Conversely, emerging low-Earth-orbit (LEO) satellite constellations enable unprecedented global coverage but face challenges such as network saturation and congestion, frequent handovers, high Doppler, user terminal constraints, short lifespan, and higher latency compared to aerial and terrestrial systems \cite{kodheli2020satellite,geraci2022integrating}. Aerial communication platforms such as High-Altitude Platform Stations (HAPS), coined as ``cell-towers-in-the-sky'', can address numerous limitations in terrestrial and satellite communication infrastructures \cite{kurt2021vision}. They offer larger coverage footprints, flexible deployment (e.g., to tackle flash crowds), short-range line-of-sight (LoS) links, resilience to certain natural disasters, seamless highway connectivity, and economical sparse coverage. HAPS are also advantageous over satellites in terms of favorable channel conditions, quasi-stationary positions with fewer handovers, reduced round-trip delay, higher area throughput, ease of maintenance and reusability, surge coverage, and direct-to-standard handset connectivity \cite{javed2023interdisciplinary}. Hence, the unified orchestration of terrestrial base stations (TBS), aerial platforms (HAPS), and LEO satellites (TBS–HAPS–LEO) through a Cooperative, Complementary, or Competitive (Triple-C) paradigm will become a cornerstone of 6G's universal connectivity. In this vision, each segment dynamically supports, reinforces, and challenges the others to enable a ubiquitous, resilient, sustainable, and inclusive communication ecosystem that can bridge the last digital divides.  

\subsection{Context: The Rise of LEO—and Its Limitations}
%%%%%%%%%%%%%%%%%%%%%%%%%%%% LEO and its limitations
NTNs leverage aerial and satellite platforms to provide coverage where terrestrial systems fall short and have re-emerged as key enablers of global connectivity, largely fueled by the rapid deployment of LEO satellite constellations and the inclusion of satellite access in recent 3GPP releases (Rel-17 and beyond) \cite{3gpp2019tr38811,3gpp2024ntn,3gpp2019tr38821}. Compared with geostationary satellites, LEO systems offer substantially reduced latency and extended broadband coverage, but the rapid motion of satellites leads to frequent handovers and stringent mobility management \cite{alansi2024handover,feng2025handover}. Mobility support is constrained by link budget limitations and dynamic beam steering \cite{alansi2024handover}, spectrum resources in Ka/Ku bands are under growing pressure \cite{inaltekin2024spectrum}, and the sheer scale of mega-constellations raises concerns related to sustainability, orbital debris, and traffic management \cite{boley2021satellite,bennett2025orbital}. Recent studies consistently point to high handover frequency, conditional handover delays, and beam/load balancing as critical barriers to maintaining quality of service, particularly for latency-sensitive and streaming applications \cite{alansi2024handover}.

\begin{figure}[t]
    \centering
    \includegraphics[width=0.99\linewidth]{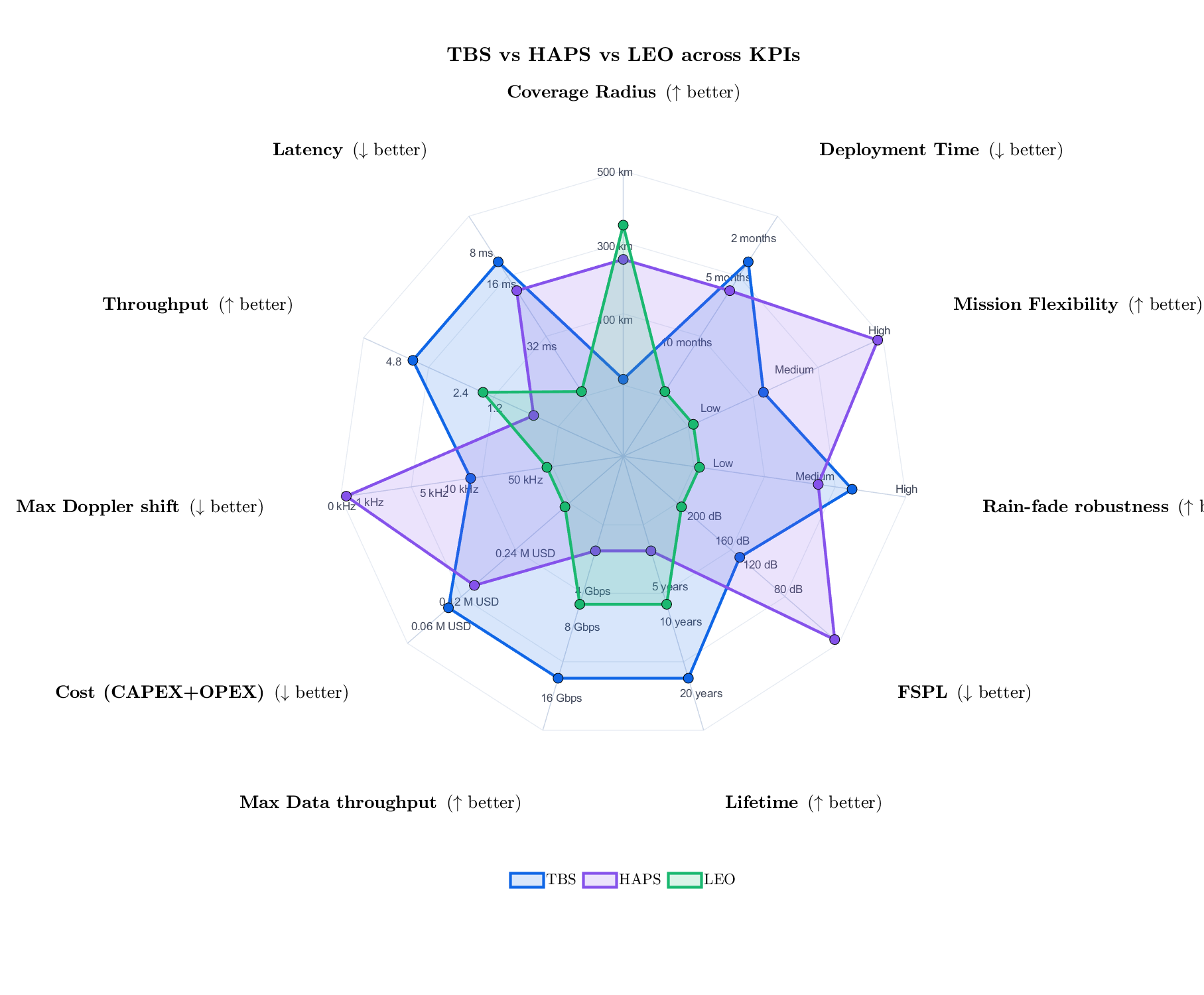}
    \caption{NTN KPIs performance radar}
    \label{fig:3Cs_KPIs}
\end{figure}

A major limitation of current LEO constellations is saturation and congestion: each satellite must pre-compensate delay and Doppler for all simultaneously served UEs, which fundamentally caps the number of active users per cell and leads to service bottlenecks in dense regions \cite{need_dopplercomp}. This has already manifested in practice, with reports of Starlink subscribers in major U.S. metropolitan areas placed on extended waiting lists due to cell-level congestion and oversubscription \cite{starlink_waitlist}. Another structural limitation is the lack of ubiquitous inter-satellite links (ISLs) in many operational shells; in their absence, traffic between distant endpoints must undergo multiple satellite–ground–gateway–satellite relays, causing avoidable delays and reducing end-to-end efficiency \cite{delay_hops_leo}. Furthermore, NTN direct-to-device connectivity requires specialized user equipment: commercial off-the-shelf 5G NTN modems \cite{doppler_5G_est_tech} or electronically steerable user terminals capable of tracking fast-moving satellites and handling Doppler pre-compensation, which increases cost and limits adoption compared to standard terrestrial handsets \cite{mobile_ue_steer}. 
Fig.~\ref{fig:3Cs_KPIs} summarizes how TBS, HAPS, and LEO compare across representative 6G KPIs such as latency, coverage, throughput, cost, and resilience. Given diverse demand and constraints, no single layer can simultaneously satisfy all key performance indicators (KPIs) motivating multi-layer integration of TN and NTN for global, economical, and robust connectivity.

\begin{table*}[t]
\centering
\caption{Representative HAPS trials and demonstrations (from literature and industry reports)}
\label{tab:haps-trials}
\setlength{\tabcolsep}{2pt}% tighter columns
\renewcommand{\arraystretch}{1.1}
\begin{tabular}{|p{3cm}|p{4cm}|p{5cm}|}
\hline
\textbf{Platform} & \textbf{Highlight} & \textbf{Capability} \\
\hline
\hline
Zephyr (AALTO)        & 26--64 day solar flights      & Direct LTE/4G link from $19$ km \cite{beast_haps} \\
SunGlider (HAPSMobile)& Fixed-wing design with MIMO   & First stratospheric video call \cite{nasa_haps} \\
PHASA-35 (BAE)        & Hydrogen power trials         & Targeting year-long persistence \cite{beast_haps} \\
Stratobus (Thales Alenia) & Solar airship design      & LTA platform \\
Loon (Google)         & 3000+ missions                & Multi-month balloon coverage \\
World View            & 10-day station-keeping (2022) & Balloon endurance record \cite{worldview_balloon} \\
HAP-alpha (DLR)       & Solar fixed-wing              & First stratospheric flight planned 2026 \cite{alpha2026} \\
\hline
\end{tabular}
\end{table*}
\subsection{Motivation: Why HAPS Remain Relevant Despite Satellite Proliferation}
HAPS are stratospheric communication platforms operating at altitudes of approximately 18–25~km, above weather and commercial air corridors, offering quasi-stationary coverage over footprints spanning tens to hundreds of kilometers \cite{haps_textbook}. 
The constraints faced by LEO constellations—frequent handovers, Doppler effects, spectrum congestion, user terminal complexity, and sustainability concerns—underscore why the near-space stratospheric layer deserves renewed attention. In this context, HAPS emerge not as a legacy idea or a simple alternative to satellites, but as an integral NTN element in their own right \cite{yanikomeroglu2023haps,lou2023haps,haps2025reference}, with dynamic and flexible deployment, horizontal repositioning according to demand (e.g., town centers by day and residential areas at night), and vertical adjustment of the capacity-versus-coverage tradeoff. Hence, HAPS are envisioned as promising candidates for surge coverage and resilient backhaul \cite{hapsAlliance2023, softbankHAPSNTN2024}. 

\textbf{1) Latency, locality, and direct-to-device connectivity:}
HAPS operate far closer to users than satellites, yielding dramatically lower communication latency and stronger link budgets. A signal round-trip via a HAPS can be on the order of $0.6 ~ms$, compared to $18 ~ms$ for a LEO satellite and $\sim 400 ~ms$ for a geostationary satellite \cite{softbanklatency}. Operating at $\sim 20 ~km$ rather than $\sim 550 ~km$ \cite{wide_coverage_LEO}, HAPS naturally support direct-to-device services, high-resolution imaging, ultra-low latency, strong link budgets, and dynamic relocation \cite{imageresgood,hapsalliance}.

%using standard 4G/5G air interfaces without requiring satellite-specific handsets, and their proximity also enables high-resolution imaging over the covered areas . This combination of ultra-low latency, strong link budgets, and on-demand locality makes HAPS ideal for latency-sensitive applications and focused coverage in areas of interest, plugging gaps that satellites might miss. 

\textbf{2) Economics, coverage, and elasticity:}
HAPS are an economical coverage solution for underserved and unserved areas with no requirement for specialized user equipment. Annualized \ac{CapEx} (about \$$0.2$M/year) and \ac{OpEx} (about \$$0.03$M/year) for platforms such as Zephyr, Sunglider, and blimps demonstrate the viability of HAPS for regional coverage \cite{wide_coverage_LEO}. Each stratospheric platform can cover a wide footprint, often up to a $\sim 200$~km diameter service area \cite{softbanklatency}, allowing one HAPS to connect many scattered ``white spot'' communities that would be uneconomical to serve with individual terrestrial towers \cite{hapsalliance}. HAPS also offer a green solution with prolonged lifespan, easy maintenance/up-gradation, coverage-capacity tradeoffs, coverage extension, emergency communications
 \cite{wide_coverage_LEO,reach2021peru}, connecting the last billion users \cite{techcrunch_blog}. %Although early HAPS business models were economically fragile, ongoing advances in solar power, batteries, and autonomy, coupled with surging demand for flexible coverage, are improving the economics and positioning HAPS as a key tool for connecting the last billion users \cite{techcrunch_blog} . 

\textbf{3) Backhaul flexibility and multi-layer NTNs:}
HAPS can serve as agile nodes in a multi-layered network, interconnecting with both terrestrial infrastructure and satellites to maximize coverage and resiliency. Unlike a standalone satellite, a HAPS can select whichever backhaul is optimal: a high-throughput feeder link to a nearby fiber-connected ground station or when no ground connectivity is available (e.g., over oceans or disaster zones) a relay via LEO/GEO satellites \cite{haps_arch_alliance}. This flexibility enables HAPS to extend service into areas far beyond line-of-sight (LoS) of any ground network while leveraging satellite networks as global backbones. 
%In broader terms, HAPS complement satellites and ground towers by filling the intermediate altitude layer; in a combined architecture, HAPS working in tandem with LEO/GEO satellites can reduce overall costs (by assigning each platform to its most suitable use case) and improve reliability through redundancy \cite{haps_arch_alliance}. Multi-layer coordination also enables intelligent load balancing—high-demand areas can be served directly by HAPS for low latency, while satellite networks handle more delay-tolerant traffic or provide backhaul. HAPS can seamlessly integrate with existing cellular cores (using standard interfaces), effectively acting as aerial base stations that tie into the network like terrestrial cells \cite{haps_arch_alliance}. By nesting HAPS into a layered connectivity stack (ground~+~sky), operators gain substantial flexibility: satellites provide wide global coverage, and HAPS provide focused capacity and rapid deployment, together delivering robust coverage ubiquity.

\subsection{HAPS Capabilities and Maturity}
Recent platform advances span solar-electric fixed-wing aircraft (e.g., Zephyr/AALTO) \cite{aalto2025zephyr}, steerable balloons (e.g., Aerostar Thunderhead; the Loon legacy) \cite{aerostar2022lte}, and emerging airships (e.g., Sceye; Thales Alenia Space’s Stratobus) \cite{haps2024certification}. State-of-the-art HAPS models can be broadly divided into Heavier-than-Air (HTA) and Lighter-than-Air (LTA) categories. HTA platforms such as Airbus/AALTO’s Zephyr, BAE Systems’ PHASA-35, and SoftBank Sunglider, and LTA platforms such as Raven Aerostar’s Thunderhead, Google Loon, Sceye, Stratobus, World View, and DLR’s HAP-alpha, have demonstrated multi-week to multi-month stratospheric persistence, station-keeping, and LTE/4G connectivity to ordinary handsets, as summarized in Table~\ref{tab:haps-trials} \cite{aalto2025zephyr,zephyr4g,phasa35,sunglider,aerostar,loon,sceyeRAN,thales5kW,worldview_balloon,alpha2026,beast_haps}. 

Crucially, the regulatory substrate has also matured. The HAPS Alliance—an industry consortium of telecom, aerospace, and tech firms—has published reference architectures and aviation guidelines to support safe, scalable HAPS operations (e.g., frameworks for certification, airspace integration, and risk management for long-endurance unmanned aircraft) \cite{hapsalliance}. At WRC-23, administrations advanced the High-Altitude IMT Base Station (HIBS) concept in sub-2.7~GHz IMT bands \cite{itu2023hibs,cept2023wrc}, complementing prior fixed-service identifications and creating clearer pathways for HAPS delivering 4G/5G waveforms directly to standard user equipment. This progress is reflected in ITU/CEPT outcomes and technical studies and summarized in industry guidance \cite{haps2024spectrum}, indicating that state-of-the-art HAPS platforms have transitioned from proof-of-concept to pre-commercial viability.

\begin{figure*}[t]
    \centering
   \includegraphics[width=0.7\linewidth]{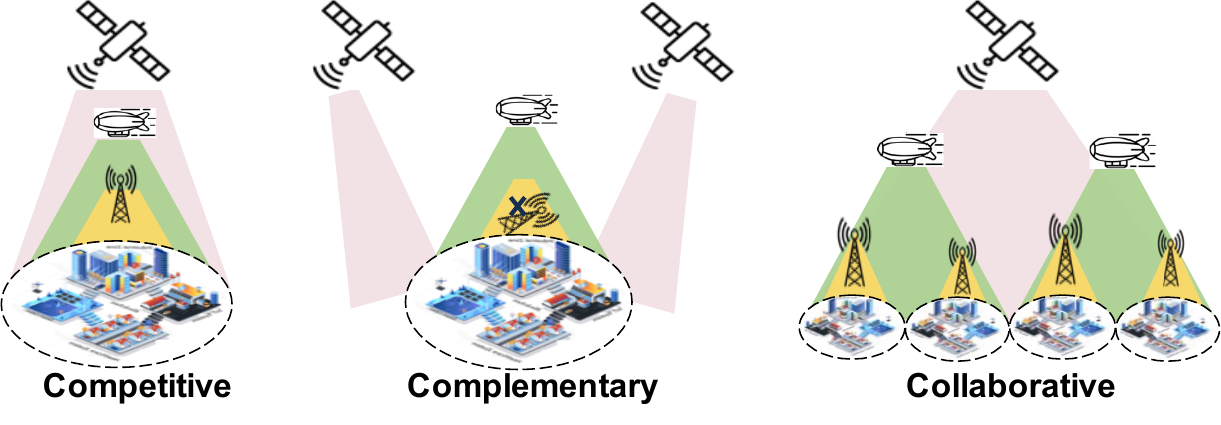}
    \caption{Triple-C Paradigm for TBS--HAPS--LEO Integration}
    \label{fig:3Cs_infra}
\end{figure*}

\subsection{Literature Review}
There have been various surveys that have addressed HAPS and NTN integration from complementary angles—
taxonomy and energy autonomy~\cite{latest2025_halim},
technological evolution~\cite{fo_ntn_survey_2022},
architectural vision~\cite{karabulut_halim},
channel and networking aspects~\cite{cao_survey},
mobility management~\cite{fo_mobility_survey_2025},
hybrid backhaul~\cite{aboelala2022hybridfso},
and early broadband frameworks~\cite{haps_old_survey}.
Taken together, these works treat HAPS/NTN mainly from energy, evolution, architecture, channel, mobility, or backhaul viewpoints, but still in a largely layer-specific manner.

From a broader NTN/\ac{SatCom} angle, \cite{kodheli2021} provides a detailed taxonomy of \ac{SatCom} architectures, air-interface/MAC/networking techniques, and 5G/NTN integration challenges, while \cite{giordani2021} offers a full-stack NTN overview covering UAV/HAPS/satellite architectures, spectrum and antenna technologies, protocol/transport issues, and an LEO–HAP–ground relaying case study; \cite{yaacoubslim2020} survey rural connectivity and ``base of pyramid'' business models, including satellites and balloons. However, all three largely treat terrestrial, aerial, and satellite options as either satellite-centric or scenario-specific technology choices, rather than as jointly coordinated tiers with explicitly shared access/backhaul roles and KPI-driven interaction.

On the terrestrial and resource-allocation side, \cite{polese2020iab} analyze NR Rel-16 integrated access and backhaul in 5G mmWave networks (3GPP IAB architecture, in-band access/backhaul multiplexing, throughput/latency trade-offs), but confine the analysis to ground gNB/IAB topologies. Closer to true multi-tier designs, \cite{alsharoaslim2020} build an integrated satellite–HAP–TBS architecture and optimize user association, OFDMA power, and HAP placement, and \cite{federated_poor} design a federated-learning-based task and resource allocation scheme for MEC-enabled high-altitude balloon networks. Yet in both cases, the non-terrestrial tiers are modeled as auxiliary resources within a fixed role and single-tier KPI focus (rate/fairness or energy–delay), without exploring when each tier should dominate, merely complement, or relay under different latency, coverage, and cost regimes.

Control- and mobility-focused works are likewise tier-centric. \cite{handoverdecideNi} proposes an ARIMA-based prediction and adaptive RSS threshold scheme tailored to a single HAPS cellular layout, while \cite{handoverdesignLi} design 5G-compatible handover procedures and a utility-based optimization strategy for a LEO–HAP system. \cite{SDNShi} propose a cross-domain SDN architecture for multi-layered satellite–aerial–terrestrial networks with separate controller domains. These contributions carefully address intra-HAPS handover, LEO–HAP handover, and control-plane virtualization, respectively, but they do not provide a KPI-driven framework that characterizes how terrestrial, HAPS, and satellite layers should share access/backhaul responsibilities or transition between competitive, complementary, and cooperative operation modes.

In summary, existing surveys and technical works span energy-efficient HAPS operation, SAGIN architectures, NTN protocol stacks, rural connectivity, integrated access and backhaul, three-tier resource optimization, stratospheric MEC, and HAPS/LEO handover and control design. However, they remain orthogonal and tier-specific: none formalize the operational interplay across terrestrial, stratospheric, and space layers in terms of \emph{how} they coordinate access and backhaul, \emph{when} they substitute or complement one another under different KPIs and cost regimes, and \emph{why} these interactions shape end-to-end performance-gaps that motivate our Triple-C paradigm for TBS–HAPS–LEO networks.

\subsection{Objective and Contributions}
The main objective of this work is to establish a unified framework for integrating TBS--HAPS--LEO into a single resilient architecture. Existing studies treat TN and NTN layers as cooperative extensions, lacking a systematic treatment of how these platforms should coordinate, suplement, or contend. To address this gap, we introduce the Triple-C framework encompassing cooperation, complementarity, and competition, positioning it as a foundational step towards globally resilient terrestrial-aerial-space integrated networks. The major contributions of this work are enumerated are:
\begin{enumerate}
    \item \textbf{Formal Unified 3C framework for TBS--HAPS--LEO:}
    We introduce a first-of-its-kind Triple-C framework across TBS--HAPS--LEO layers. Unlike traditional static integration, this work explicitly defines the conditions, triggers, and transitions for heterogeneous cooperation, complementarity, and competition, establishing a structured behavioural grammar for TN-NTN integration.  
    
    \item \textbf{Systematic Architecture for Operationalizing 3C:} We present systematic architecture that specifies pre-conditions for each operational mode, formulates KPI/topology/policy based decision triggers, and defines measurable deliverables (QoS compliance, load balancing, coverage recovery, resilience uplift, etc.). This transforms the Triple-C paradigm from a conceptual idea into a practical/implementable system. 
 
     \item \textbf{Triple-C view of HAPS in multi-layer NTNs:}
    We present a fresh perspective on HAPS that goes beyond the conventional ``gap-filler'' narrative by framing their role with respect to terrestrial and satellite layers from a Triple-C (Competition/Complementarity/Cooperation) lens.

    \item \textbf{Multi-layer enabler stack for TN-NTN integration}
  We propose a multi-layer enabling technology taxonomy spanning the physical (spectrum and resource allocation, network architecture, power and beam management), logical (routing, orchestration, federation, SBA/MEC placement), and novel cognitive layer (AI predictions, federated learning, distributed inference). This unprecedented layered mapping provides a clear roadmap for future NTN design. 

    \item \textbf{Use cases and focus applications:}
    We map canonical use cases—such as rural/remote coverage, disaster relief, and low-altitude economy services—to appropriate Triple-C configurations, showing how 3C solves real bottlenecks under techno-economical or environmental conditions. It is the first structured linkage between operational 3C modes and real-world TN-NTN scenarios. 

    \item \textbf{Performance comparison and emulation insights:}
    We compare TBS, HAPS, and LEO performance using analytical models and channel-emulations, illustrating the latency, Doppler, path loss, and delay profiles for evidence-based engineering. 

    \item \textbf{Cost--benefit and sustainability analysis:}
    We contrast CapEx/OpEx, platform lifetime, deployment elasticity, and sustainability aspects of TBS, HAPS, and LEO, and discuss how Triple-C orchestration can yield cost-effective and energy-aware connectivity strategies.

    \item \textbf{Safety and privacy implications:}
    We discuss safety, security, and privacy implications of persistent TBS--HAPS--LEO integration, including airspace safety, ground-risk considerations, and concerns around continuous aerial observation and data handling.

    \item \textbf{Standardization and regulatory aspects:}
    We synthesize ongoing standardization and regulatory efforts relevant to HAPS and multi-layer NTNs (e.g., higher-airspace management, HIBS concepts, and HAPS Alliance guidance). This broadens the value of Triple-C framework beyond theory, showing its feasibility and policy relevance. 

    \item \textbf{Gap analysis and research directions for Triple-C NTNs:}
    Our gap analysis reveals unresolved issues across layers and forward-looking research map specifies open challenges in mobility management, spectrum and airspace coordination, platform and payload constraints, and AI-native orchestration.

\end{enumerate}

% \textcolor{red}{ this section should only talk about our objective and contributions--DONE THIS PART LEAVING THE COMMENTS AND REPEAT CONTNENT AS IS FOR REVIEW LATER} 
% This paper fills that gap by introducing the \textbf{Triple-C Paradigm},
% a unified framework capturing \emph{cooperation, complementarity, and competition}
% across TBS–HAPS–LEO networks.

% \begin{enumerate}
% \item The objective of this work is to present a fresh perspective on HAPS’ role beyond traditional narratives from the prism of the 3Cs (Competition/Complementarity/Collaboration):

% \item \textcolor{red}{3C operational modes are just one contribution, we have many more. Please list all those here in the form of bullets to show a thorough list of our contributions. }

% \item \textcolor{red}{e.g., emerging synergies, write the main contribution of this section in 2 lines}
% \item \textcolor{red}{e.g., use cases and focus applications, write the main contribution of this section in 2 lines}
% \item \textcolor{red}{e.g., performance comparison and emulations}
% \item \textcolor{red}{e.g., cost-benifit analysis}
% \item \textcolor{red}{e.g., safety and privacy implications}
% \item \textcolor{red}{e.g., standardization and regulatory aspects}
% \item \textcolor{red}{so on}

% \end{enumerate}
\begin{table*}[t!]
\centering
\caption{Competition, complementarity, and cooperation among HAPS, LEO, and TBS.}
\label{tab:HAPS_LEO_TBS}
\renewcommand{\arraystretch}{1.2}
\begin{tabular}{|p{3cm}|p{4.2cm}|p{4.2cm}|p{4.2cm}|}
\hline
\textbf{Domain} & \textbf{HAPS} & \textbf{LEO Satellites} & \textbf{Terrestrial BSs} \\ \hline \hline

\textbf{Competition} 
& Low latency ($\sim$1--5 ms) \cite{TedHAPS}; regional coverage; cost-effective for hotspots \cite{HAPScoverageyork}
& Higher latency ($\sim$30--50 ms) \cite{30mslatencyleo}; costly mega-constellations \cite{30mslatencyleo}; global reach including oceans \cite{globalreachleo} 
& Ultra-low latency ($<1$ ms) \cite{ultralowlatencyBS}; very high capacity \cite{ultralowlatencyBS}; limited coverage, infrastructure-intensive \cite{costlyBS} \\ \hline

\textbf{Complementarity} 
& Fills coverage gaps in suburban/rural zones \cite{haps_coverage_gap}; offloads LEO and TBS traffic \cite{haps_offload_traffic}
& Ensures global access in remote or maritime regions \cite{30mslatencyleo}; provides resilience \cite{HAPScoverageyork}
& Supports ultra-dense urban areas \cite{HAPScoverageyork}; backbone for massive connectivity and high throughput \cite{ion_report} \\ \hline

\textbf{Cooperation} 
& Acts as agile backhaul/mid-layer \cite{not_promised}; rapid deployment in emergencies \cite{not_promised}; integrates in multi-layer NTN \cite{HAPScoverageyork}
& Provides global backbone \cite{30mslatencyleo}; enables seamless integration with HAPS/TBS via orchestrated handovers \cite{seamless_integrationLEO} 
& Provides edge computing, local caching, and user proximity \cite{quiMobile}; anchors NTN integration with HAPS/LEO \cite{zhangMobileEdgeCaching2022} \\ \hline
\end{tabular}
\end{table*}

%\textcolor{red}{Dr. Baha said he cannot finalize the figures, but looking at this again I think these flowchart-like figures I have placed as drafts(THERE ARE 3 of them) may not be necessary at all as it feels repetitive. If you suggest to keep it, I can try making a better preliminary version... The FBCs and MBCs I made prelim version adn pasted laready - I prefer to keep their improved versions even if they are repetitive to text. They improve the readability and can summarize the main points that we want to highlight.}

\section{TBS-HAPS-LEO: From Cooperation to Complementarity \& Competition}

The evolution of ubiquitous connectivity in 6G communications heavily relies on the convergence of heterogeneous TBS--HAPS--LEO networks into an integrated adaptive ecosystem. This work crystallizes the Triple-C paradigm through complementary, cooperative, and competitive functional modes as highlighted in Table \ref{tab:HAPS_LEO_TBS} . This section elaborates ``how'' TBS--HAPS-LEO entities interact with each other to realize three operational modes. We formalize the \emph{Triple-C} paradigm by viewing TBS, HAPS, and LEO satellites as layers that can \emph{cooperate} to jointly serve users and backhaul, \emph{complement} each other by specializing in distinct KPI regimes or geographic footprints, or \emph{compete} for the same users, spectrum, and revenue. Using representative KPIs (latency, coverage, throughput, cost, and reliability) and deployment scenarios, we map out these three interaction modes, clarify the associated access/backhaul role allocations, and set the stage for the more detailed architectural and performance discussions in the subsequent sections. A high-level schematic of these three interaction modes and their access/backhaul roles is shown in Fig.~\ref{fig:3Cs_infra}.

\begin{figure}[t]
    \centering
   \includegraphics[width=1\linewidth]{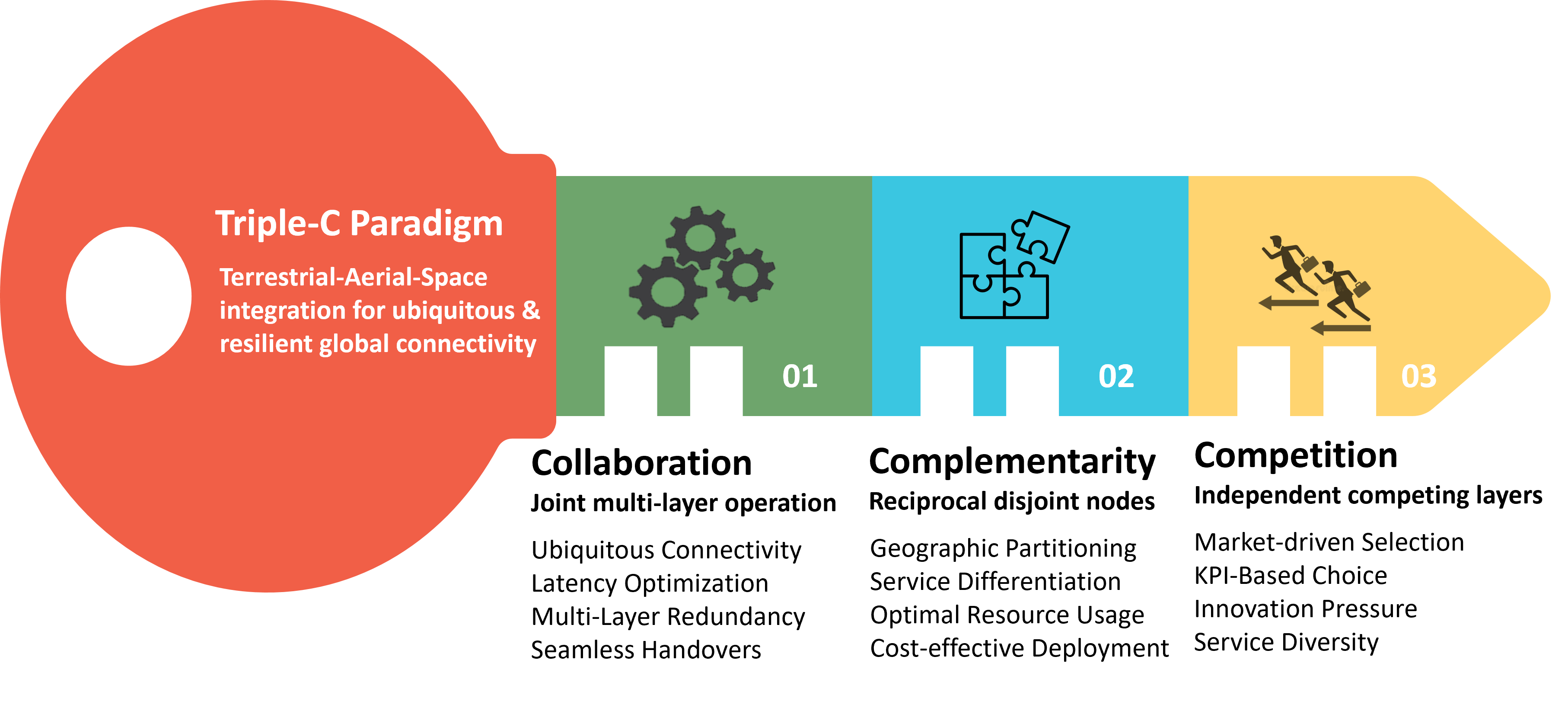}
    \caption{Unlocking the Triple-C paradigm (competition, complementarity, and cooperation) among TBS, HAPS, and LEO layers. }
    \label{fig:3C_Key}
\end{figure}

Having established the RF/FSO channel behavior per link and 3C mode, the payload architecture determines how a HAPS actually participates in those links whether as a passive relay, an active base-station node, or a sensing platform feeding the network. Table~\ref{tab:haps-payload-3c} maps these payload types to concrete roles within the Triple-C paradigm, indicating which combinations of TBS, HAPS, and LEO are feasible or preferred under different KPI and channel conditions. In what follows, we unpack these interaction modes in more detail. Fig.~\ref{fig:3C_Key} provides a compact visual summary of how the three layers unlock competition, complementarity, and cooperation under different KPI and deployment regimes.

\subsection{Cooperation (Joint, co-designed multi-layer operation)}

Cooperative integration reveals a paradigm shift where TBS, HAPS, and LEO collaborate to provide direct-to-device connectivity. They can create a layered structure to \textit{collaborate} and offer services to the same region either by distributing the load or sharing the responsibilities of access and backhaul links. For instance, HAPS can provide the access link to the ground users, while LEO satellites provide backhaul to the HAPS. Alternatively, HAPS can act as an intermediate relay node between ground users and satellites.
Such a cooperative strategy reaps the benefits of each layer involved in the infrastructure. Beyond the traditional lens of competition between terrestrial and non-terrestrial elements, this technological leap will provide a resilient network capable of performing under varying operational stresses (e.g., jamming, natural disasters, cyber disruptions, or congestion). 

\begin{table*}[!t]
\footnotesize
\setlength{\tabcolsep}{2pt}
\renewcommand{\arraystretch}{1.08}
\caption{HAPS payload types mapped to the 3C paradigm (Competition–Complementarity–Cooperation).}
\label{tab:haps-payload-3c}
\centering
\begin{tabular}{%
|p{0.26\textwidth}|p{0.21\textwidth}|p{0.21\textwidth}|p{0.26\textwidth}|}
\hline
\textbf{Payload type} \&
\textbf{Functional definition (role)} &
\textbf{Competition} &
\textbf{Complementarity} &
\textbf{Cooperation} \\
\hline
\hline
\textbf{Transparent (``bent-pipe'')} 
Bent-pipe RF/FSO repeater; no baseband on board. Operates as high-altitude remote radio head; gateway/gNB on ground\cite{seamless_integrationLEO}. &
Relay-based coverage competing with LEO/TBS in rural/disaster/shadowed areas; can replace failed ground cells. &
Fills white-spots; hotspot offload; backhauls isolated cells using overlapping TN-NTN bands\cite{JAPCC_website}. &
Relay in Ground$\leftrightarrow$HAPS$\leftrightarrow$LEO chains; supports UE multi-connectivity. Inter-platform routing/ISLs generally need \emph{regenerative} payloads\cite{seamless_integrationLEO}. \\
\hline
\textbf{Regenerative (on-board gNB/eNB)}  demod–decode–route–remod. Acts as aerial cell/IAB; cuts RTT vs.\ distant gateways\cite{seamless_integrationLEO,sunglider-demo-regen}. &
Stratospheric cell competes with TBS and satellite user links for broadband/coverage. &
Offloads/extends LTE/5G; serves standard handsets over large, sparse regions. &
Enables inter-HAPS/LEO links, multi-hop backhaul, device multi-connectivity, and cooperative aerial meshes; ISLs/regenerative relaying noted in NTN docs\cite{seamless_integrationLEO}. \\
\hline
\textbf{Surveillance (remote-sensing)} 
EO/IR cameras, SAR/radar, SIGINT; persistent ISR/monitoring \cite{zephyr-another}.&
Alternative to ISR satellites/UAVs; higher persistence, rapid tasking from $\sim$20\,km. &
Complements satellites with continuous local monitoring between passes; higher-resolution regional views.&
Backhauls sensor data via SAT/TBS; supports ISAC and multi-HAPS roles (one sensing, one backhauling), e.g., Raven Aerostar\cite{haps-key-issues} \\
\hline
\end{tabular}
\end{table*}

\textbf{1) Methodology:} It can be achieved by multi-layer cooperation among terrestrial, aerial, and space layers. We can stack (a) TBS for dense urban coverage for high capacity and low-latency local service, (b) HAPS for rapid coverage expansion and mobility augmentation, and (c) LEO satellites for global backhaul and resilience. This framework enables cross-layer orchestration using AI-driven software-defined networking (SDN) or network function virtualization (NFV), and coordinated scheduling to dynamically route traffic across layers for optimal load sharing and seamless handovers. It also supports inter-layer frequency coordination and interference mitigation for efficient spectrum management.  
Beyond loose complementarity, cooperation also allows “layered access–backhaul” designs, where HAPS serve as access nodes for direct handset connectivity while LEO satellites provide the backhaul to the core network \cite{novelNTN}. Recent research has actively explored these concepts using software-defined HAPS relays to assist satellite communication for aircraft in flight \cite{TedHAPS}, optimizing joint resource allocation when HAPS and LEO share spectrum\cite{novelNTN}, and even running collaborative computing or learning tasks across HAPS and LEO node \cite{novelNTN}. Cooperative operation may also involve task sharing and load balancing: HAPS and LEO jointly serve the same region, splitting or handing off traffic in real time \cite{novelNTN}.

 \textbf{2) Pre-requisites:} The multi-layer nodes would require standardized protocols (3rd Generation Partnership Project (3GPP) NTN, International Telecommunication Union (ITU) frameworks) for seamless interoperability. The infrastructure also needs to be updated to introduce NTN compatibility in TBS, regenerative payloads in HAPS, and phased-array antennas on LEO satellites. Moreover, efficient resource management algorithms will empower predictive routing, congestion control, latency optimization, and quality of service (QoS) assurance.  
Enabling technologies include hybrid free-space optical (FSO) and radio-frequency (RF) inter-platform links. A concrete cooperative design is a HAPS–Earth Station Optical \ac{RIS}-routed FSO backbone coupled with a Simultaneous Transmitting and Reflecting RIS ground panel for dual indoor–outdoor access; this five-hop SAG topology has been analyzed with accurate closed-form performance and refined RF approximations in \cite{Shang2025HybridSTAROIRS}. RF backhaul (e.g. Ka/Q/V-band links) is robust to weather but has limited bandwidth and may require large antennas. Meanwhile, FSO links offer gigabit-per-second capacity and ultra-low latency (light travels faster and with less protocol overhead), potentially under 1–2 ms one-way \cite{TedHAPS}. Because HAPS operate above most atmospheric turbulence, LEO–HAPS FSO links avoid many impairments faced by direct satellite–ground optics. However, precision pointing and tracking are required between fast-moving LEOs and quasi-stationary HAPS \cite{TedHAPS}, and hybrid RF/FSO designs are often used for reliability\cite{muratOutage}. A related advance is the use of Optical Intelligent Reflecting Surfaces (OIRS) for cooperative non-terrestrial FSO links. In \cite{Shang2025OIRS}, a three-hop \ac{OGS}--HAPS--OIRS--user architecture was analyzed, showing that passive optical panels can dynamically redirect beams to maintain alignment and extend coverage under blockage or turbulence.

 \textbf{3) Deliverables:} This framework will deliver ubiquitous connectivity from dense urban to remote regions ensuring latency optimization (HAPS for time-critical applications and LEO for high-capacity backbone), improved availability (hybrid free space optics (FSO)/radio frequency (RF) backhaul), mobility support (high-speed users connect to HAPS instead of TBS), service continuity (seamless handovers), compatibility (standard handsets), and cost effectiveness (no dense TBS for hard-to-reach /sparse areas). The network resilience in this mode guarantees the service continuity even if one layer is unavailable (e.g., damaged TBS, congested HAPS), with multi-layer backup support.  
%more detials about the regen architectures with diagrams could be helpful here
In practice, this cooperation supports flexible payload configurations: transparent “bent-pipe” relays for low-complexity forwarding, regenerative payloads that decode and re-encode signals to function as aerial base stations \cite{TedHAPS}, and surveillance payloads for Earth observation and monitoring \cite{surveillancewebsite}, making them extremely versatile components in 6G NTN scenarios.

\subsection{Complementarity (distinct roles, loosely coupled)}
In the complementarity framework, inherent strengths of one segment overcome the weaknesses of other layers. TNs can offer high-throughput and low latency in dense regions but suffer from limited reach and costly expansion. Aerial platforms can fill coverage voids in underserved regions with rapid, flexible deployment but suffer from short endurance. Meanwhile, LEO satellites extend global footprints for unserved regions, but experience frequent handovers and user saturation. The complementary framework suggests strategic reorientation and disjoint coexistence  
for optimal performance. Each node is chosen to play the distinct roles in either geographic zones (urban-TBS, suburban-HAPS, rural/remote-LEO) or service zones (TBS for Ultra-reliable low latency communications (URLLC)/ internet-of-things (IoT), HAPS for flash crowds/monitoring, and LEO for navigation/surveillance/broadcasting), subject to its specialization. This synergy is reinforced by the 3GPP Release 19 NTN evolution and ETSI-ISG NTN, which advocate unified interfaces for terrestrial and non-terrestrial integration.  

 \textbf{1) Methodology:} It can be achieved by the location-based needs of a user as well as the demand profile in a particular area, and assigning the layer that performs the best as described earlier. Likewise, service differentiation between each node would allow for matching consumer requirements with the most suitable layer.  
HAPS and LEO can also complement each other by playing to their unique strengths in a coordinated way. Rather than both serving the exact same users, each platform can focus on what it does best, covering distinct regions or use cases that the other cannot optimally handle \cite{wide_coverage_LEO}. For example, a HAPS is ideal to provide intensive coverage over a small high-demand area (such as an urban hotspot, a dense event venue, or a disaster zone) where low latency and real-time responsiveness are crucial, while surrounding sparsely populated areas can be serviced by LEO satellites designed to blanket large swaths of the Earth \cite{hapsalliance}. In this way, a hand-off or partitioning can occur: the HAPS delivers connectivity in the core area (or during peak times), and LEO fills in everywhere else. HAPS can also be rapidly repositioned or deployed on-demand to areas with temporary needs (e.g., post-disaster connectivity or special events), complementing the always-on global presence of a LEO constellation \cite{JAPCC_website}.

 \textbf{2) Pre-requisites: }The complementarity requires the knowledge of users/service demands to carry out the cloud-based management for traffic routing and zone allocation (distinct geographic or service zones for node deployment/assignment). The user devices demanding specific services need to be compatible with multi-band/multi-mode support (TBS/HAPS/LEO). Moreover, the spectrum and policy alignment are essential to coordinate across TN and NTN to prevent overlap and interference between layers \cite{wang2025non}.  
The means to enable this complementarity include dynamic network orchestration: the system must intelligently direct users to either the HAPS or the satellite depending on signal quality, load, and service needs. This raises research challenges in handover coordination, multi-tier resource allocation, and interference management. Ensuring a seamless user experience as devices move in and out of a HAPS’s footprint (or as the HAPS itself moves) requires careful planning. Regulatory alignment is also required so that HAPS and LEO spectrum use can be harmonized or non-interfering in complementary operation \cite{regulationZhou}.

\begin{figure}[t]
    \centering
   \includegraphics[width=1\linewidth]{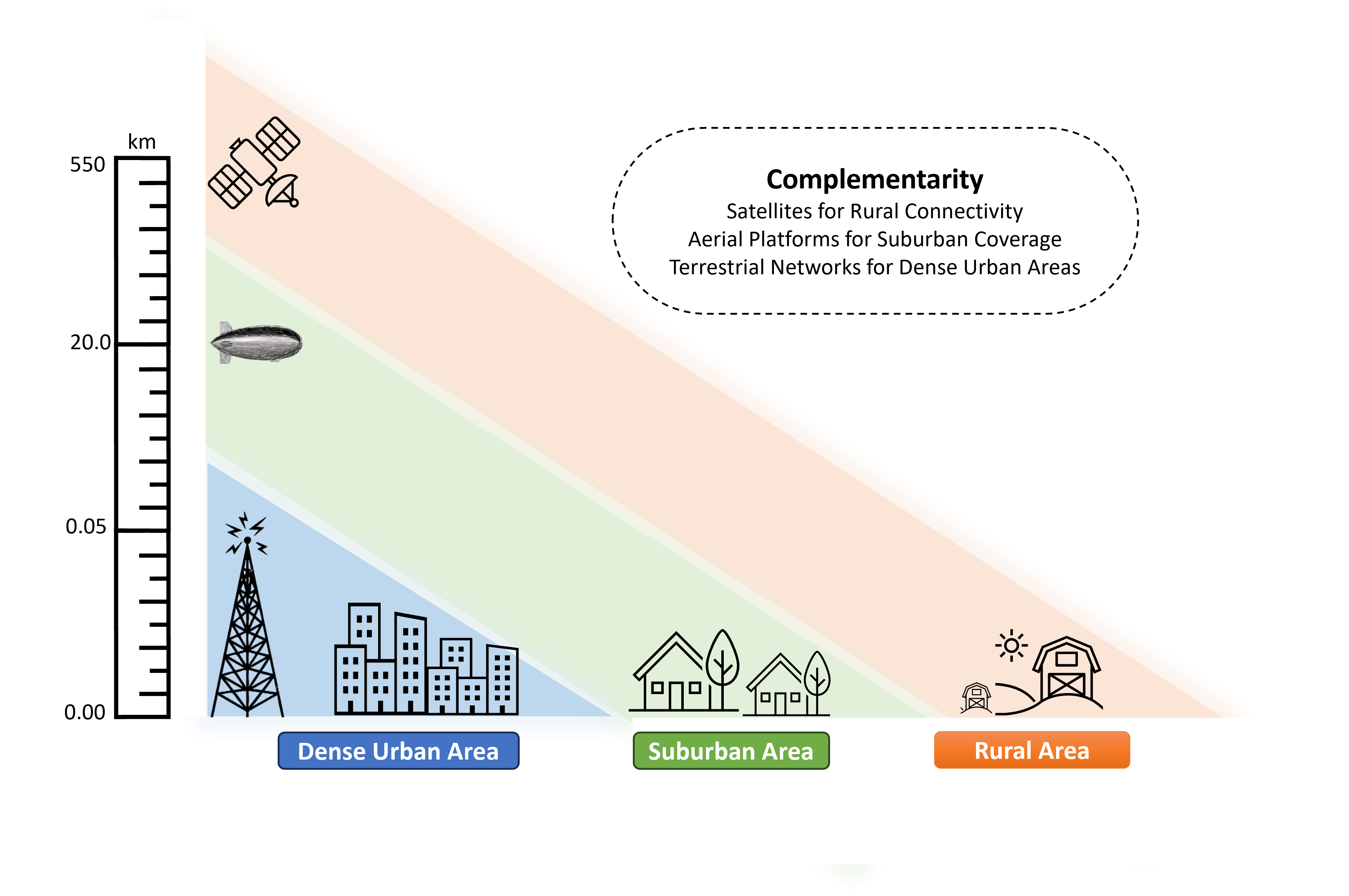}
   \vspace{-1.2cm}
    \caption{Illustration of Geographic Complementarity}
    \label{fig:3Cs_Geographic}
\end{figure}

\textbf{3) Deliverables:} The complementary framework ensures optimal utilization of each layer instead of overbuilding infrastructure by letting each node handle its niche \cite{wide_coverage_LEO}. This reduces the redundancy and deployment cost in rural/remote regions while maximizing returns in dense zones. In particular, users experience enhanced QoS by receiving the right service from the right platform. This approach promises to exploit performance diversity (low-latency links from HAPS and high-capacity links from LEO) while maintaining economic pragmatism by using the right tool for the right job. It increases overall network resilience if a satellite service is weak or overloaded in one spot, a HAPS can be dispatched to assist, and vice versa. Studies confirm that combining tiers often outperforms either alone and integrating LEO and HAPS yields better throughput and user coverage than using only one platform \cite{tier_is_better,wide_coverage_LEO}.
A representative example of geographic complementarity between urban TBS, suburban HAPS, and deep-rural or maritime LEO deployments is sketched in Fig.~\ref{fig:3Cs_Geographic}.

\subsection{Competition (either HAPS or LEO for the same use case)}
\begin{figure}[t]
    \centering
   \includegraphics[width=1\linewidth]{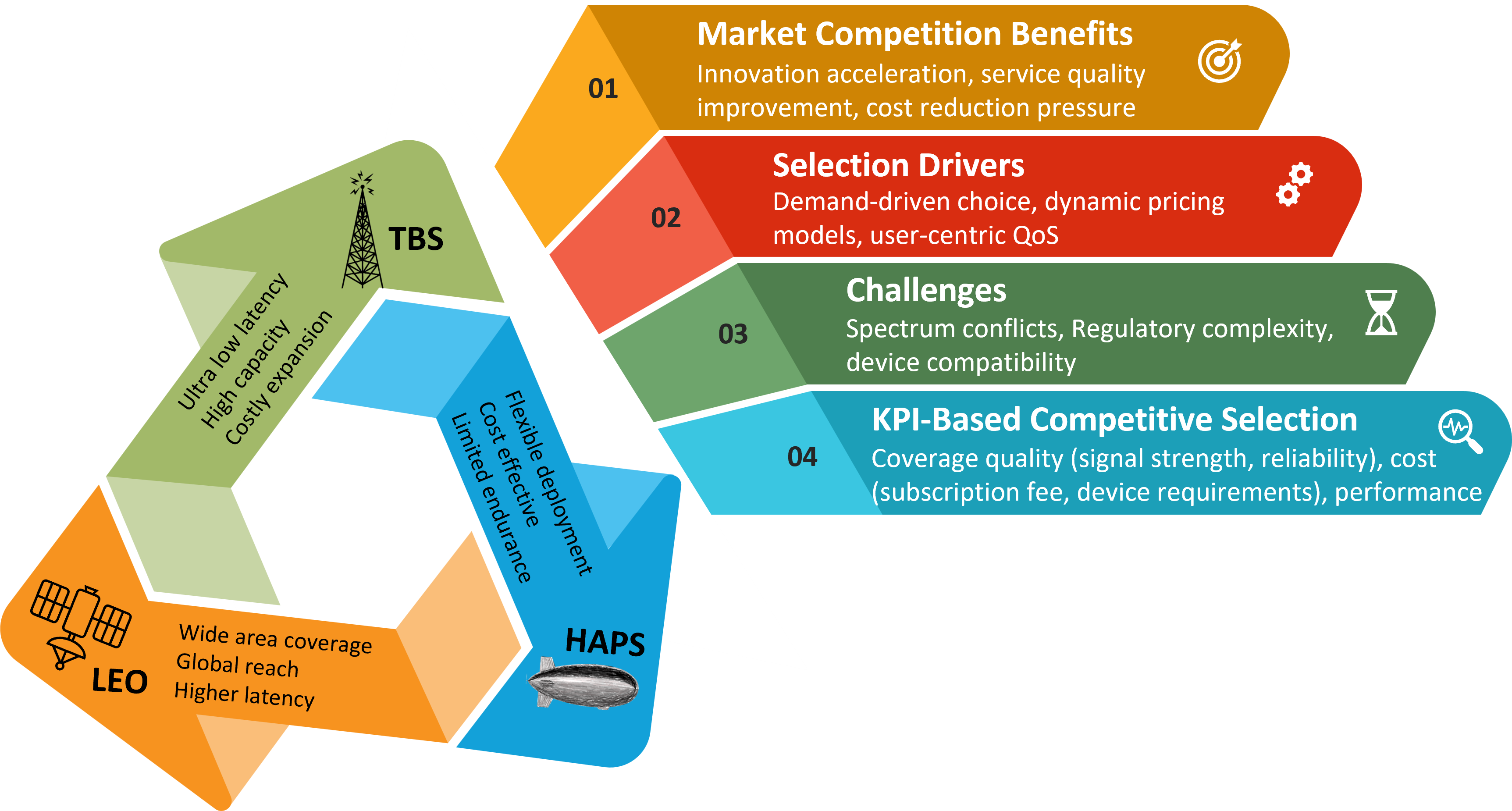}
    \caption{Suburban Coverage Competition and Selection Criteria}
    \label{fig:compFig}
\end{figure}

The competition within the Triple-C framework does not imply conflict but adaptive performance. It can be seen as a dynamic interplay where each entity autonomously negotiates access to limited spectrum or coverage opportunities while maintaining global network harmony. 
TBS densification is the preferred choice for reliable connectivity in dense urban areas. On the other hand, LEO is suitable for deep rural areas because of its wider footprint. However, the main competition arises in suburban areas where TBS renders low-latency communications yet a costly expansion, LEO satellites offer wider coverage but suffer from over-subscription (leading to lower user capacity and higher charges). The competition arises from HAPS as an optimal KPI-compliant option in terms of latency, cost, coverage, throughput, and direct-to-device reach. In essence, HAPS are the preferred choice from the prism of latency and cost-sensitivity, whereas LEO satellites are favorable for wide-area coverage. Such competitive yet mutually beneficial behavior enhances spectrum utilization, energy efficiency, and service continuity, which are the key enablers of sustainable 6G communications. Fig.~\ref{fig:compFig} summarizes this suburban competition, showing how TBS, HAPS, and LEO options feed into a KPI-based selection process driven by market benefits, selection drivers, challenges, and KPI-based criteria.

 \textbf{1) Methodology:} This framework relies on the demand-driven selection, where users/operators choose among TBS, HAPS, or LEO based on the coverage, cost, and QoS requirements. It allows them to select the best-performing layer, which is competing for the same service region without imposed coordination. In practice, a rural broadband user could be served either by a HAPS-based platform or by a LEO satellite constellation, with both systems striving to meet the same connectivity needs \cite{wide_coverage_LEO}. The proximity of HAPS also allows it to provide higher data rates and lower latency in many cases \cite{hapsalliance}. On the other hand, LEO satellites excel in wide-area reach and capacity, covering entire countries or oceans continuously, whereas a single HAPS cannot \cite{wide_coverage_LEO,JAPCC_website}.

 \textbf{2) Requirements:} Network operators need to design pricing and policy models to attract users with competitive prices and service-level agreements. Healthy inter-layer competition demands spectrum policy and independent investments from mobile/HAPS/satellite operators. Additional challenges include spectrum conflicts if HAPS and LEO operate in similar bands, as well as regulatory hurdles (HAPS airspace falls under aviation rules, unlike satellites in outer space) \cite{JAPCC_website}. Performance benchmarking of real-world conditions is also required to determine when a HAPS vs. LEO link is advantageous \cite{hapsalliance}.

 \textbf{3) Deliverables:} Competition introduces a user-centric approach and market competition (spanning innovation, service diversity, cost reduction, performance improvement). Each region receives the most efficient service based on the target KPIs, pushing each operator to improve its strengths. HAPS compete with lower cost and regional agility—a techno-economic study estimated HAPS ownership costs at about \$4.4M, compared to \$1.5M per LEO satellite (multiplied across hundreds of satellites for global service) \cite{wide_coverage_LEO}. By contrast, LEO offers higher aggregate throughput and global scaling, with commercial constellations (e.g., Starlink) delivering hundreds of Mbps per user \cite{wide_coverage_LEO,hapsalliance}. Ultimately, this competitive tension drives innovation: LEO reduces latency with optical inter-satellite links, while HAPS boosts capacity with advanced antenna designs benefiting users through better and cheaper connectivity options.

In summary, this section has recast TBS–HAPS–LEO integration through the lens of Triple-C, showing that HAPS are neither simple gap-fillers nor mere satellite auxiliaries, but active players whose role shifts with latency, coverage, capacity, and cost requirements. By explicitly characterizing when each layer should dominate, when it should support or offload others, and when it inevitably competes for users and spectrum, the Triple-C view turns a loosely coupled stack of segments into an operational design space.

\section{Multidimensional Enablers for Triple-C Interoperability}
Previous generations treated TBS, HAPS, and satellites as isolated systems, emerging 6G/NTN designs increasingly view them as components of a single space–air–ground fabric.
Realizing the Triple-C (Complementary, Cooperative, and Competitive) interaction across terrestrial, aerial, and space segments hinges on an enabling multidimensional framework spanning across physical, logical, and cognitive layers. These enablers will create a policy-driven self-organizing adaptive ecosystem for sustained interoperability across heterogeneous networks.  

\subsection{Physical and Topological Enablers}
The physical topology and interaction among heterogeneous platforms are defined by the infrastructure skeleton of link-level inter-connectivity among different nodes. 

\begin{enumerate}
\item \textbf{Ad-hoc and Mesh Networking} are the elementary architectures for TBS--HAPS--LEO connectivity. The infrastructure-independent ad-hoc connectivity allows nodes to forward traffic to the nearest stable link through self-organizing Space--Air--Ground Integrated Networks (SAGIN) \cite{chengSAGIN2023, lyuLEOMesh2024}. In mesh networking, a structured multi-hop topology connects all layers to form a self-healing network. 
In a cell-free NTN paradigm, ground stations, HAPS, and LEO satellites jointly serve users without rigid cell boundaries.  Early studies demonstrate that applying strategies such as rate-splitting or network coding to joint HAPS--LEO--terrestrial resource allocation can significantly enhance throughput compared to single-layer systems \cite{wide_coverage_LEO}.
Researchers have analyzed HAPS operation in
mesh-style topologies, ell-free coordinated deployments, or integrated access and backhaul (IAB) frameworks \cite{lou2023}.
For example, space-air-ground (SAG) architectures with HAPS interconnected by optical terminals are projected to provide terabit-per-second throughput for 6G traffic \cite{singh2024,samy-fso}, underscoring the role of stratospheric nodes in extending satellite throughput and resilience.
The ad-hoc meshes sustain terrestrial outages through adaptive link discovery and distributed routing, which is critical for complementary resilience and competitive self-healing.

\item \textbf{Integrated Access and Backhaul (IAB)} is a 3GPP-defined framework and can be extended to NTN for resilience \cite{3gpp38874}. It allows terrestrial, HAPS, and LEO nodes to dynamically share access and backhaul responsibilities using the same radio resources. In case of backhaul failure, terrestrial gNBs can reroute to HAPS or LEO IAB for resilient backhaul. Likewise, a HAPS serving as a super-macro relay can bridge isolated terrestrial nodes by offering backhaul to LEO \cite{ghasemi2024}. Moreover, regenerative LEOs offering D2C connectivity to ground users can backhaul to ground gateways. Multi-hop IAB chains (e.g., gNB → HAPS → LEO → gateway) are critical for dynamic multihop routing, spectrum efficiency, load distribution, cooperative capacity scaling, and seamless offloading in a resilient hierarchical architecture. 

\item \textbf{Multi-band, Multi-RAT Spectrum Framework} allows adaptive spectrum management of sub-6GHz, Ka/Q/V bands, mmWave, THz, and optical wavelengths as per capacity and coverage trade-offs \cite{ituM2541, wildTHzProcIEEE2023}. Lower frequencies offer wider coverage, whereas higher frequencies deliver higher throughput for complementary synergies. THz are anticipated to be feasible for co-design of ultra-capacity backhaul and sensing-communications \cite{ituM2541}.
Likewise, optical links offer multi-Gbps capacity and low latency for competitive high-throughput data transport in space–air–ground FSO networks \cite{samy-fso}, while RF ISLs ensure service continuity under turbulent atmospheres \cite{kaushalFSO2017, rfc8986}. Thus,  hybrid RF/FSO schemes offer resilience against weather while delivering high capacity \cite{singh2024}. Multi-band backhaul solutions using millimeter-wave and free-space optical (FSO) links are critical. Adhoc multi-RAT switching between LEO/HAPS/TBS nodes allows self-organization based on CSI, topology changes, or mission requirements \cite{chengSAGIN2023, uavSAGINreview2025}.

\item \textbf{Spectrum Sharing and Resource Management} requires dynamic sensing, allocation, and trading of radio spectrum, bandwidth, and power amongst multifaceted tiers. Dynamic spectrum sharing empowered by real-time federated spectrum sensing, AI-driven optimization, and shared databases can reassign channels across TN and NTN bands for complementary mitigation of congestion/interference
\cite{flIotSurvey2024}. Spectrum resources can also be automatically coordinated following the Citizens Broadband Radio Service (CBRS) exemplar \cite{fccCBRS} and cross-band agility across FR1/FR2 and Ka/Q/V bands \cite{ituM2541}. Another key enabler could be the cognitive radio scheme and spectrum-as-a-service APIs. They introduce context-aware leasing models, where idle terrestrial bands can be temporarily allocated to HAPS/LEO beams under cooperative sharing agreements.
Lastly, blockchain-enabled spectrum markets can enforce transparent auctions and auditable spectrum trading contracts, shifting competition into an efficient and regulated process \cite{blockchainDSS2024}.

\item \textbf{Direct-to-Device (D2D/D2C)}  connectivity permits a standard user device to communicate directly to HAPS/LEO, bypassing terrestrial dependence \cite{3gpp38811, 3gpp38821, hapsAlliance2023}. Terrestrial and aerial connectivity assumes static and quasi-stationary gNBs. However, LEO D2C connectivity requires precise delay/Doppler pre-compensations owing to its high mobility. These pre-compensations are currently carried out at the LEO satellites for standard user equipment (UE), leading to overheads while limiting the number of active subscribers. However, future UEs with NR-NTN capabilities are expected to compute and precompensate for delay and Doppler shifts with LEO mobility. In essence, D2C connectivity enables service reach to remote users and enables traffic offloading during terrestrial congestion. 

\item \textbf{Antenna Design Configurations} are vital for service continuity despite the high mobility of NTN nodes. Dynamic beamforming and beam-handover coordination are enabled by the advancements in electronically steerable phased antenna arrays and predictive tracking algorithms, respectively, during fast-paced LEO passes or HAPS station-keeping flights \cite{ts38300, joBeamMobility2021}. These capabilities permit LEO/HAPS/TBS nodes to dynamically adjust their beams in real-time and precoordinate the handovers before a beam drifts out of range. This achieves cooperative mobility management, preserves service continuity, and facilitates aggressive spatial-frequency reuse across three tiers.   
Demonstrations already exist of Ka-band phased arrays (26--31 GHz) with substrate-integrated waveguide elements and multi-channel beamformers tailored for HAPS operation \cite{planarSi}. Distributed MIMO techniques, where multiple HAPS form a cooperative virtual array, are also proposed to enhance coverage and spectral efficiency \cite{dong2015}. These antenna innovations enable dynamic spectrum reuse and agile beam steering across cooperative, complementary, and competitive modes of integration.

\begin{figure}[!t]
    \centering
    \includegraphics[width=1\linewidth]{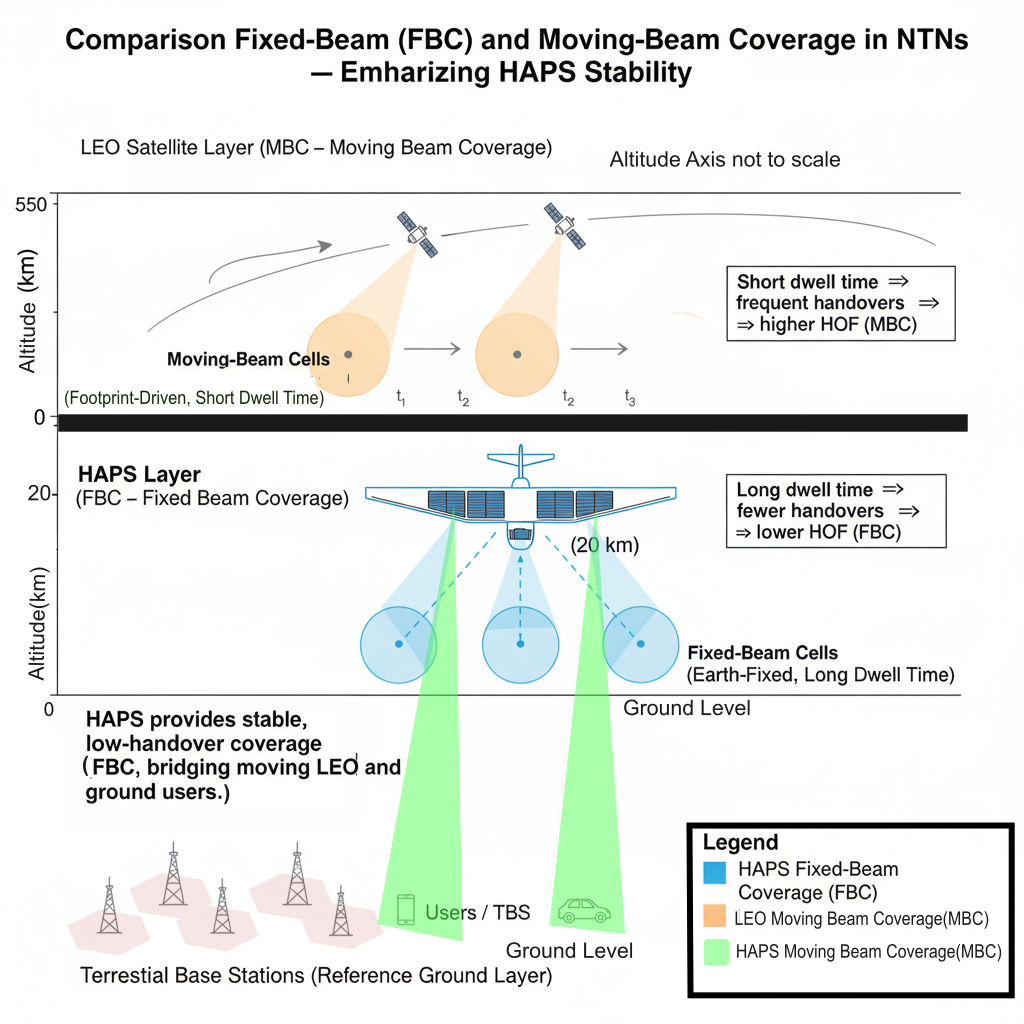}
    \caption{Comparison of fixed-beam and moving-beam coverage. }
    \label{fig:FBC_MBC}
\end{figure}

\item \textbf{Fixed/Moving Beam Cells:} Recent very Low Earth Orbit (vLEO) work distinguishes \emph{Fixed-Beam Cells (FBCs)}—earth-fixed virtual cells realized by dynamic beam steering and fast inter-satellite coordination—from \emph{Moving-Beam Cells (MBCs)}, in which the cell moves with the satellite footprint \cite{fo_mobility_survey_2025}. In the proposed cooperation strategy,  FBCs are natural anchors for multi-layer handovers (LEO$\leftrightarrow$HAPS$\leftrightarrow$TBS). Earth-fixed cells decouple access mobility from satellite motion, simplifying inter-layer orchestration and shortening dual-connectivity (DC) handovers.
For complementarity, we employ FBCs over high-demand zones (urban cores, events) while MBCs blanket sparse surroundings. Cell boundaries then align with service zones rather than satellite tracks. The contrasting footprints and handover behavior of FBCs and MBCs are illustrated in Fig.~\ref{fig:FBC_MBC}.
While competing to offer services to the same device, expected dwell time and handover failure probability (HOF) differ across FBC (longer, earth-fixed) vs.\ MBC (shorter, footprint-driven). Thus, users/operators can select the layer/cell type with the best KPI trade-off. Although \cite{fo_mobility_survey_2025} targets vLEO, the concepts can be applied to HAPS. The quasi-stationarity natively realize earth-fixed beams (FBC-like) over tens–hundreds of km, serving as cooperative anchors for LEO DC handovers; MBC-like operation is less needed at HAPS due to negligible footprint motion.

\begin{table*}[t!]
\centering
\caption{Enabling Technologies in 3Cs Paradigm}
\label{tab:HAPS_LEO_TBS}
\renewcommand{\arraystretch}{1.2}
\begin{tabular}{|p{2.5cm}|p{4.2cm}|p{5cm}|p{4.5cm}|}
\hline
Enabling Technologies	& Cooperative Mode  (integrated layers)	& Complementary Mode (layer covers its niche) &	Competitive Mode (Layers compete) \\
\hline
\hline
Frequencies (FSO/RF/mmWave) &	Multi-band operation \cite{multi_band,nokia_multiband}, dynamic spectrum sharing \cite{dynamicspectrum_sharing_website,AI_sharing}, hybrid RF/FSO backhaul \cite{hybrid_transmitter} 	& Frequency partitioning by domain (TBS = sub-6GHz, HAPS = mmWave, LEO = Ka/Ku/FSO)& 	Independent spectrum allocations \cite{intelligent_spectrum}, KPI-driven frequency selection for cost/latency advantages \\
\hline
Antenna Design	& Phased-array, steerable multi-beam antennas for seamless inter-layer handovers \cite{phased_aray_beamforming}	& Adaptive antennas optimized for specific urban/rural/sparse zones. (LEO=Wider beams, HAPS=Narrower Beams) \cite{5g_array,beam_tracking} 	& Antenna systems optimized per platform (TBS = small cells, 
HAPS = steerable beams, LEO = phased arrays) \\ \hline
Architectures (Cellular, Ad-hoc, Cell-free, IAB) &	Integrated cellular + integrated access and backhaul (IAB) with SDN/NFV orchestration across all layers \cite{6gnet_vision,6g_service} &	Partitioned coverage: 
TBS = urban grids, 
HAPS = hotspots, 
LEO = sparse/remote \cite{dicandia2022}, possible cell-free overlays & Independent cellular networks, no enforced integration, market/KPI-driven access \\ \hline
Payloads &	Regenerative payloads on HAPS/LEO for onboard processing and inter-layer traffic management \cite{flexSAN}	& Specialized payloads per platform (HAPS = Environmental sensors, 
LEO = Surveillance/Imaging Modules \cite{leo_imaging} TBS = metro capacity communication module)&	Differentiated payloads used as competitive advantage (HAPS = LoS direct-to-device \cite{haps_los}, LEO = inter-satellite routing \cite{leo_routing}, TBS = dense MIMO)
\\ \hline
\end{tabular}
\end{table*}
\item \textbf{Power Management:} TNs enjoy abundant power supply, while NTNs require immaculate power management to complete their operation. Satellites are mostly solar powered and utilize it for payload with negligible propulsion requirements due to Newtonian orbital motion. On the other hand, HAPS require high power for propulsion and payloads which cannot be solely harvested from solar irradiance due to limited on-boeard storage and long night cycles relative to fast-paced orbiting LEOs (completing 16 day-night cycles per day). Suitable power sourcing and mnaagement is essential for long-endurant HAPS operation. Solar photovoltaics and batteries help to store excess daytime energy (~200 W) for nocturnal operation of HAPS \cite{javed2023interdisciplinary}. 
Liquid hydrogen fuel cells can supplement solar energy to deliver higher sustained power (~20 kW) to support larger payloads and  eliminate the night-time power gap \cite{beast_haps}. 
Laser or wireless power-beaming has also been proposed as a third approach to sustain flight reducing onboard energy storage requirements \cite{laser_feasibility_RAND}. Startups such as Orbiting Grid are advancing optical wireless power transmission for space, developing scalable high-power laser emitters and planning the deorbit and attitude demonstrator (a 200 kW laser LiDAR) to validate space-based laser operations and realize the “Skylight” vision of an orbital energy grid that beams solar power to electric aircraft and propulsion tugs \cite{orbital_fso}.
 \end{enumerate}

Table \ref{tab:HAPS_LEO_TBS} presents a clear crystallization of physical enabling technologies across Cooperative, Complementary, and Competitive modes of TBS–HAPS–LEO integration.
In particular, breakthroughs in antenna designs, architectures, payloads, and multi-RAT operations have been instrumental in realizing Triple-C for complex applications \cite{karabulut_halim}.

\subsection{Network, Control, and Management}
Management and orchestration of heterogeneous networks require more than isolated configuration mechanisms. They demand a standardized, interoperable, and intelligence-driven coordination framework. The intrinsic diversity of TBS--HAPS--LEO systems (operational characteristics, mobility patterns, resource constraints, coverage footprints, and administrative ownership) makes integrated orchestration a fundamental challenge to achieve seamless end-to-end connectivity.  

\begin{enumerate}
\item \textbf{Network Slicing and Network Function Virtualization} across Terrestrial--Aerial--Space dedicates mission-specific virtual networks to each layer. For example, one slice for emergency communications from HAPS, another slice for maritime broadband from LEO. It promotes cooperative resource management, service orchestration, lifecycle management, performance control and fault detection for end-to-end management and orchestration among competitive tenants \cite{ts28530}. Moreover, virtualized core functions on HAPS/LEO cloud platforms enable redundancy and rapid recovery during terrestrial failures \cite{etsiNFV002, etsiNFVSOL}. 

\item \textbf{Software-Defined Networking (SDN) Controllers} with decoupled control plane (decision-making) and data plane (packet forwarding), centrally/distributedly manage heterogeneous networks with global visibility of the network states (e.g., topology, congestion, link conditions, energy levels, LEO ephemeris, HAPS beams). This equips controllers to perform cross-domain path optimization and policy coherence across TBS--HAPS--LEO segments through open programmable interfaces \cite{etsiNFV002, etsiNFVSOL}. 
SDN can facilitate seamless mode transitions among cooperative, complementary, and competitive operational modes. Programmable SDNS can orchestrate multi-hop overlays and coordinated handovers to guarantee end-to-end QoS objectives under rapid NTN topology variations. 

\item \textbf{Service-Based Architecture (SBA)} expands 3GPP extension to NTN for cooperative API-driven service composition of control-plane functions across layers \cite{3gppNEF29522}. Through the service exposure framework, SBA authorizes each segment (terrestrial gNBs, HAPS controllers, and LEO gateways) to exchange their service information, i.e., capabilities, constraints, mobility events, and QoS states in real-time \cite{ts38300}.  For instance, in cooperative mode, segments can exchange their load, availability, congestion, and predictions to jointly compose multi-tier services for consolidated session management, multi-path anchoring, and harmonized policy enforcement. In complementary mode, SBA allows a degraded/congested entity to request fall-back routing, augmented capacity, or emergency QoS boosts from other entities for seamless service migration. In competitive mode, SBA preserves operators' autonomy while enabling controlled interworking for fair competition. 

\item \textbf{Integrated Multi-Access Edge Computing (MEC)} can revolutionize the integration of gNBs, HAPS, and LEO gateways. It provides a computational substrate for localized decision-making 
for latency-critical tasks \cite{etsiMEC003}. The computational and storage resources on TBS/HAPS/LEO MECs accommodate cross-segment coordination (e.g., cooperative caching, multipath routing optimization, cross-layer interference estimation, mobility predictions, and distributed AI inference), ensuring cooperativeness with seamless end-to-end performance. 
During the complementary phase, MEC allows one entity to temporarily augment/substitute another during unforeseen events. For example, LEO MEC substituting degraded ground core functionalities or HAPS MEC as a buffer for task offloading and mobility anchoring, preventing service discontinuity. In competitive mode, MECs can host differentiated services, proprietary optimization engines, enhanced security functions, etc, to gain/maintain a competitive advantage.
Furthermore, end-to-end service mesh fabric links MECs through east-west control channels for consistent policy enforcement, telemetry exchange, and load balancing between domains \cite{etsiMEC003, nistZTA2023}.

\item \textbf{Cross-Domain Orchestration \& Federation} defines the unified management of resources, services, policies, and network functionalities across multi-tiers. Orchestration framework treats TBS--HAPS--LEO as a federated policy-driven ecosystem to realize Triple-C architecture, while guarding multi-operator sovereignty \cite{etsiZSM}. Orchestrator acts as a semantic bridge translating high-level intents (latency targets, coverage expansion, QoS requirement) into domain-specific actions through cooperative policy dissemination \cite{etsiZSM, etsiMEC003}. In cooperative mode, federation allows joint optimization of 
routing, spectrum, caching, and service placement for coordinated policies across segments. In complementary mode, the orchestrator dynamically steers traffic to healthier segments during unfavorable events (surge capacity or emergency backhaul), ensuring core requirements (continuity, redundancy, and sustainability) of resilient Triple-C networks. In competitive mode, orchestration and federation allow multi-operator coexistence with autonomy over their own resources, but the dissemination of policy-driven capabilities. This prevents monopolization while enabling fair market-based competition. The coordination layer transforms three disjoint segments into an economically viable Triple-C ecosystem, enabling dynamic cooperation, graceful complementarity, and structured competition across terrestrial, aerial, and orbital layers. 
\end{enumerate}

\subsection{Cognitive and AI-Native Enablers}
The triple-C paradigm may be seen as a static framework, however,  cognitive and AI-native enablers can realize it as a dynamic switching protocol depending on the network functionalities and user demands. 
The maximum benefits of cross-layer integration can only be reaped with an intelligent Triple-C framework.  This will enable continuous perception, reasoning, predictions, network adaptation, autonomous decision-making, and knowledge exchange across domains.

\begin{enumerate}
%intent-to-action (network intelligent adaption)
\item \textbf{AI-Native Network Cognition Plane} embeds distributed machine-learning (ML) and artificial intelligence (AI) agents at the terrestrial-gNBs, HAPS platforms, LEO gateways, and sometimes on-board satellite processors. They analyze raw telemetry data (link quality, network congestion, failures, orbital dynamics, interference maps, user mobility, and resources) to provide actionable insights. The network data analytics functionality provides real-time and context-aware performance monitoring, resource allocation, predictive maintenance, and autonomous optimization \cite{liAINative2023}. For cooperation, it supports shared intelligence among segments, allowing coordinated routing, synchronized beam steering, cooperative load balancing, and multi-hop path stitching across TBS--HAPS--LEO segments beyond siloed control. From a complementary perspective, AI agents can detect faults, jamming, outages, weather degradation, or traffic surges in one node to autonomously trigger assistance from other nodes. For instance, predicted terrestrial overload can plan proactive offloading to HAPS/LEO; emerging rain fade on a HAPS link can prompt support from nearby TBS. The cognition plane offers a technological leap by providing anticipatory resilience, instead of reactive fallback. In competition, AI agents can support each operator to optimize performance while respecting federated constraints, slice boundaries, and service level agreements (SLA) isolation \cite{hexaXAIOrch2024}. Overall, the cognitive plane acts as an intelligent core to sense, interpret, and adapt network behaviour across heterogeneous entities. 

\begin{figure}[!t]
    \centering
    \includegraphics[width=0.65\linewidth]{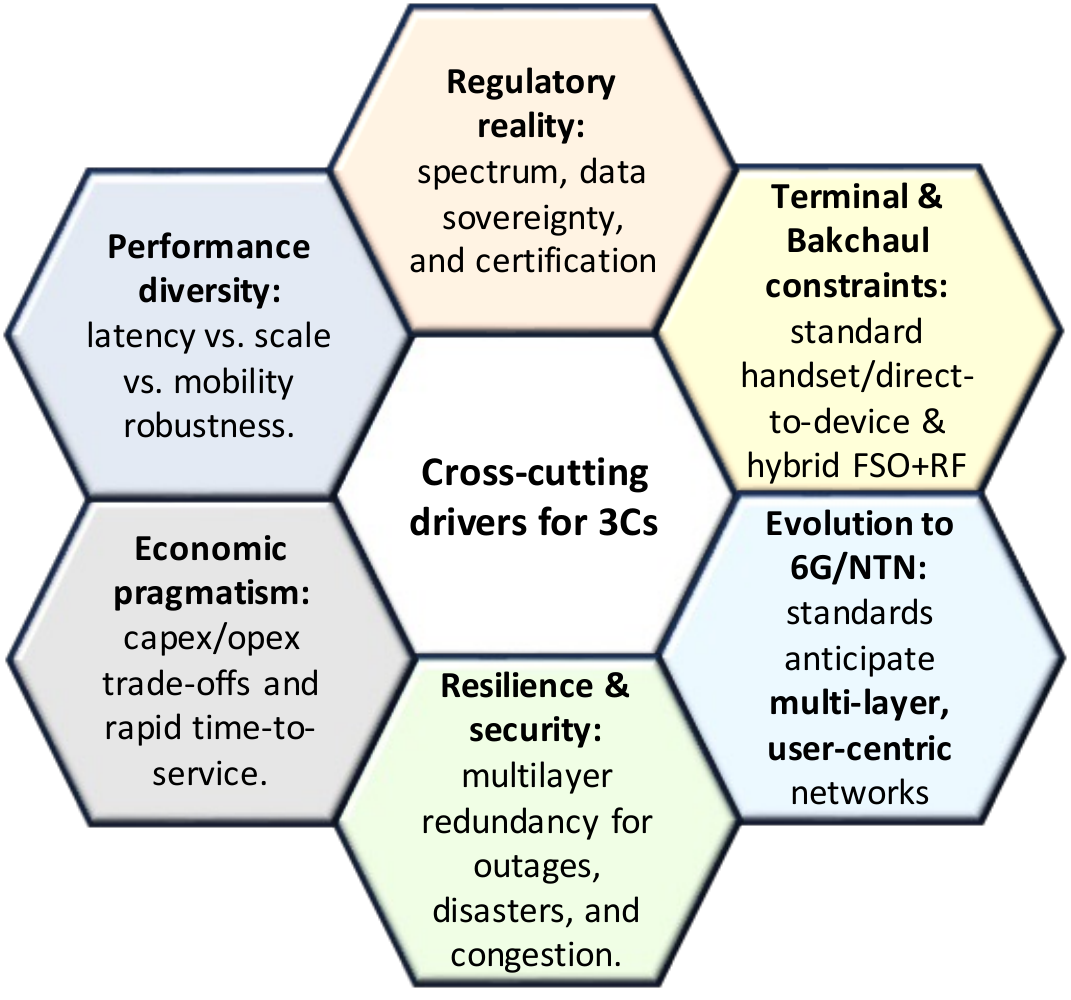}
    \caption{Key Drivers of Triple-C Paradigm}
    \label{fig:3C_drivers}
\end{figure}

%Knowledge Exchange
\item \textbf{Federated Learning (FL) \& Distributed Inference} enables TBS, HAPS, and LEO nodes to locally and independently train their models, exchanging model parameters across domains \cite{flIotSurvey2024}. This approach preserves privacy, sovereignty, and regulatory compliance. The cross-domain knowledge is shared using a unified semantic representation of network entities, topologies, environmental conditions, spectrum usage, and operational constraints for shared situational context \cite{serviceOrientedSAGIN2022}. In cooperative function, FL enables segments to collectively improve prediction models for situational awareness by aggregating insights from geographically and topologically diverse nodes. Moreover, nodes with limited compute can seek inference support from other capable nodes. In the complementary state, FL supports the transfer of robust intelligence from healthy segments to impaired ones. In case of terrestrial degradation, HAPS/LEO can infer fallback routing, schedule emergency capacity, or even recommend service migration. FL can also balance competitive autonomy with constrained interoperability, aligning perfectly with the Triple-C philosophy. 

\item \textbf{AI-Enhanced Resource Allocation} provides predictive intelligence to coordinate proactive resource allocation  \cite{liAINative2023, flSpectrum2023}. The deep forecasting models transform multi-layer telemetry (interference trends, link quality, congestion probabilities, HAPS energy trajectories, and LEO beam drifts) into anticipatory resource control for distributed power/bandwidth/routing using Multi-Agent Reinforcement Learning (MARL) \cite{marlSAGIN2024, marlOffload2025}. For a cooperative strategy, AI predictions align resources to tackle an anticipated event. AI agents can design a globally optimal, not locally reactive, resource coordination, i.e., spectrum reuse patterns, optimal transmit power budgets, and multi-hop routes under latency–throughput–energy trade-offs. In complementary operation, AI predictions support load distribution and fallback routes for proactive resilience. If predictions show that HAPS is entering a low-energy window or LEO suffers atmospheric degradation, the remaining nodes can proactively allocate resources to maintain service continuity. Moreover, forecasting cross-layer contention, future congestion hotspots,  power limitations, and potential spectrum collisions, operators can make informed decisions about traffic offloading, spectrum occupancy, slice allocation, and bidding for healthy competition. 
\end{enumerate}
In summary, emerging hybrid architectures show that satellites, HAPS, and terrestrial systems can already be woven into interoperable space--air--ground fabrics, from ad-hoc relaying to tightly integrated IAB and cell-free operation. Advanced enablers such as multi-band RF/FSO backhaul, electronically steered multi-beam arrays, SDN/NFV-based orchestration, and cognitive plane intelligence provide the knobs to shape how each layer participates in end-to-end service delivery. 
These ingredients collectively demonstrate that TBS, HAPS, and LEO need not be designed in isolation; instead, through the interplay of mutually reinforcing enablers that span physical, logical, and cognitive layers. This enables a self-adaptive hybrid fabric of the terrestrial–aerial–space continuum to achieve the key 6G objectives of global, intelligent, and sustainable connectivity.

\section{Use Cases and Focus Applications }
Advancements in the aforementioned enabling technologies will unlock the true potential of Triple-C paradigm, paving the way for a broad spectrum of transformative and futuristic use cases. Far beyond conventional TN-NTN interoperability, their integration will support emerging digital, industrial, and mission-critical applications.  This section presents diverse use cases and focus applications unlocking new capabilities. Each scenario demonstrates how coordinated, adaptive, and policy-driven interactions across TBS--HAPS--LEO translate to measurable service improvements. It reinforces Triple-C as a foundational design principle for the next generation communications. 

\subsection{Use Cases}
The employment strategy describes the use-case driven analysis to decide the favorable integration role from 3Cs. The role selection criterion is dictated by KPIs. We present numerous use cases and suggest  the most suitable 'C' from 3Cs to fulfil its most critical requirement. These use cases are visually presented in Fig. \ref{fig:UseCases}.

\begin{enumerate}
\item	\textbf{Emergency communications (demands rapid deployment):} Complementarity - HAPS deliver broad, direct-to-device, and pop-up coverage within hours/days enabling rapid capacity injection for events or disasters ensuring operational agility. Aerial links  also improve energy-per-bit efficiency as opposed to the long orbital links \cite{not_promised, haps_arch_alliance}. In contrast, LEO satellites can provide backhaul when TNs are impaired. Each platform complements the role of the other to ensure continuous services.
\item	\textbf{Tactile networks (demands ultra-low latency):} Complementarity – This use case requires ultra-low latency and reliability from localized and trusted nodes. HAPS can provide consistent and regionalized links while TBS handles dense local access. LEO may provide encrypted long-haul links, but they are not suitable for ultra-low latency and reliability needs. Load sharing allows tactile-grade performance.
\item	\textbf{Flash crowds (demands temporary link):} Complementarity – HAPS can offload congested terrestrial cells and saturated satellite beams, adding on-demand capacity and improving service continuity where LEO satellites have limitations due to orbital dynamics \cite{joint_aol}; operating above weather also helps reliability on feeder/access links (with appropriate band selection and fallbacks) \cite{not_promised}.
On the other hand, LEOs can support broadcast or traffic offloading beyond the HAPS footprint. Complementarity allows scaling without overbuilding TBS.

\begin{figure}[t]
    \centering
   \includegraphics[width=1\linewidth]{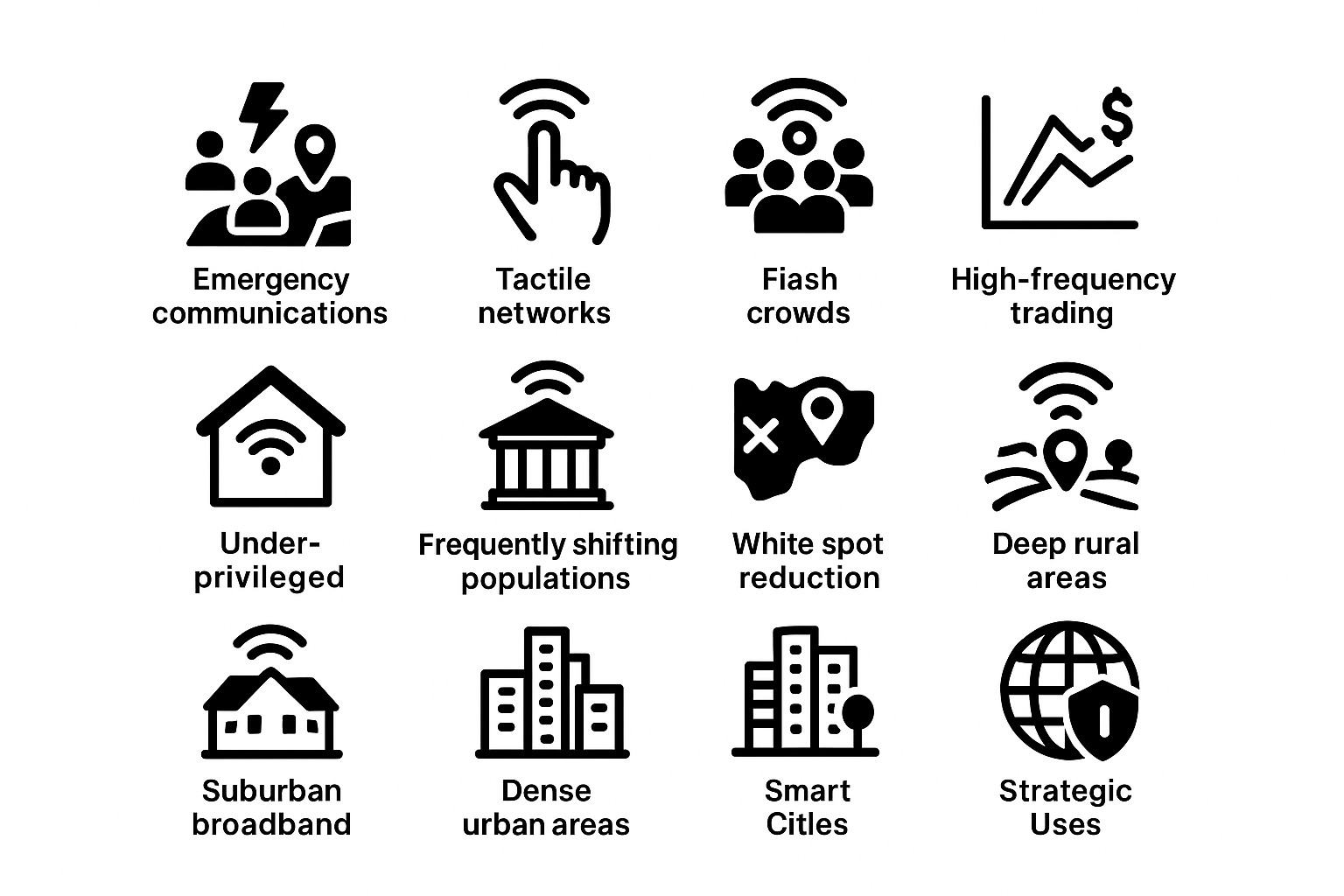}
    \caption{Triple-C Use Cases}
    \label{fig:UseCases}
\end{figure}

\item	\textbf{High-frequency trading (demands ultra-low-latency):} Complementarity – TBS is the preferred choice for local access in trading hubs, while HAPS operate as near-fiber latency relays across regional corridors. LEO is the least effective due to latency constraints. 
\item	\textbf{Under-privileged (demands economic coverage):} Competition – TBS is too costly to roll out in low-\ac{ARPU} villages. However, HAPS and LEO compete to provide low-cost coverage, despite the higher terminal costs for LEO connectivity. 
\item	\textbf{Frequently shifting populations (demands flexibility/mobility):} Competition – HAPS is the preferred choice in dynamic settings with shifting populations (near town center during the day and towards residential areas during the night) owing to its dynamic mobility versus costly terrestrial densification throughout the area. 
\item	\textbf{White spot reduction (demands Lack of infrastructure):} Complementarity – Remote locations such as maritime, aviation, mountainous, or desert can be covered by the global LEO satellites, while HAPS covers the regional corridors, and TBS offers the local access in coasts and airports. 
\item	\textbf{Deep rural areas/greenfield coverage (demands wider coverage):} Complementarity –  Satellites can cover deep rural areas due to the larger coverage area and minimum capacity requirements, where the TBS exhibits limited or no feasibility. Complementary layering fills the rural gaps efficiently. 
\item	\textbf{Suburban broadband (demands QoS assurance):} Competition – TBS can offer fiber-like speeds but requires dense ground rollout. Alternatively, HAPS competes as a regional broadband provider with cheaper deployment rates. In contrast, LEO competes by advertising high throughput, but the capacity saturates quickly in the suburbs. 
\item	\textbf{Dense urban areas (demand blind spots coverage):} Collaborative - Metropolitan cities can utilize the integrated TBS-HAPS-LEO layers to defeat coverage holes collaboratively. TBS leaves blind spots due to blockages and shadowing, whereas HAPS can cover these due to their LoS links.
\item	\textbf{Smart Cities (demands URLLC and resilience):} Cooperation – TBS handles dense IoT and vehicular communications for IoT connectivity and autonomous driving applications. HAPS adds elevation diversity and cell-free macro-panels for city-wide coordination, while LEO supplies resilient backhaul and synchronization across the metro area. The cooperation from all three nodes ensures zero blind spots and ultra-reliability for delay-sensitive applications. 
\item	\textbf{Strategic Uses (demands secure and assured communications):} Competitive – TBS is vulnerable to infrastructure damage (through a deliberate act or natural disaster), whereas LEO satellites are susceptible to jamming, spoofing, and even anti-satellite inventory because of their predictable orbits. However, HAPS competes with the existing technologies by offering secure and assured regional/ wide area communications in such scenarios. 
\item \textbf{High-Capacity Backhaul:} HAPS can act as aerial relay stations, offering high-capacity backhaul for dense urban or aerial base stations. For example, due to line-of-sight links, a HAPS is very appropraite for wireless backhauling of TBSs using high-frequency (mmWave/FSO) links \cite{haps_backhaul_zhu}. Similarly, HAPS can offer focused links for aerial base station backhaul(e.g., UAV base stations) \cite{haps_backhaul_zhu}, overcoming the limitations of fiber or microwave backhaul in crowded networks.
\item \textbf{Seamless mobility support in multi-layer NTN}
LEO requires frequent handovers as satellites move rapidly, which can introduce short interruptions; HAPS are quasi-stationary, reducing churn. In multi-layer operation, a HAPS can bridge LEO handover gaps and anchor sessions until the next pass, minimizing outages \cite{joint_aol,not_promised}.
\end{enumerate}

Combined space–air–ground architectures have been identified for use cases including rural/remote connectivity, maritime coverage, and augmentation of TNs \cite{wide_coverage_LEO}. In such SAGIN systems, LEO satellites enable broad coverage, and HAPS provide regional high-capacity service. These integrate to serve underserved areas (e.g., remote or disaster-affected regions) and boost capacity in hotspots, aligning with 3GPP’s NTN use cases of service continuity and ubiquity \cite{ntn_3gpp}.

\begin{figure}[t]
    \centering
   \includegraphics[width=1\linewidth]{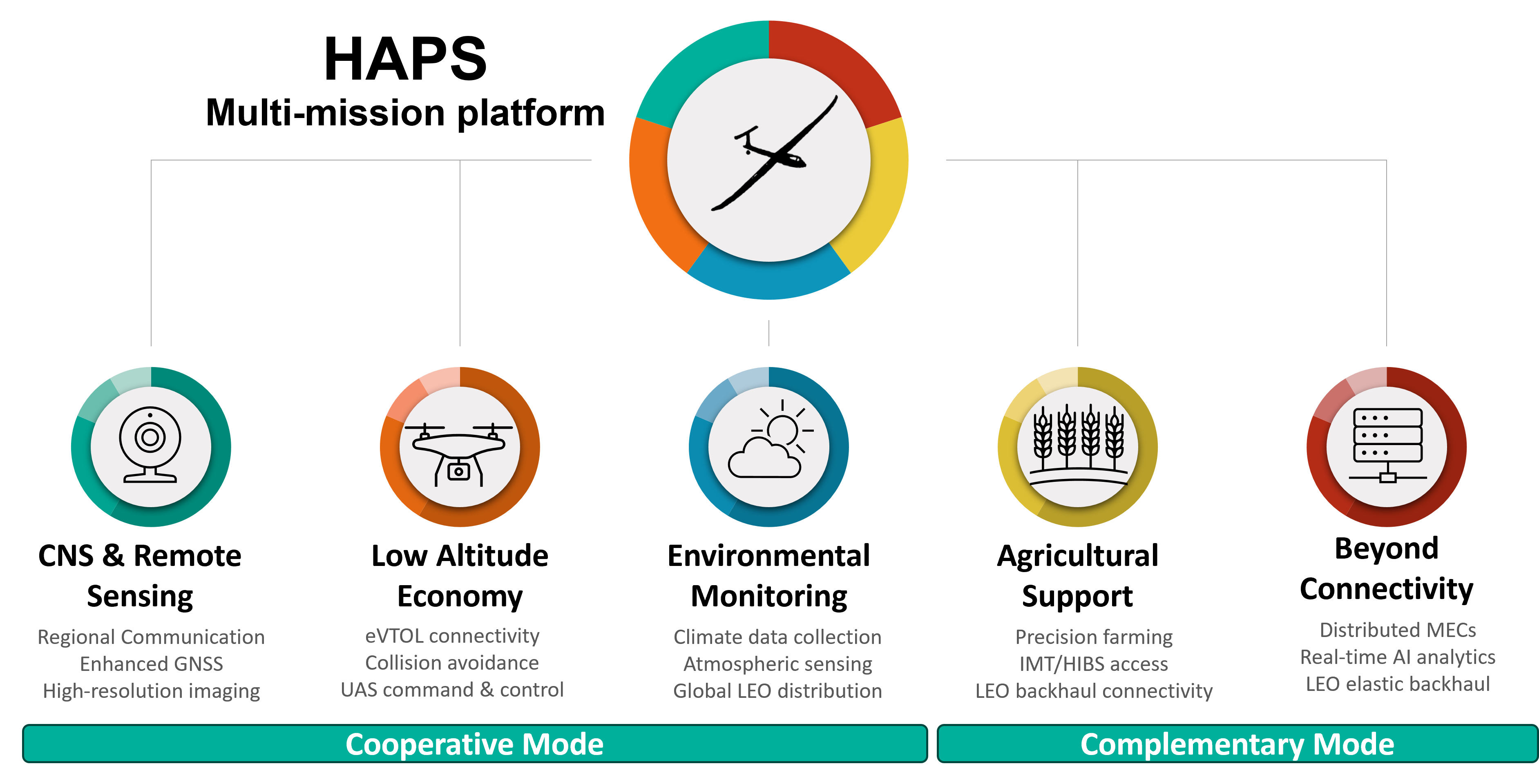}
    \caption{HAPS Enabling Focus Applications with Diverse Payload Capabilties}
    \label{fig:FOCUS}
\end{figure}

\subsection{Focus Applications}
Focus applications highlight the role of HAPS well beyond basic communications. Fig.~\ref{fig:FOCUS} provides an overview of these focus applications with diverse payload capabilities. HAPS – being unique – is capable of rendering the following dividends:

\begin{enumerate}
\item \textbf{CNS \& Remote Sensing/Imaging (Cooperation)}
In a collaborative setting, HAPS can provide persistent regional communication, navigation, and surveillance with low-latency and reduced bandwidth (with edge pre-processing). It can also support high-resolution regional imaging with existing optical technologies. However, it cannot replace the Global Navigation Satellite System (GNSS) offered by satellites.  
\item \textbf{Beyond Connectivity (Complementarity)}
Distributed Edge Processing can be carried out on HAPS, while LEO offers elastic backhaul. The dual-stack backhaul balances capacity (FSO) and weather resilience (RF).
\item \textbf{Environmental Monitoring (Cooperation)}
The climate monitoring and remote sensing can be retrieved from HAPS sensors and sent to LEO, ensuring global distribution/storage. This fills the gap beyond the HAPS footprint. 
\item \textbf{Agricultural Support (Complementarity)}
In regions with sustained demand for precision farming, a few HAPS can outperform LEO on cost/latency using IMT/HIBS access and standard handsets; where fiber is sparse, HAPS access and LEO backhaul keep opex low.
\item \textbf{Low-Altitude Economy (Cooperation)}
The connectivity, collision avoidance, and trajectory planning of electric Vertical Take-Off and Landing (eVTOL) and unmanned aircraft systems (UAS) can be carried out using HAPS regional command and control powered by the multi-access edge computing (MEC). LEO extends backhaul connectivity or inter-city/oceanic continuity; terrestrial small cells add street-level capacity where available. 
\end{enumerate}
\begin{figure}[t]
    \centering
   \includegraphics[width=1\linewidth]{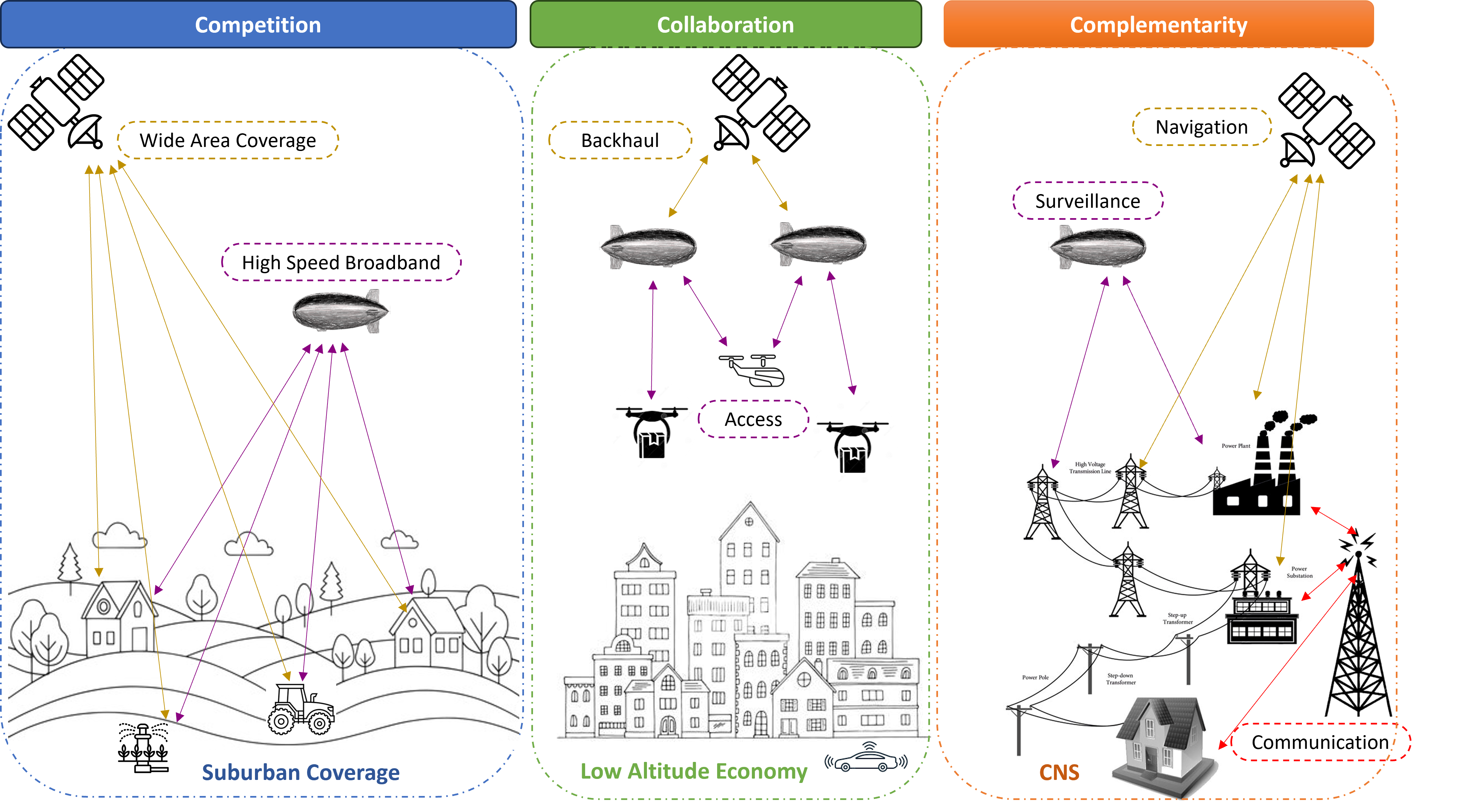}
    \caption{Triple-C integration to realize potential applications}
    \label{fig:3Cs_Apps}
\end{figure}

It is important to highlight that the preferred choice of the Collaboration/Complementarity/Competition framework in any use case scenario is based upon the key requirements, and it may opt for other modes as a fall-back position. Fig.~\ref{fig:3Cs_Apps} maps representative applications to their dominant Triple-C operating modes, highlighting where cooperation, complementarity, or competition is most natural. For most of the scenarios, these technologies are more complementary than competitive, each addressing unique challenges and opportunities in the quest for seamless global connectivity.

\section{Performance Comparison}

The effectiveness of the proposed 3C framework is underpinned by the performance of each participating TBS–HAPS–LEO platform. This section quantifies how different nodes perform under realistic channel conditions, QoS-demands, and network constraints from a link- and mobility-centric perspective.  Rather than a full techno-economic study, we focus on a few key technical levers that directly impact user experience: path loss, propagation, and atmospheric impairments, resulting SNR and outage trends, propagation delay, and mobility-induced handover behavior. We first establish simple, platform-consistent RF/FSO channel models and use them to evaluate outage and latency across TBS, HAPS, and LEO links. Next, we analyze mobility-aware metrics such as handover success rate, dwell time, and dual-connectivity (LEO$\rightarrow$HAPS) effects to highlight how a quasi-stationary HAPS layer can reduce handover failures and stabilize association compared to satellite-only or dense-terrestrial operation. Finally, PropSIM FS16 emulations over two locations ground these comparisons in realistic delay, Doppler, path-gain, and range-rate traces for the involved NTN nodes. The underlying propagation and performance metrics governing each segment of the network are analyzed for a fair comparison.
 
 %Table~\ref{tab:3C-channel-models} summarizes the representative channel models and enabling formulations adopted for the 3C paradigms across RF (2\,GHz)  and FSO (1550\,nm) links. The “comprehensive” RF model adds ITU-R gaseous, rain, and cloud attenuation to free-space loss, while the FSO model includes Beer–Lambert absorption, turbulence (Gamma–Gamma), and pointing effects.
%illustrating in concrete terms when HAPS act as a cooperative anchor, a complementary gap-filler, or a competitive access tier relative to LEO and TBS.
% \textcolor{blue}{
% \begin{itemize}
% \item 	Comparison of performance metrics (Complementary/Competitive): congestion, coverage gaps, capacity, latency, deployment/operational cost, handover 
% \item	Perks that can be achieved with collaboration: New Latencies, achievable data rates, minimized outages during Satellite handovers with HAPS/without HAPS
% \end{itemize}}

\textbf{1) Path loss} under line-of-sight conditions is modeled by the free-space path loss (FSPL): $L_{\mathrm{FS}} = 20\log_{10}(4\pi d f / c)$, where $d$ is the path length, $f$ is frequency, and $c$ is the speed of light. In practice, additional losses are included via a log-distance model: $L(d) = L_0 + 10n \log_{10}(d/d_0) + X_\sigma$, where $n$ is the path-loss exponent and $X_\sigma \sim \mathcal{N}(0,\sigma^2)$ accounts for shadowing. For TBS in urban environments, $n\approx3\text{--}4$ and $\sigma\sim5$–10 dB; for HAPS/LEO, $n\approx2$ with minimal $\sigma$ at high elevation angles.

\textbf{2) Atmospheric attenuation} due to rain is modeled by ITU-R P.838: $\gamma_R = k R^\alpha$ (dB/km), where $R$ is rain rate in mm/h and $k,\alpha$ are frequency-dependent constants. For instance, at 20GHz, heavy rain ($R\sim50$mm/h) yields $\gamma_R \sim 4.5$ dB/km. Cloud and fog losses follow ITU-R P.840: $\gamma_C = K_\ell(f,T)\,\rho_\ell$ (dB/km), with $\rho_\ell$ being the liquid water density. For moderate fog ($\rho_\ell \sim 0.05$ g/m³), $\gamma_C$ may reach several dB/km at 100GHz.

\textbf{3) Fading models} differ by platform. TBS often experiences Rayleigh fading in NLOS conditions and Rician fading in LOS scenarios. HAPS and LEO generally undergo Rician fading due to persistent LOS, with negligible delay spread. Shadowing in TBS is more severe ($\sigma_{\mathrm{SF}} \sim 5$–12dB), while in HAPS/LEO it is much smaller unless operating near the horizon.

\textbf{4) Signal-to-noise ratio (SNR)} is given by $P_r(dBm) = P_t + G_t + G_r - L_{\mathrm{FS}} - L_{\text{atm}}$, and the Shannon capacity becomes $C = B\log_2(1 + \text{SNR})$ (bps), where $B$ is bandwidth. Outage probability under Rayleigh fading is $P_{\text{out}} = 1 - \exp(-\gamma_{\text{th}} / \overline{\gamma})$.

\textbf{5) Latency} is dominated by propagation delay $t = d/c$: HAPS (~20 km) yields ~0.067 ms, LEO (~1000 km) yields ~3.3 ms, and GEO (>35,000 km) yields >100 ms. Hence, HAPS provide ultra-low latency and LEO offers a balance between reach and responsiveness. 
This review is not exhaustive. Beyond channel models and propagation impairments, there is substantial work on HAPS mobility and station-keeping/stationarity. For example, earlier ITU guidance constrained a HAP to a 400 m horizontal radius and $\pm700$ m altitude band~\cite{old_itu_based_paper}, and the HeliNet program proposed two service cylinders 2.5 km$\times$1 km (99.9\% availability) and 4 km$\times$3 km (99\%)~\cite{haps_old_survey}.

\begin{figure}[!t]
    \centering
   \includegraphics[width=1\linewidth]{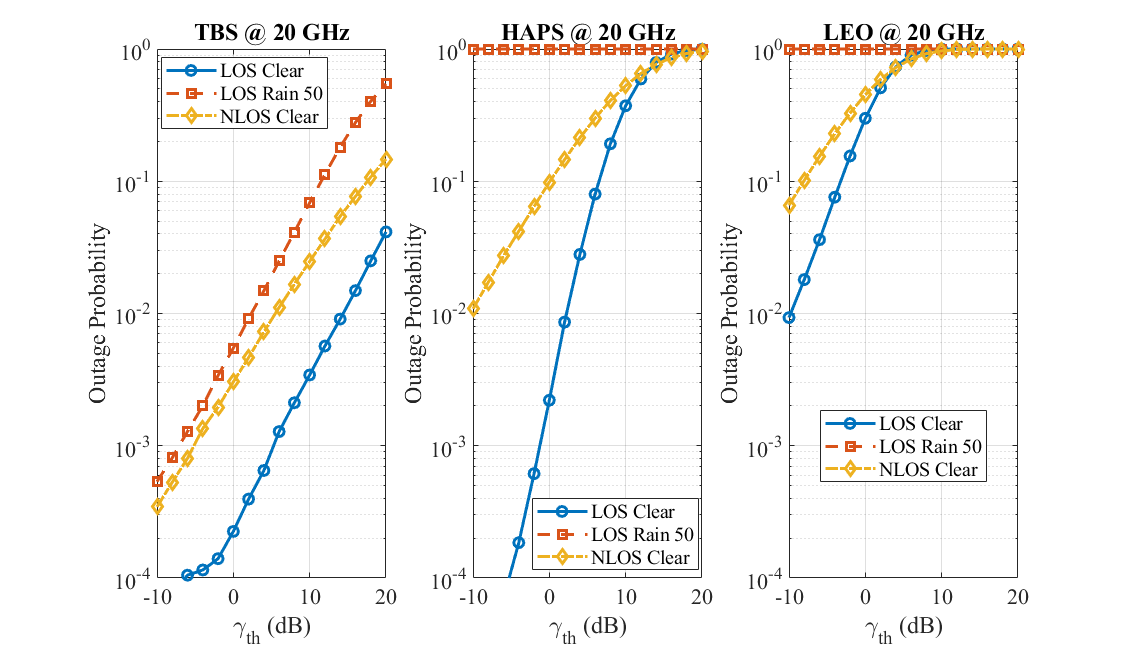}
    \caption{Outage probability $P_{\mathrm{out}}=\Pr[\gamma<\gamma_{\mathrm{th}}]$ 
as a function of SNR threshold $\gamma_{\mathrm{th}}$ for TBS, HAPS, and 
 LEO downlinks at 20~GHz. For each platform, the curves compare clear-sky 
LOS (Rician fading), LOS with 50~mm/h rain (additional attenuation), 
and NLOS (Rayleigh) conditions.
}
    \label{fig:outage}
\end{figure}

\textbf{6) Link Outages:} All the three participating nodes are analyzed at prospective higher frequencies in the FR3 band ($20$GHz frequency) as illustrated in  Fig. \ref{fig:outage}.
While LoS TBS links exhibit low \emph{conditional} outage in the plot, it does not include the platform-specific probability of LoS (HAPS vs.\ TBS vs.\ LEO). Those LoS probabilities and hence true link availability are determined by mobility/geometry models treated in other studies, so the figure should be interpreted with caution. 
From Fig.~\ref{fig:outage}, we observe that in clear LoS conditions the HAPS curve drops fastest with $\gamma_{\text{th}}$. Under 50~mm/h rain, only the TBS downlink avoids near-certain outage across the considered thresholds; both HAPS and LEO operate essentially at $P_{\text{out}}\approx 1$. In NLoS clear conditions, the LEO link again suffers the largest outage probability over the whole range, with HAPS in between TBS and LEO. In other words, only TBS survives in such heavy rainfall, and LEO experiences the most outages in all three scenarios.

\textbf{7) Inter-layer channel models:}
In collaborative mode, inter-layer channels (such as HAPS–LEO and TBS–LEO) characterize the access/feeder/relay links. They typically enjoy favorable LoS conditions but are mainly affected by the Doppler and pointing errors due to high-mobility. 
 In contrast, for complementary/competitive modes, nodes directly communicate with the end-users without needing any inter-layer data plane structure. These links (namely LEO–mobile, HAPS–mobile, and TBS–mobile) are dominated by stronger multipath, shadowing, and weather-induced fading. In a nutshell, cooperation relies on aerial/orbital relay channels whereas complementarity/competition is governed by D2C propagation statistics. Joint HAPS and LEO operation enables lower end-to-end latencies (stratospheric relay vs.\ long satellite hops), higher effective data rates (optical/RF split across feeder and access), and reduced outages during satellite handovers compared to satellite-only operation \cite{joint_aol}. 

\textbf{8) Handovers in Triple-C framework}
Traffic offloading across TBS--HAPS--LEO layers depends on the underlying 3C architecture. The \textit{cooperative mode} expects dual-connectivity (DC) of UE with multiple nodes as opposed to the \textit{complementary/competitive modes} with no dual-connectivity (non-DC). The DC handover (HO) implies make-before-break strategy, whereas Non-DC HO denotes break-before-make capability. Expectedly, the DC handover takes less time as compared to the non-DC handover i.e., $T_{\text{HO}}^{\text{(DC)}}<T_{\text{HO}}^{\text{(non-DC)}}$, with higher success rate. The handover failure probability in DC cooperative mode is given by
\[
\mathrm{P}_{\text{HOF}}^{\text{(DC)}}
~\lesssim~
\Pr\!\left(\frac{4R_b}{\pi v_{\text{rel}}} < T_{\text{HO}}^{\text{(DC)}}\right)
+
\Pr\!\big(\min\{\gamma_{\text{HAPS}},\gamma_{\text{LEO}}\}<\gamma_{\text{th}}\big),
\]

On the other hand, the complementary/competitive non-DC HO success rate can be defined as $\mathrm{HSR}\triangleq 1-\mathrm{P}_{\text{HOF}}^{\text{(non-DC)}}$, with handover failure probability modeled as
\[
\mathrm{P}_{\text{HOF}}^{\text{(non-DC)}}
~\approx~
\Pr\!\big[T_{\text{dwell}} < T_{\text{HO}}^{\text{(non-DC)}}\big]
~+~
\Pr\!\big(\gamma(t_{\text{HO}})<\gamma_{\text{th}}\big),
\]
where $T_{\text{HO}}$ is the handover duration, $T_{\text{dwell}}$ is the remaining dwell time in the serving cell at HO trigger, $\gamma(t_{\text{HO}})$ is the instantaneous post-HO SINR, and $\gamma_{\text{th}}$ is radio-link threshold.

This signifies the importance of both time-budget and SINR risks in the success of a HO event. Considering the satellite's MBC and aerial FBC, the MBCs offer 
shorter dwell times and hence a higher risk of handover failure. In contrast, FBC for HAPS/TBS experience much lower relative motion, allowing
longer dwell times and higher HSR. This implies that the cooperation via a quasi-stationary HAPS (FBC-like anchor) is preferred over satellite-only MBC operation~\cite{fo_mobility_survey_2025}.
% For a circular cell of radius $R_b$, an average (random-chord) dwell-time model gives
% \[
% \mathbb{E}[T_{\text{dwell}}]
% ~\approx~
% \frac{4R_b}{\pi\,v_{\text{rel}}},
% \qquad
% v_{\text{rel}}
% ~=~
% \|\mathbf{v}_{\text{user}}-\mathbf{v}_{\text{footprint}}\|.
% \]
% Thus,
% \[
% \mathrm{P}_{\text{HOF}}
% ~\approx~
% \Pr\!\left(
% \frac{4R_b}{\pi\,v_{\text{rel}}}
% <
% T_{\text{HO}}
% \right)
% +
% \Pr\!\big(\gamma<\gamma_{\text{th}}\big).
% \]
% \emph{MBC vs.\ FBC:} In MBC (satellite-attached cells), $v_{\text{footprint}}$ is dominated by ground-track speed; in LEO this yields large $v_{\text{rel}}$ and smaller $\mathbb{E}[T_{\text{dwell}}]$, increasing $\mathrm{P}_{\text{HOF}}$. In FBC (earth-fixed cells), $v_{\text{footprint}}\!\approx\!0$; for HAPS/TBS this gives $v_{\text{rel}}\!\approx\!\|\mathbf{v}_{\text{user}}\|$ and larger dwell times, lowering $\mathrm{P}_{\text{HOF}}$.

% \medskip
% \noindent\textit{Dual connectivity (LEO$\rightarrow$HAPS) effect.}
% Dual connectivity can further improve continuity by enabling make-before-break operation. When a LEO link is complemented by a quasi-stationary HAPS anchor, the system can initiate a new connection before the previous one is released, effectively reducing both timing constraints and post-handover signal degradation. This cooperative mechanism substantially increases reliability compared with satellite-only MBC operation \cite{fo_mobility_survey_2025}.

\textbf{9) Network-selection criteria:}
In \emph{complementarity/competition} mode, a simple KPI-driven cell selection rule identifies suitable destination platform $p\!\in\!\{\text{TBS},\text{HAPS},\text{LEO}\}$, depending on the 
capacity $C_p\!=\!B_p\log_2(1+\gamma_p)$
and HSR$_p$ for service continuity.
\[
p^\star
~=~
\arg\max_{p}
\left\{
\underbrace{\mathbb{E}[C_p]}_{\text{rate}}
\cdot
\underbrace{\mathrm{HSR}_p}_{\text{continuity}}
\right\},
\]
The main challenge is a distinct balance between achievable data rate and handover reliability in each partcipating platform TBS--HAPS--LEO. Therefore, a practical network-selection policy prioritizes the layer providing the best joint trade-off between throughput and continuity. In cooperative operation, this translates into a joint optimization that favors dual connectivity anchored at an earth-fixed node (HAPS or TBS), ensuring smooth transitions across tiers of the integrated TBS–HAPS–LEO network \cite{fo_mobility_survey_2025}.

\begin{figure}
    \centering
    \includegraphics[width=1\linewidth]{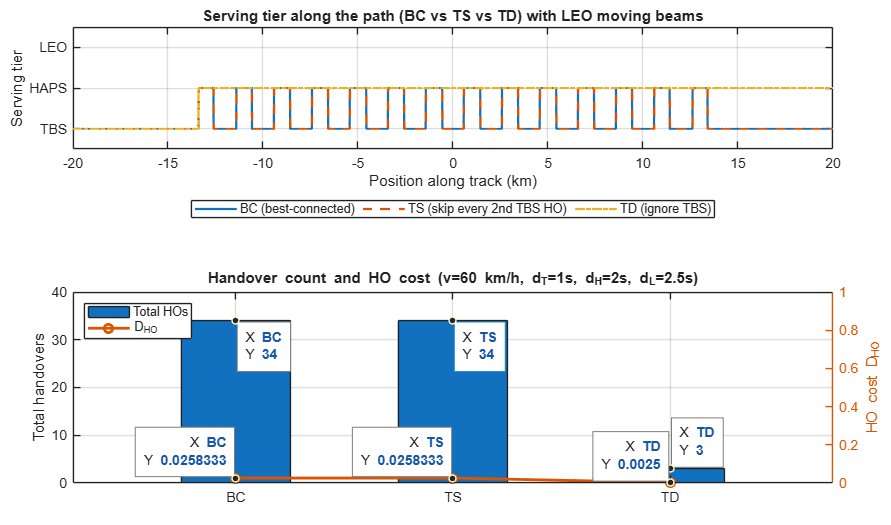}
    \caption{Serving-tier evolution, total handovers, and normalized handover cost under different association schemes in a representative TBS--HAPS--LEO network.}
    \label{fig:handover_demo}
\end{figure}

Fig.~\ref{fig:handover_demo} illustrates the serving-tier dynamics and corresponding HO statistics evaluated for a three-tier terrestrial--HAPS--LEO network.
Three association schemes are compared following \cite{handover_tareq}: 
\textbf{Best Connected (BC)}---the aerial user always attaches to the cell providing the highest average received power; 
\textbf{TBS Skipping (TS)}---similar to BC but every second TBS-related HO is skipped to mitigate frequent terrestrial switching; and 
\textbf{TBS Disabled (TD)}---the TBS tier is ignored, and association is allowed only to HAPS or LEO cells.
The upper subplot shows the serving-tier sequence along a 40~km trajectory, where dense TBS deployments lead to frequent handovers while HAPS coverage spans wider regions with longer dwell times.
The lower subplot indicates that BC and TS yield nearly identical HO counts (around 34) and a normalized HO cost of $D_{\mathrm{HO}}\!\approx\!0.026$, 
whereas TD by excluding the TBS layer reduces both the HO count and cost by nearly an order of magnitude.
These results highlight the \emph{stabilizing advantage of the HAPS layer}, which extends connectivity intervals and lowers HO overhead compared with dense terrestrial tiers, 
a property especially beneficial when coordinating with fast-moving LEO beams~\cite{handover_tareq}.    The LEO tier does not appear in the serving-tier trace because, under the considered link budgets, its large slant range ($\sim\!550$~km) and moderate antenna gain result in a significantly lower average received power compared with HAPS and TBS tiers; hence, the LEO layer never dominates the association in this configuration.

\textbf{10) Emulation Results}
The channel emulation results for the HAPS flights and LEO satellite passes are analyzed using Keysight’s FS16 PropSIM channel emulator (product F8820B) and Channel Studio GCM Tool. We import the two-line elements of the live STARLINK orbital passes over King Abdullah University of Science and Technology (KAUST) and the University of Bristol (UoB). We also emulate HAPS circular station-keeping flights over these regions at an altitude of 20km. We investigate the channel parameters for a comparative analysis of the LoS connectivity provided to the same user equipment (UE) from HAPS and LEO satellite.
 
Considering the HAPS flight and LEO pass over the KAUST campus as illustrated in Fig. \ref{fig:haps-leo-labeled}(a) and \ref{fig:haps-leo-labeled}(b), respectively. We assume HAPS with a ground speed of $110$km/h and $5$km flying radius.  At this
speed and radius, the HAPS completes one full circular orbit in approximately $16$
minutes. We analyze the delay, Doppler, path gain, and range-rate profiles for the
channel between the ground UE and HAPS, assuming a regenerative gNB on HAPS serving UE using LTE band $1$. We have observed a maximum Doppler of $\pm 22$ Hz, delay of $58 \mu$s, path gain of $-0.55$dB, and range-rate of $3.2$m/s from HAPS circular station keeping flight as detailed in Fig. \ref{fig:haps-leo-labeled}(c),\ref{fig:haps-leo-labeled}(d),\ref{fig:haps-leo-labeled}(g), and \ref{fig:haps-leo-labeled}(h), respectively. On the other hand, we consider a bent-pipe LEO satellite passing over KAUST with UE located near the Al-Khwarizmi building, KAUST while the ground-station is assumed to be located in King Abdullah Economic City (KAEC). For $550$km altitude and $7.8$km/s speed of LEO satellite, we observe the service link for the same LTE band for direct-to-device connectivity.

\begin{strip}
\centering
\begingroup
\setlength{\tabcolsep}{2pt}
\renewcommand{\arraystretch}{1}

\newlength{\mapH}\setlength{\mapH}{0.25\textheight}
\newlength{\metH}\setlength{\metH}{0.18\textheight}

% ========== KAUST BLOCK ==========
% Row 1: two maps
\begin{minipage}[t]{0.49\textwidth}\centering
  \includegraphics[width=\linewidth,height=\mapH,keepaspectratio]{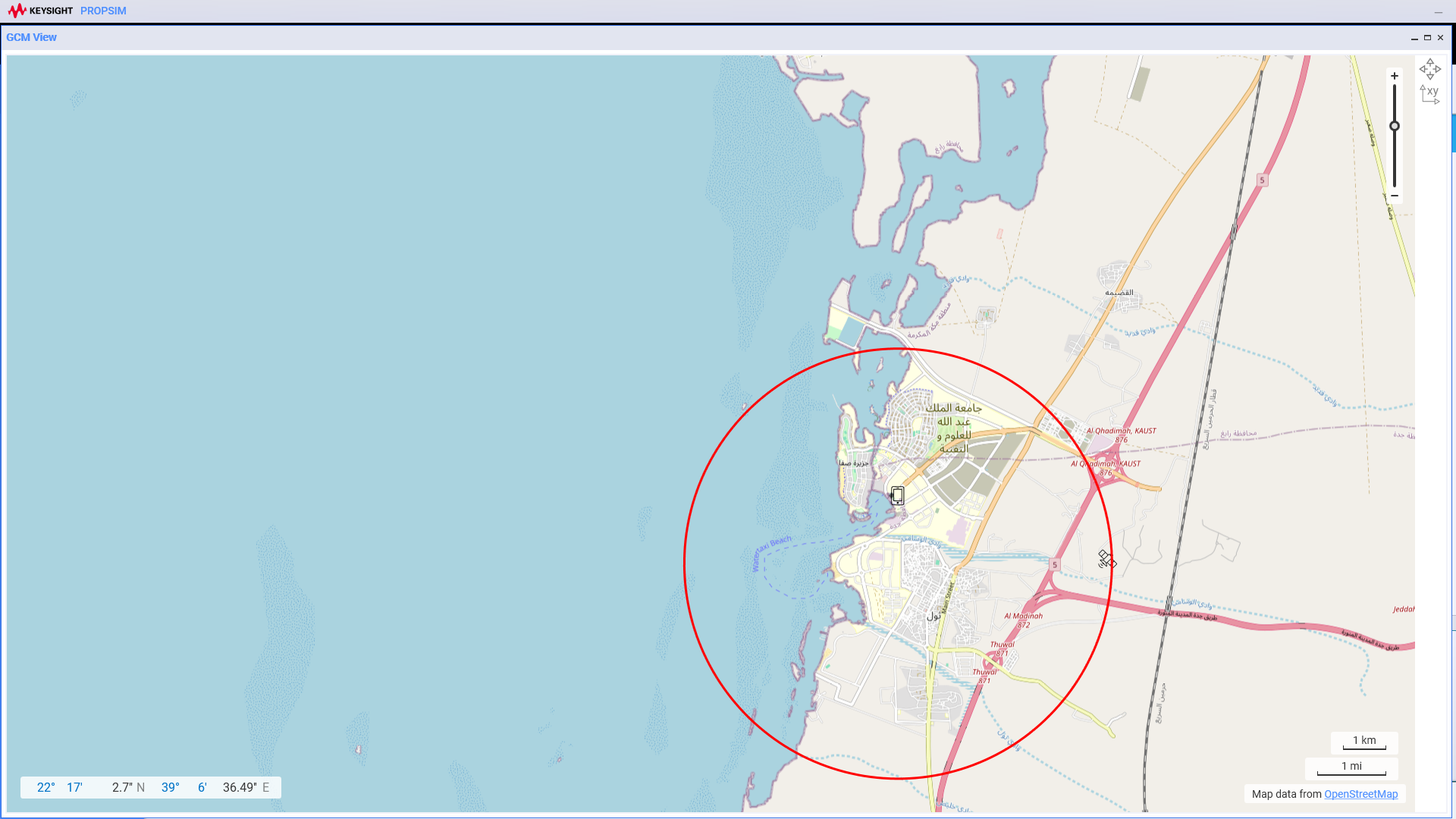}
  \par\small (a)
\end{minipage}\hspace{2pt}%
\begin{minipage}[t]{0.49\textwidth}\centering
  \includegraphics[width=\linewidth,height=\mapH,keepaspectratio]{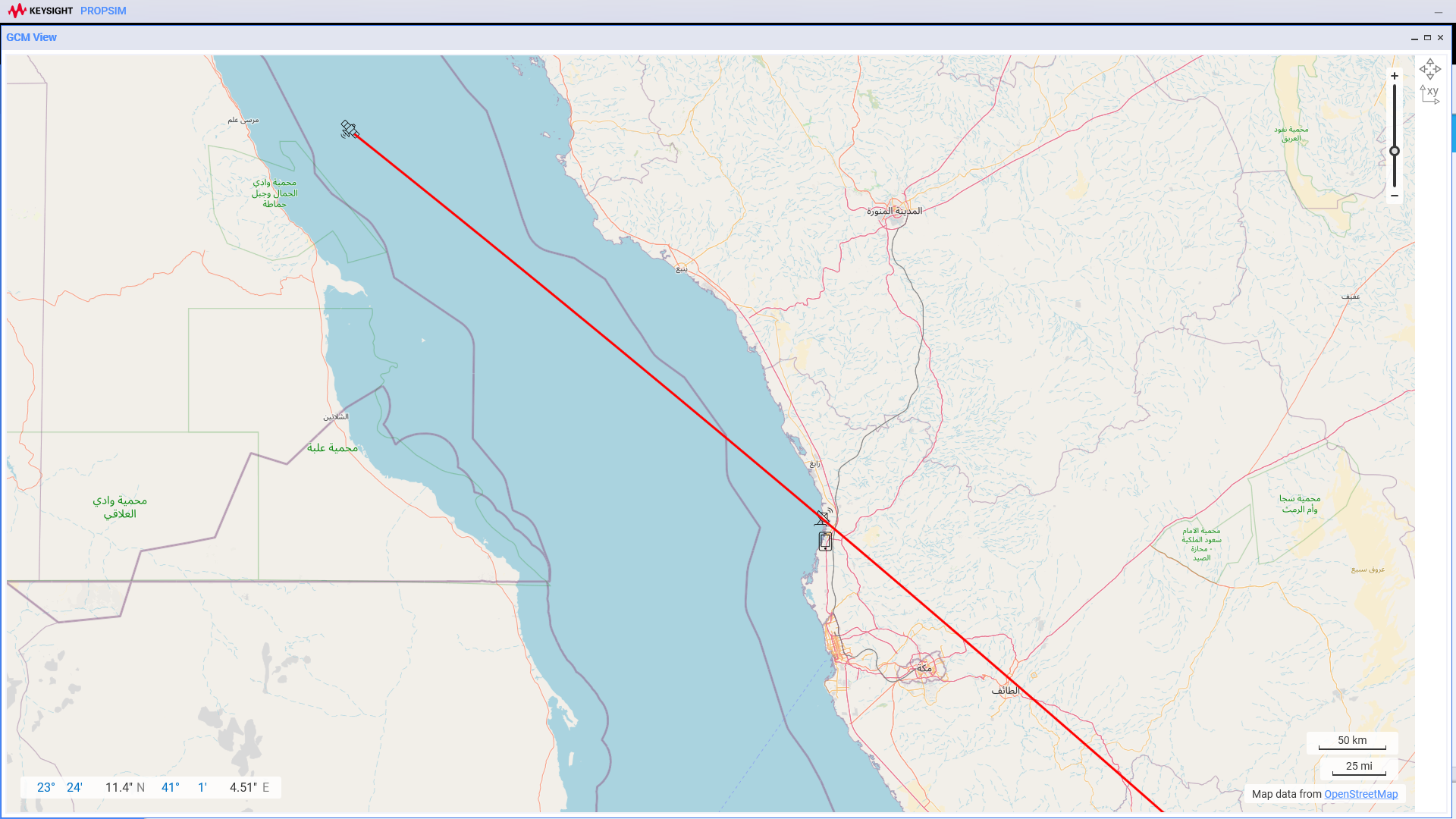}
  \par\small (b)
\end{minipage}

\vspace{2pt}

% Row 2: Delay & Doppler
\begin{minipage}[t]{0.24\textwidth}\centering
  \includegraphics[width=\linewidth,height=\metH,keepaspectratio]{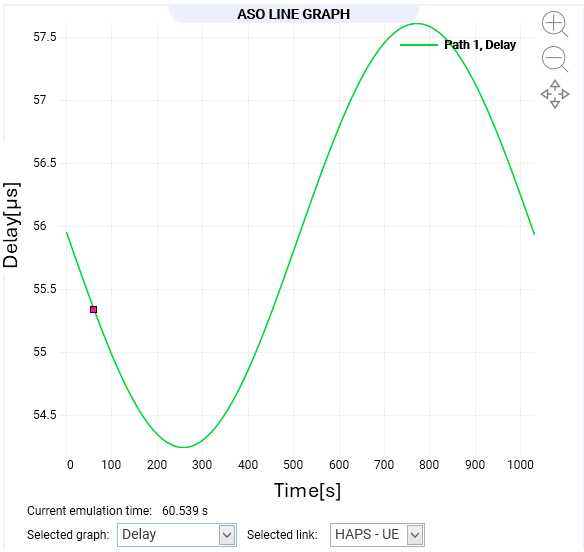}
  \par\small (c)
\end{minipage}\hspace{2pt}%
\begin{minipage}[t]{0.24\textwidth}\centering
  \includegraphics[width=\linewidth,height=\metH,keepaspectratio]{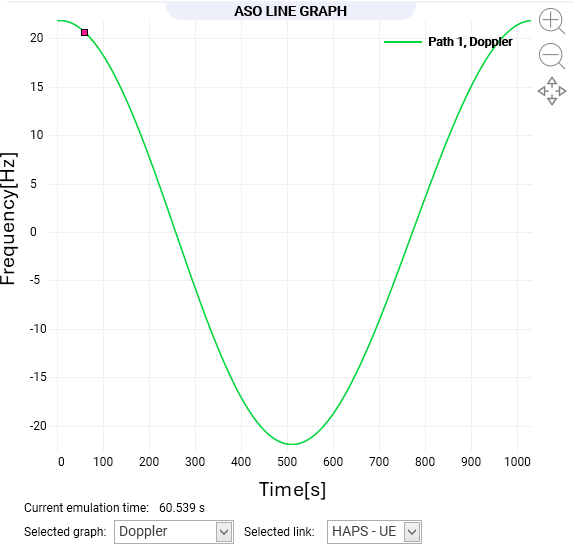}
  \par\small (d)
\end{minipage}\hspace{2pt}%
\begin{minipage}[t]{0.24\textwidth}\centering
  \includegraphics[width=\linewidth,height=\metH,keepaspectratio]{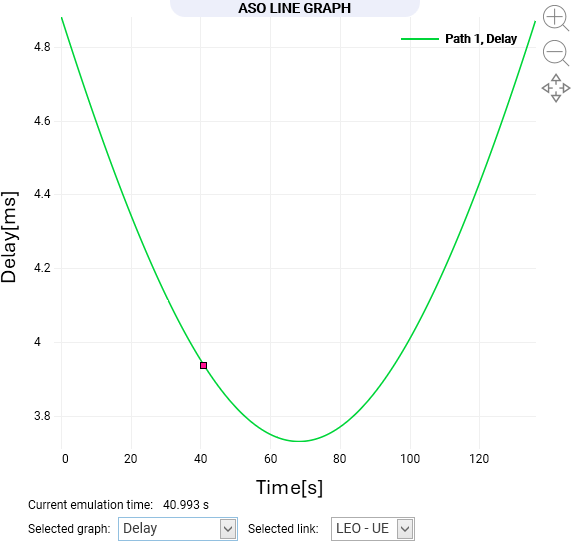}
  \par\small (e)
\end{minipage}\hspace{2pt}%
\begin{minipage}[t]{0.24\textwidth}\centering
  \includegraphics[width=\linewidth,height=\metH,keepaspectratio]{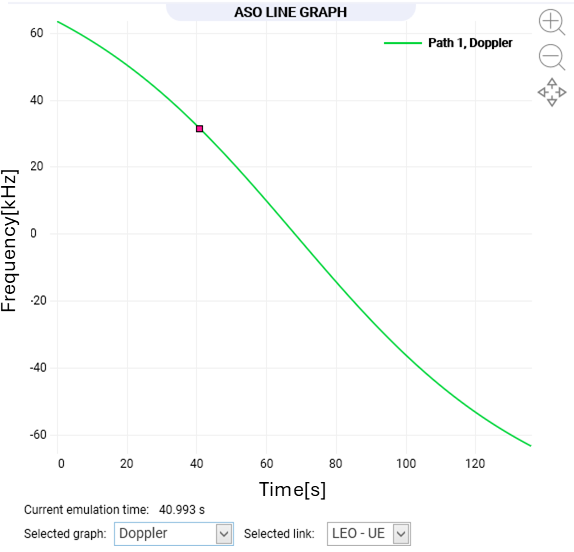}
  \par\small (f)
\end{minipage}

\vspace{2pt}

% Row 3: Gain & Range-rate
\begin{minipage}[t]{0.24\textwidth}\centering
  \includegraphics[width=\linewidth,height=\metH,keepaspectratio]{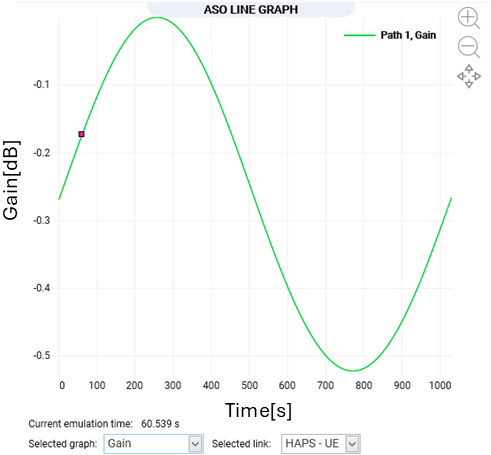}
  \par\small (g)
\end{minipage}\hspace{2pt}%
\begin{minipage}[t]{0.24\textwidth}\centering
  \includegraphics[width=\linewidth,height=\metH,keepaspectratio]{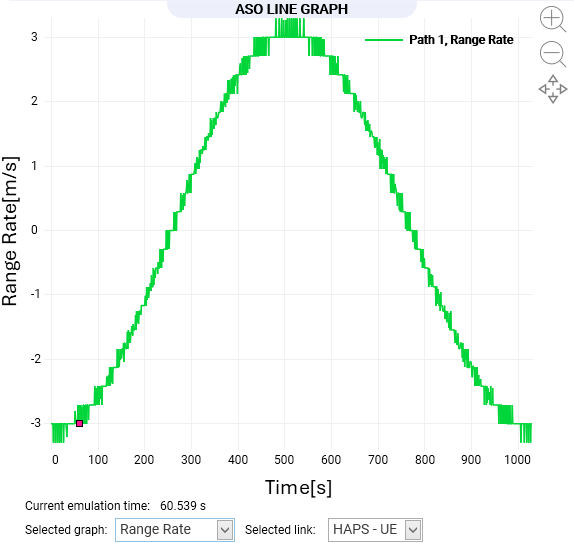}
  \par\small (h)
\end{minipage}\hspace{2pt}%
\begin{minipage}[t]{0.24\textwidth}\centering
  \includegraphics[width=\linewidth,height=\metH,keepaspectratio]{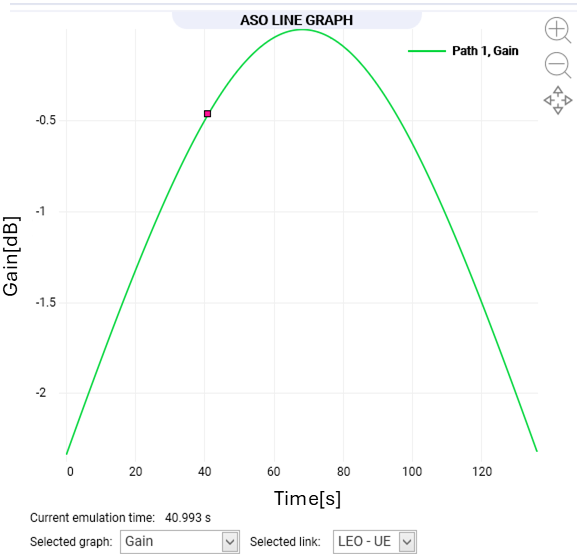}
  \par\small (i)
\end{minipage}\hspace{2pt}%
\begin{minipage}[t]{0.24\textwidth}\centering
  \includegraphics[width=\linewidth,height=\metH,keepaspectratio]{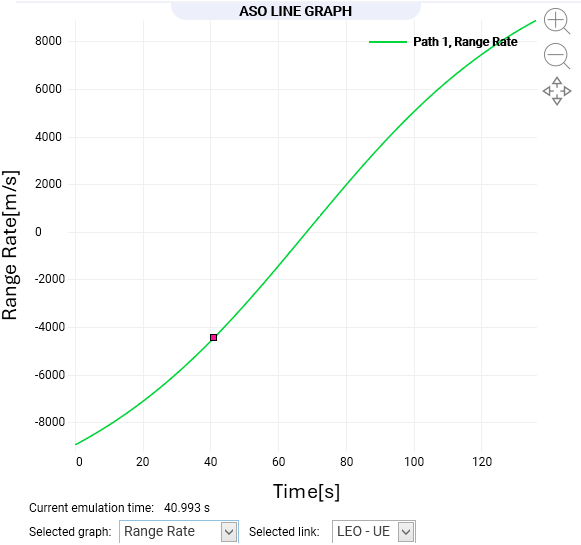}
  \par\small (j)
\end{minipage}

\captionof{figure}{PropSIM FS16 channel emulations for HAPS and LEO passes over KAUST, KSA.
(a) HAPS trajectory over KAUST. 
(b) LEO trajectory (Starlink pass). 
(c) HAPS–UE delay. 
(d) HAPS–UE Doppler. 
(e) LEO–UE delay. 
(f) LEO–UE Doppler. 
(g) HAPS–UE path gain. 
(h) HAPS–UE range-rate. 
(i) LEO–UE path gain. 
(j) LEO–UE range-rate.}
\label{fig:haps-leo-labeled}

\vspace{1.2em} % space between KAUST and Bristol blocks

% ========== BRISTOL BLOCK ==========
% Row 1: two maps
\begin{minipage}[t]{0.49\textwidth}\centering
  \includegraphics[width=\linewidth,height=\mapH,keepaspectratio]{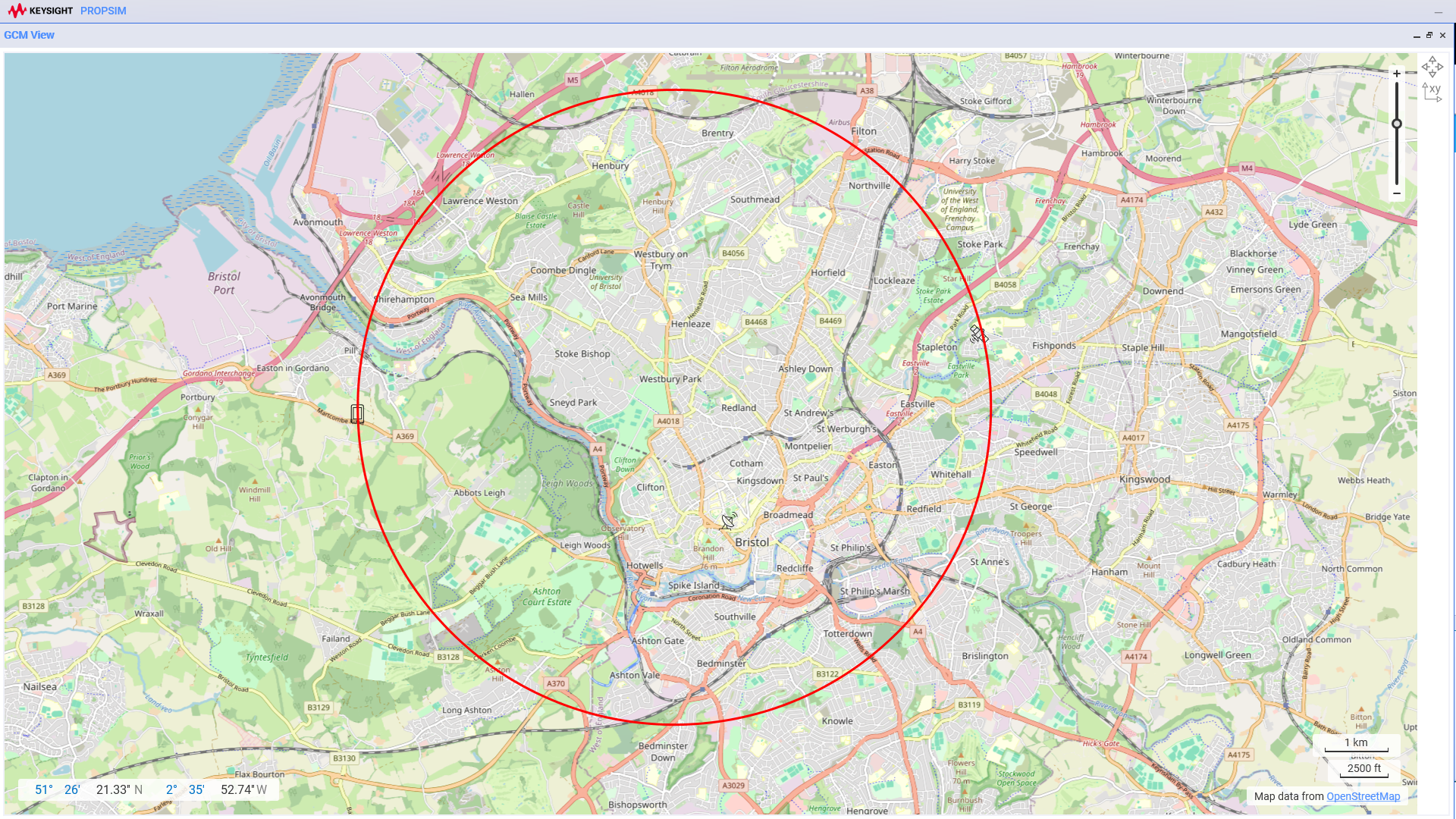}
  \par\small (a)
\end{minipage}\hspace{2pt}%
\begin{minipage}[t]{0.49\textwidth}\centering
  \includegraphics[width=\linewidth,height=\mapH,keepaspectratio]{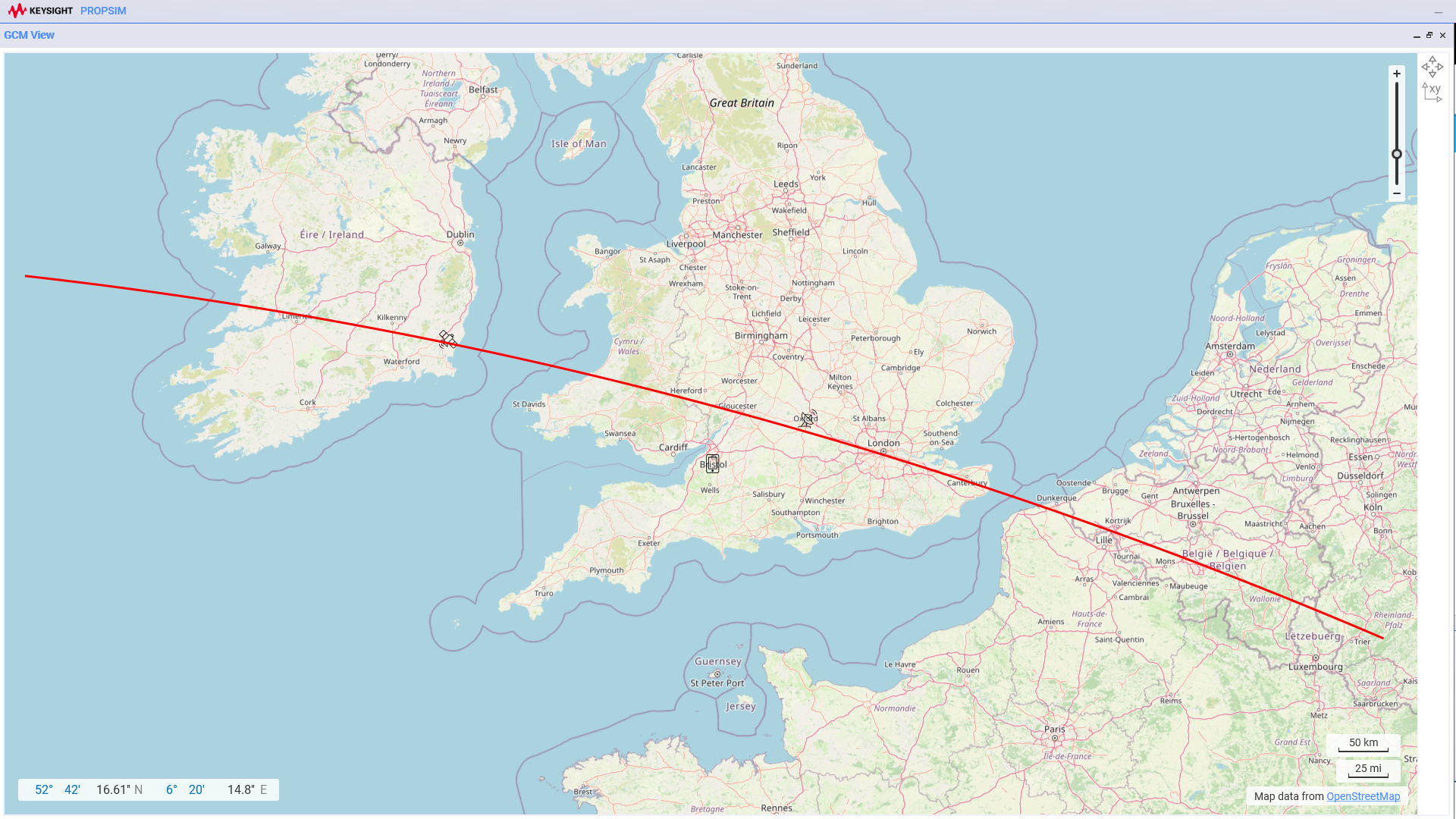}
  \par\small (b)
\end{minipage}

\vspace{2pt}

% Row 2: Delay & Doppler
\begin{minipage}[t]{0.24\textwidth}\centering
  \includegraphics[width=\linewidth,height=\metH,keepaspectratio]{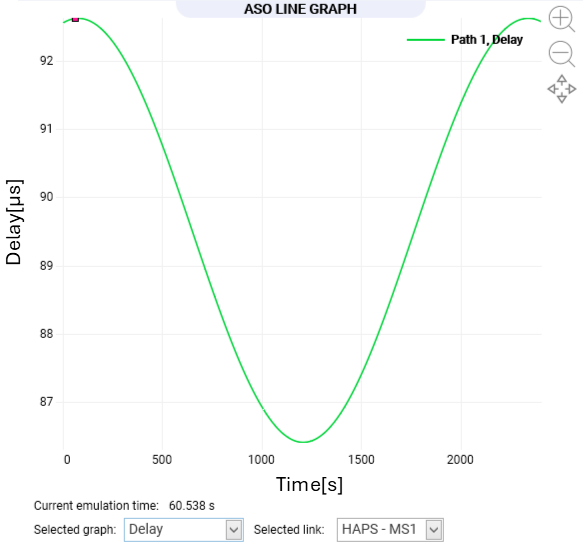}
  \par\small (c)
\end{minipage}\hspace{2pt}%
\begin{minipage}[t]{0.24\textwidth}\centering
  \includegraphics[width=\linewidth,height=\metH,keepaspectratio]{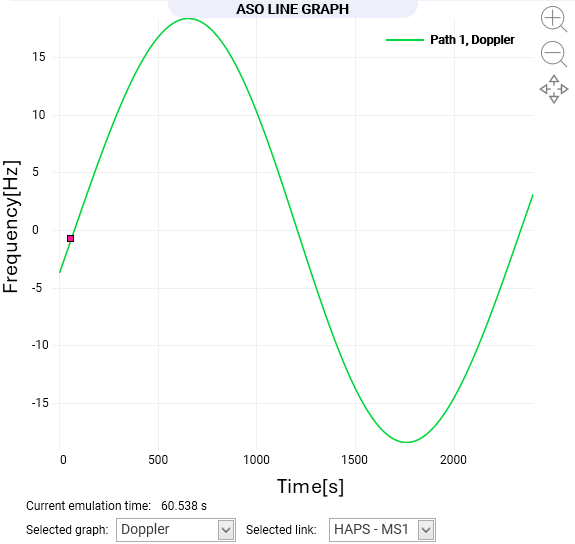}
  \par\small (d)
\end{minipage}\hspace{2pt}%
\begin{minipage}[t]{0.24\textwidth}\centering
  \includegraphics[width=\linewidth,height=\metH,keepaspectratio]{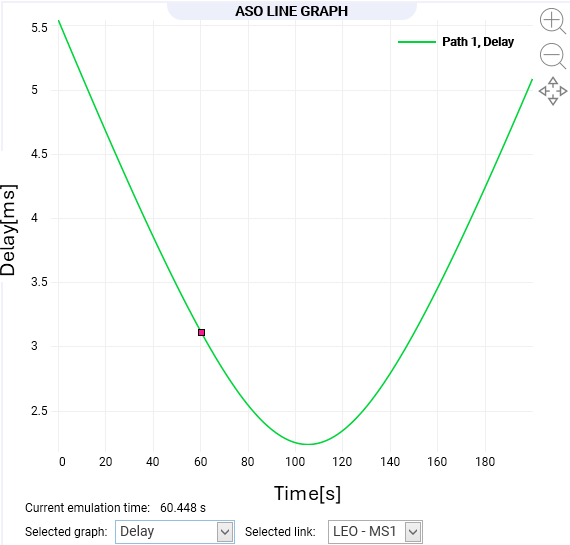}
  \par\small (e)
\end{minipage}\hspace{2pt}%
\begin{minipage}[t]{0.24\textwidth}\centering
  \includegraphics[width=\linewidth,height=\metH,keepaspectratio]{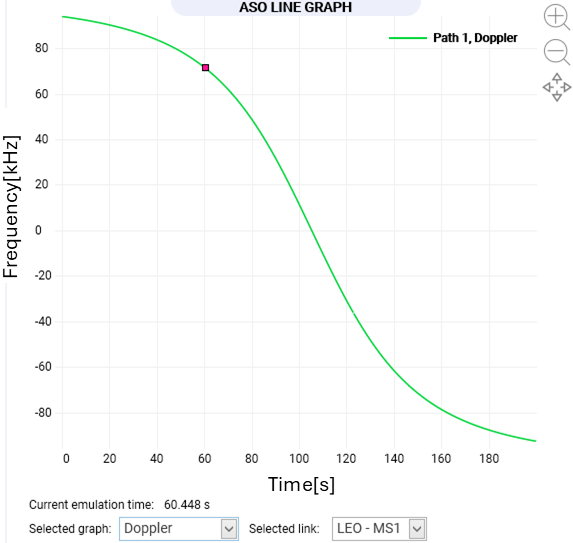}
  \par\small (f)
\end{minipage}

\vspace{2pt}

% Row 3: Gain & Range-rate
\begin{minipage}[t]{0.24\textwidth}\centering
  \includegraphics[width=\linewidth,height=\metH,keepaspectratio]{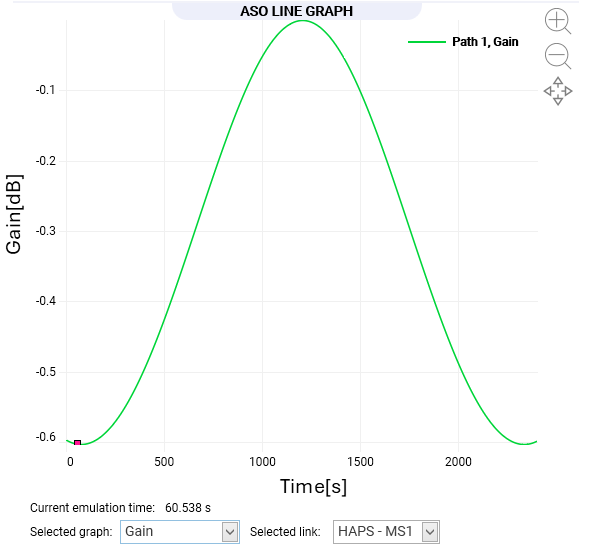}
  \par\small (g)
\end{minipage}\hspace{2pt}%
\begin{minipage}[t]{0.24\textwidth}\centering
  \includegraphics[width=\linewidth,height=\metH,keepaspectratio]{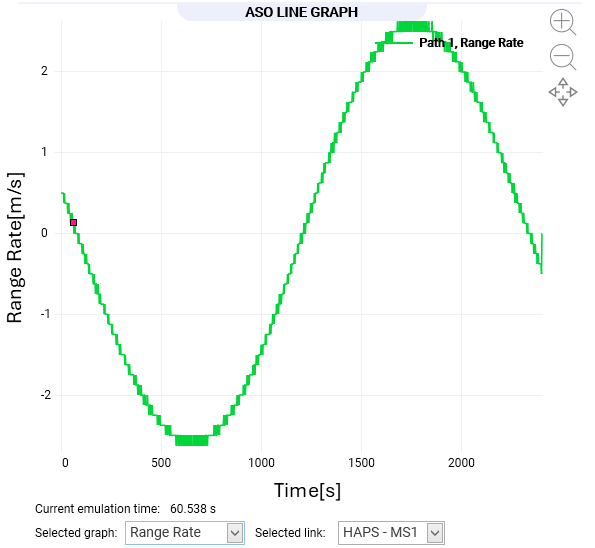}
  \par\small (h)
\end{minipage}\hspace{2pt}%
\begin{minipage}[t]{0.24\textwidth}\centering
  \includegraphics[width=\linewidth,height=\metH,keepaspectratio]{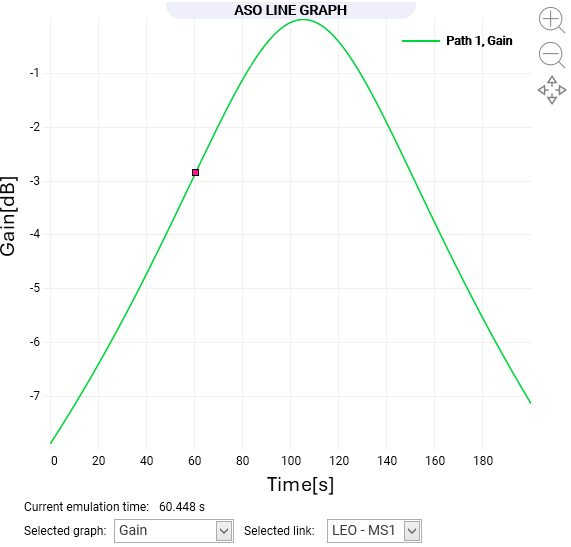}
  \par\small (i)
\end{minipage}\hspace{2pt}%
\begin{minipage}[t]{0.24\textwidth}\centering
  \includegraphics[width=\linewidth,height=\metH,keepaspectratio]{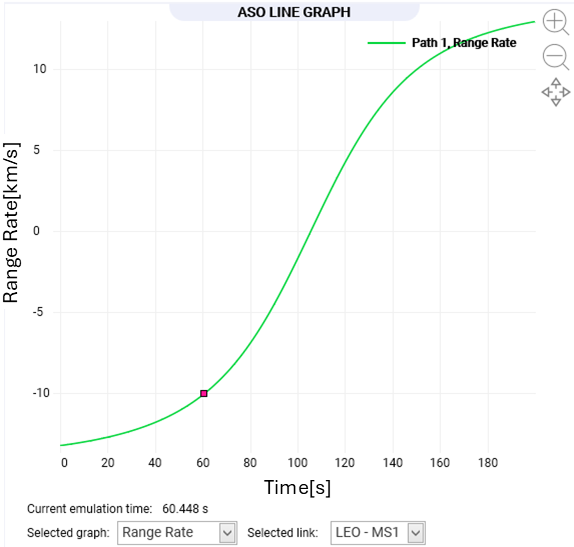}
  \par\small (j)
\end{minipage}

\captionof{figure}{PropSIM FS16 channel emulations for HAPS and LEO passes over Bristol, UK. 
(a) HAPS trajectory over Bristol. 
(b) LEO trajectory (Starlink pass). 
(c) HAPS–UE delay. 
(d) HAPS–UE Doppler. 
(e) LEO–UE delay. 
(f) LEO–UE Doppler. 
(g) HAPS–UE path gain. 
(h) HAPS–UE range-rate. 
(i) LEO–UE path gain. 
(j) LEO–UE range-rate.}
\label{fig:haps-leo-labeled-bristol}

\endgroup
\end{strip}

During the observation window of around $3$minutes with maximum visibility, we
observe the maximum Doppler of $\pm 63$ kHz, delay of $5$ms, path gain of $-2.5$dB, and range-rate of $8.5$km/s as presented in Fig. \ref{fig:haps-leo-labeled}(e),\ref{fig:haps-leo-labeled}(f),\ref{fig:haps-leo-labeled}(i), and \ref{fig:haps-leo-labeled}(j), respectively. Evidently, the emulation analysis reveals the superior link performance from HAPS to the UE relative to the one from LEO. For instance, we observe that the LEO path has about 100 times greater latency than the HAPS link depending on their slant-range distances. The Doppler experienced from HAPS is limited to tens of hertzs, which is negligible for most modern communication systems. In contrast, LEO
produces approximately three-order-of-magnitude increase in Doppler due to its high
orbital velocity requiring precise delay/Doppler pre-compensations. The difference in path-loss (~2dB) is moderate, mainly due to the higher transmission power and
directional antennas on LEO satellites compensating for its greater range. Moreover,
the fast range-rate change in LEO is the root cause of high Doppler derivates and
rapid link parameter variations. Eventually, the relatively short visibility period of LEO per pass would require frequent handovers due to its high orbital speed and limited pass geometry.

We also performed a similar HAPS emulation over Bristol to study the impact of HAPS flying speed, UE placement, and atmospheric losses as shown in Fig. \ref{fig:haps-leo-labeled-bristol}(a). The HAPS fly over Bristol with $50$km/h ground speed to complete $5$km radial flight in approximately $38$ minutes. Moreover, we assume a mobile station (MS1) at the edge of station-keeping flight as opposed to the middle in the previous case. In addition, we also account for the atmospheric losses i.e., $10$ mm/h rainfall in Bristol due to its humid and rainy climate as opposed to the KAUST scenario with an arid desert climate. Our analysis reveals a relatively higher Delay of $92.6 \mu$s for HAPS over Bristol in Fig. \ref{fig:haps-leo-labeled-bristol}(c) as opposed to $58 \mu$s for HAPS flight over KAUST in Fig. \ref{fig:haps-leo-labeled}(c).  It is mainly due to the UE position at the cell edge in Bristol, contrasting the central location of UE in the previous example of HAPS flying over KAUST.

We identify slightly reduced Doppler with peaks at $\pm 18$ Hz for HAPS-Bristol in Fig. \ref{fig:haps-leo-labeled-bristol}(d) as opposed to $\pm 23$ Hz in HAPS-KAUST in Fig. \ref{fig:haps-leo-labeled}(d). The circular flight with ground speed $50$km/h induces $\pm 5$Hz less Doppler as opposed the ground speed of $110$km/h. The lowest path gain recorded in HAPS-Bristol scenario is $-0.6$ dB for HAPS-Bristol in Fig. \ref{fig:haps-leo-labeled-bristol}(g) as opposed to $-0.52$ dB in HAPS-KAUST in Fig. \ref{fig:haps-leo-labeled}(g). The rain fade of $10$mm/h in Bristol induces additional $-0.8$ dB path loss relative to the dry climate in KAUST.  Furthermore, we record the range-rate of $\pm 2.5$ m/s for HAPS-Bristol in Fig. \ref{fig:haps-leo-labeled-bristol}(h) as opposed to $\pm 3.2$ m/s in HAPS-KAUST in Fig. \ref{fig:haps-leo-labeled}(h) due to different ground speeds. This performance comparison between HAPS and LEO under different atmospheric conditions, flying speeds, and geographical locations are summarized in Table \ref{tab:haps_leo_summary}.

Next, we analyze the impact of orbital inclination on UEs at different latitudes. The considered STARLINK orbits have inclination around $53^o$, which implies that they will be closer to Bristol ($51^o$N) than KAUST ($21^o$N). This proximity will result in wider visibility window at higher elevation angles at higher latitudes i.e., it will spend more time above $30^o$ elevation in Bristol as compared to KAUST. However, the delays can increase up to $5.5$ms in LEO-Bristol (Fig. \ref{fig:haps-leo-labeled-bristol}(e))  as opposed to  $5$ms in LEO-KAUST (Fig. \ref{fig:haps-leo-labeled}(e)). Although LEO enters the visibility window earlier but at that instance the distance between LEO and MS1 is significantly large. This also leads to increased Doppler of $\pm 90$kHz in LEO-Bristol (Fig. \ref{fig:haps-leo-labeled-bristol}(f)) relative to $\pm 63$kHz in LEO-KAUST (Fig. \ref{fig:haps-leo-labeled}(f)) at the start and end of emulation within the visibility period. Moreover, the rainfall in Bristol due to its temperate maritime climate leads to significant path losses e.g.,  $-7.5$dB (Fig. \ref{fig:haps-leo-labeled-bristol}(i)) in contrast to $-2.5$dB in LEO-KAUST (Fig. \ref{fig:haps-leo-labeled}(i)). Furthermore, we notice the peak range-rate of $\pm 14$km/s in LEO-Bristol (Fig. \ref{fig:haps-leo-labeled-bristol}(j)) relative to $\pm 8.5$km/s in LEO-KAUST (Fig. \ref{fig:haps-leo-labeled}(j)). 

\begin{table}[!t]
\centering
\caption{HAPS vs.\ LEO: Channel–emulation summary}
\label{tab:haps_leo_summary}
\setlength{\tabcolsep}{4pt}
\renewcommand{\arraystretch}{1.4}
\begin{tabular}{|p{1.3cm} | c| c| c| c| c|}
\hline
\textbf{Region} &
\textbf{Entity} &
\textbf{Delay} &
\textbf{Doppler} &
\textbf{Pathloss} &
\textbf{Range-Rate}  \\
\hline
\hline
KAUST & HAPS & $58\,\mu\text{s}$ & $\pm 22\,\text{Hz}$   & $-0.52\,\text{dB}$   & $3.2\,\text{m/s}$    \\
&  LEO  & $5\,\text{ms}$    & $\pm 63\,\text{kHz}$ & $-2.5\,\text{dB}$  & $8.5\,\text{km/s}$  \\
\hline
UoB & HAPS & $92.6\,\mu\text{s}$ & $\pm 18\,\text{Hz}$   & $-0.6\,\text{dB}$   & $2.5\,\text{m/s}$    \\
&  LEO  & $5.5\,\text{ms}$    & $\pm 90\,\text{kHz}$ & $-7.5\,\text{dB}$  & $14\,\text{km/s}$  \\
\hline
\end{tabular}
\end{table}

\begin{figure}[t]
    \centering
   \includegraphics[width=1\linewidth]{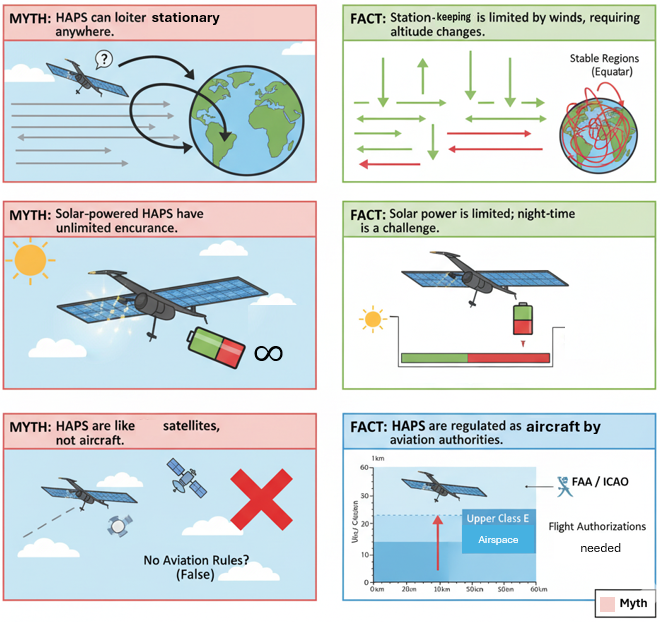}
    \caption{Illustration of common myths versus facts about HAPS, addressing misconceptions on station-keeping, solar endurance, and aviation regulation.}
    \label{fig:haps_myth}
\end{figure}

To ensure that the subsequent performance analysis is interpreted correctly and not influenced by common misconceptions, we briefly address some prevailing myths surrounding HAPS operations as illustrated in Fig.~\ref{fig:haps_myth}.

\label{sec:myths}
\noindent\textbf{Myth 1:} Solar HAPS have “unlimited” power. \textit{Fact:} PV yields are limited; night cycles cap payload power, motivating FC/hybrid and careful duty-cycling.

\noindent\textbf{Myth 2:} HAPS can loiter stationary anywhere. \textit{Fact:} Station-keeping depends on seasonal stratospheric winds; altitude selection and latitude matter.

\noindent\textbf{Myth 3:} HAPS are regulated like satellites. \textit{Fact:} They are aircraft in higher airspace; licensing, airworthiness, and upper Class-E traffic management apply.

\noindent\textbf{Myth 4:} HAPS are posed to replace towers and satellites. \textit{Fact:} Best viewed as a middle layer (3Cs) that complements TBS/LEO.

\section{Economical, Safety, and Environmental Implications}
The cross-domain integration extends far beyond the technical performance gains. Its adoption carries profound economical, safety, privacy, and environmental considerations that are bound to shape the feasibility and sustainability of future communications ecosystem. 
With shared infrastructure, adaptive resource pooling, and flexible service provisioning across heterogeneous domains, the proposed architectures promise significant cost reductions, network resilience, and ubiquitous connectivity. However, airspace safety, users privacy, and environmental footprints need to be carefully evaluated to align these advances with the UN sustainability development goals. This section explores the economical, safety, and environmental  considerations highlighting the opportunities and trade-offs that must guide the Triple-C adoption. 

\subsection{Economic and Commercial Viability}
Cost-performance analysis indicates that LEO constellations are effective for wide-area global connectivity, whereas HAPS are more suitable for localized, regionally focused deployments such as supplementary metropolitan service, sub-urban coverage, or disaster recovery \cite{wide_coverage_LEO}.

\textbf{1) Cost--Benefit Analysis:}
%AI-44
From an economic standpoint, HAPS offer a highly favorable cost–benefit profile compared to both satellites and terrestrial infrastructure. Their capital expenditure is relatively modest typically a few million USD per platform, whereas LEO and GEO satellite missions demand investments on the order of 0.1–1 billion USD \cite{not_promised,haps_cheap}. HAPS also avoid the substantial cost and risk of launch campaigns, and unlike satellites, they can be retrieved, serviced, and upgraded, preventing premature obsolescence \cite{haps_cheap}.
Operational expenditure is similarly advantageous. Powered primarily by solar energy with onboard storage, HAPS incur minimal fuel costs and limited maintenance requirements. In contrast, TNs require continual spending on energy supply, site upkeep, and security, while satellite constellations face ongoing ground-segment and regulatory costs. In sparsely populated or remote regions—where fiber deployment or tower installation is economically infeasible. HAPS can provide wide-area coverage using minimal gateway infrastructure. Their combination of low CAPEX, low OPEX, and scalable coverage makes HAPS particularly attractive for rural or disaster-prone environments where conventional infrastructure is cost-prohibitive \cite{not_promised}.

\textbf{2) Economic Advantages in Hybrid Roles.}
%AI-65
When integrated with satellites, HAPS act as economical intermediaries that reduce total system cost and improve spectrum efficiency. A HAPS can aggregate local LTE/5G or IoT traffic from users using standard devices, and forward this aggregated data via a single high-capacity feeder link to a satellite or ground gateway \cite{joint_aol}. This architecture minimizes the number of high-power satellite terminals and transponders needed to serve a given region, thereby improving spectral utilization and reducing satellite bandwidth costs. Similarly, HAPS backhaul can reduce the need for fiber deployment or microwave relay infrastructure in mountainous or maritime areas.  For navigation or GNSS augmentation, deploying HAPS as ``pseudo-satellites'' can deliver enhanced positioning precision in dense urban or mountainous environments where satellite signals degrade, offering a cost-effective solution compared to launching additional satellites \cite{not_promised}.  
% AI-0
HAPS also provide resilience and redundancy benefits at relatively low incremental cost. When terrestrial or satellite infrastructure is disrupted due to disasters, congestion, or maintenance, HAPS can be quickly deployed to restore connectivity or offload traffic. This dynamic scalability offers measurable financial advantage: operators avoid over-provisioning satellite capacity for rare demand peaks and can instead dispatch HAPS on demand.

\textbf{3) Market Opportunities and Sectors.}
 
% AI-0
The economic and commercial potential of HAPS spans multiple domains. In telecommunications, HAPS can extend broadband and mobile services to rural, island, or underserved regions that deliver IMT-compliant 4G/5G coverage to standard user equipment. This directly supports government's universal-service objectives and private-sector's rural connectivity programs, where traditional infrastructure rollouts remain uneconomical.  Interestingly,  HAPS technology is highlighted among the Top 10 Emerging Technologies of 2024 by the World Economic Forum, underscoring its anticipated societal impact \cite{haps_cheap}. Moreover, major aerospace and telecom companies (Airbus, SoftBank, Alphabet, etc.) are actively participating in HAPS development, and an industry alliance (HAPS Alliance) is pushing standards and trials, indicating a growing commercial ecosystem. While challenges remain (regulatory hurdles, airspace coordination, and ensuring a profitable business model), the market outlook is increasingly positive.

% \section{ Economic and Commercial Viability}
% \textcolor{blue}{
% \begin{itemize}
% \item 	Cost-Benefit Analysis: Assess the economic feasibility of deploying HAPS compared to satellites and terrestrial infrastructure.
% \item	Economic Advantage: HAPS in collaboration with satellites as a relay or HAPS used for surveillance/GPS to complement satellites
% \item	Market Opportunities: Identify potential markets and sectors where HAPS could offer competitive advantages.
% \end{itemize}}

\begin{figure}[!t]
    \centering
   \includegraphics[width=1\linewidth]{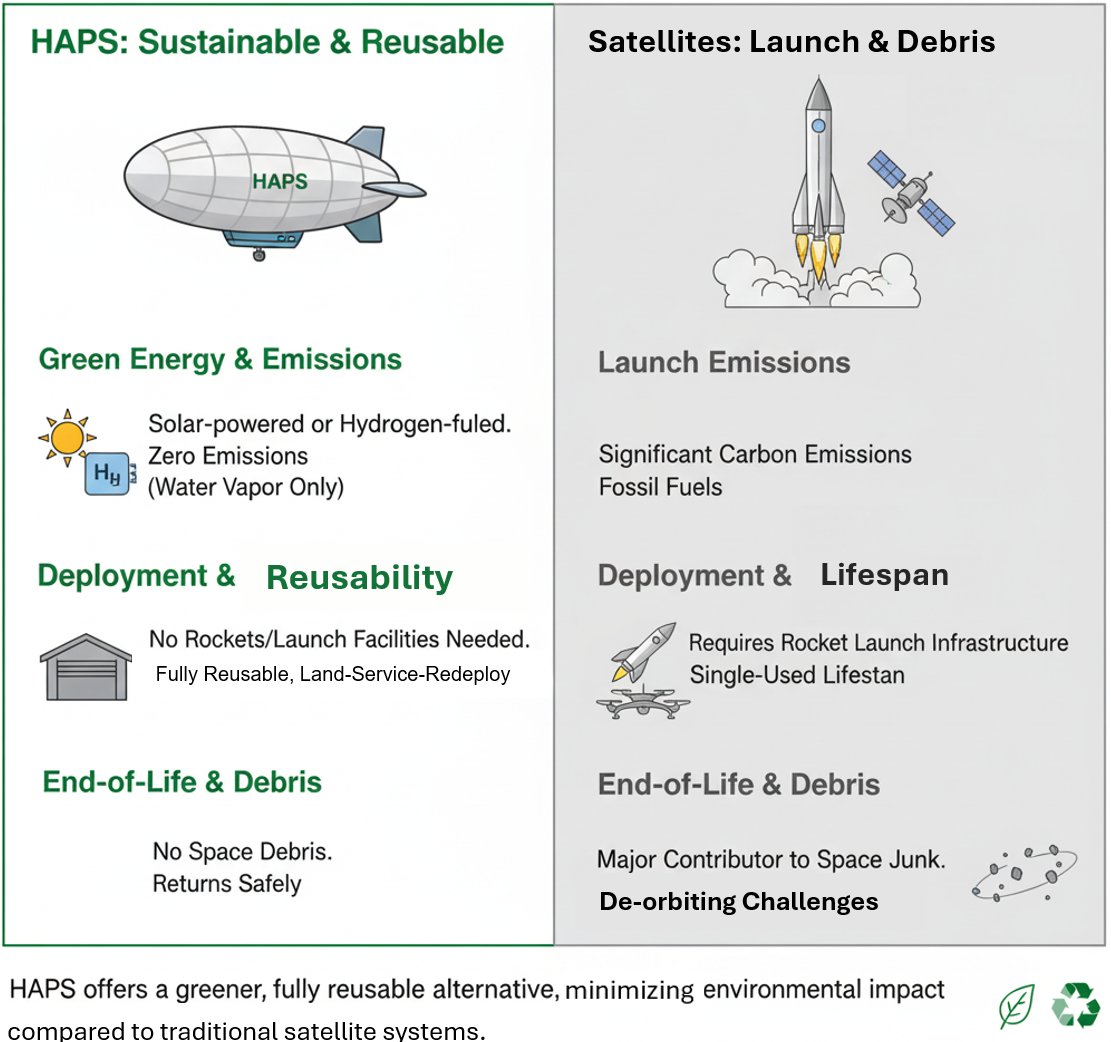}
    \caption{Environmental Impact of HAPS and LEO satellites.}
    \label{fig:haps_green}
\end{figure}

\subsection{Privacy and Safety Concerns} HAPS introduce notable surveillance and privacy considerations. Their ability to loiter over a region near 20~km with high-resolution sensing means HAPS can deliver satellite like acuity at much closer range, intensifying concerns about persistent observation and civil liberties \cite{haps_spy}. Unlike satellites governed by established space law, HAPS operate in a comparatively under-specified regulatory space at lower altitudes, which heightens the need for safeguards. Advocacy groups emphasize that the public interest requires balancing potential benefits of HAPS with clear limits against intrusive or indiscriminate surveillance \cite{haps_spy}. On the other hand, HAPS wireless laser power transfer and airborne flights raise some safety concerns for public health and airspace management, respectively. HAPS fly well above commercial flights ensuring collision avoidance. However, multi-HAPS deployment would require air control and management in the stratosphere. On the other side, while conventional lasers are harmful for the eye, recent long-range Optical Wireless Power Transfer used eye-safe 1.55-µm sources and compact receivers (20cm PD) in daylight conditions, aligning with public safety constraints \cite{infrared_use}.

\begin{figure*}[t]
    \centering
   \includegraphics[width=0.95\linewidth]{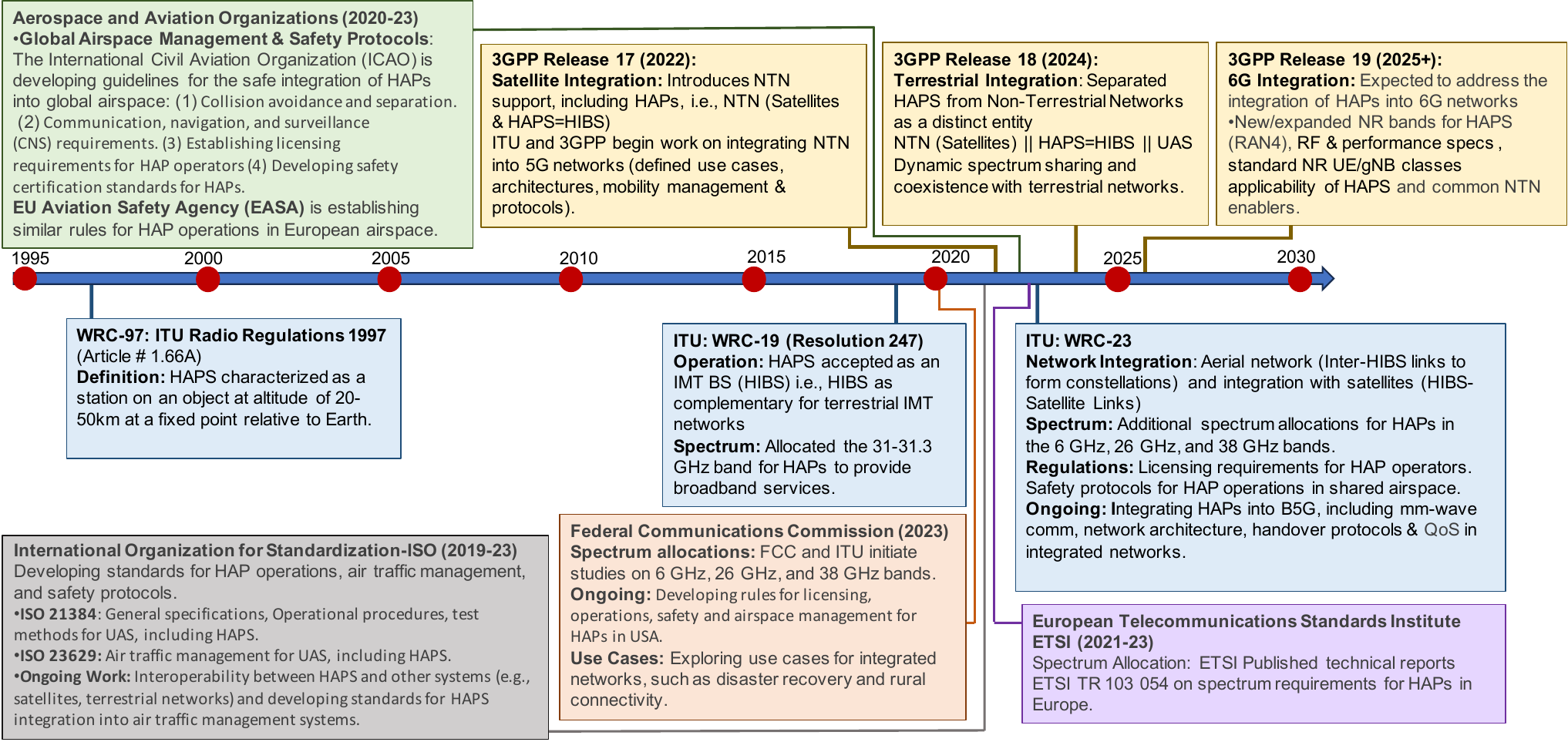}
    \caption{Standardization Activity for HAPS NTN Integration}
    \label{fig:Std}
\end{figure*}

\subsection{Environmental Impact} Overall, the environmental burden of HAPS is typically lower than that of orbital assets. Solar-electric or hydrogen fuel-cell propulsion can limit operational greenhouse emissions, with hydrogen systems producing only water vapor as the primary by-product. Because HAPS do not rely on rocket launches, they avoid the stratospheric exhaust and ozone-relevant byproducts associated with access to space. Current evidence points to minimal stratospheric impact from routine HAPS operations beyond localized effects, whereas launch activities inject soot and reactive compounds into upper layers of the atmosphere. In addition, HAPS platforms can be recovered and safely landed upon completing their lifetime, while satellites burn up on re-entry, potentially releasing alumina and other materials aloft. Taken together, electrically powered HAPS can achieve very low in-flight emissions, and continuing advances in batteries and solar technologies further improve their sustainability profile \cite{nodebris_haps}. On the other hand, initiatives of companies like Orbiting Grid and Celestia are notable for pledging to mitigate orbital debris – Orbiting Grid’s first laser mission will help clear space junk \cite{orbital_fso}, and Celestia vows to remove an equivalent mass for every satellite it launches \cite{celestia_removemass}. Fig.~\ref{fig:haps_green} contrasts the environmental footprint of HAPS and dense LEO constellations. The HAPS competitive environmental advantages over satellites can be enumerated as:

\begin{enumerate}
    \item \textit{No Space Debris (Reusability):} Decommissioned satellites add to orbital debris, whereas HAPS are retrievable and reusable, avoiding long-lived space-junk accumulation \cite{nodebris_haps}.
    \item \textit{No Launch Infrastructure Needed:} HAPS reach operational altitude without rockets or heavy launch campaigns, eliminating the environmental and logistical footprint of space launch \cite{haps_nolaunch}.
    \item \textit{Green Communications:} Solar and hydrogen power enable near-zero-emission flight, with hydrogen producing only water vapor during operation.
    \item \textit{Fully Reusable \& Sustainable:} HAPS can land for maintenance, upgrades, and redeployment, reducing waste and extending useful life compared to single-use satellites \cite{nodebris_haps}.
    \item \textit{Eco-Friendly User End (No Special Ground Equipment):} HAPS can deliver service to standard cellular devices, limiting the need for specialized user terminals and the associated manufacturing footprint \cite{haps_arch_alliance}.
    \item \textit{Efficient 6G Backhaul (Fewer Nodes):} A single HAPS can cover very large areas (potentially substituting for a large number of ground base stations) reducing the land use and energy overhead of dense terrestrial deployments \cite{softbank_haps_hugecoverage}.
\end{enumerate}

\section{Strategic Vision: Regulations, Legislation, and Policy Standardizations}
The HAPS Alliance and standardization bodies stress that existing frameworks do not fit HAPS. In essence, HAPS will require dedicated spectrum licensing, traffic-management rules, aviation regulations, operational legislation, and cross-domain standardizations. Nevertheless, there have been efforts in: 

\textbf{1) Standardization Efforts:} As HAPS mature from trials to commercial service, standardization aims to ensure seamless interworking with terrestrial and satellite networks. 3GPP has incorporated HAPS within its NTN specifications (Rel-17) and continues to expand inter-layer capabilities in Rel-18/19, while ITU has formalized HIBS with relevant spectrum identifications \cite{itu_hibs}, as detailed in Fig. \ref{fig:Std}. Industry groups such as the HAPS Alliance complement these activities with technical guidance and reference architectures to support certification and interoperability.
\begin{table*}[t]
\centering
\small
\setlength{\tabcolsep}{6pt} % optional: adjusts spacing
\renewcommand{\arraystretch}{1.2} % optional: adds line spacing
\begin{tabular}{|p{5.2cm}| p{11.4cm}|}

\hline \textbf{Standardization Initiative} & \textbf{Key Developments Relevant to HAPS} \\
\hline \hline
\textbf{3GPP Release 17 (2022)} & Introduced NTN-specific improvements that included HAPS \cite{haps_ondemand} \\ \hline
\textbf{3GPP Release 18 (2024)} & Strengthened interoperability and includes AI-driven enhancements \cite{haps_ondemand}. \\ \hline
\textbf{ITU WRC-23 / HIBS} & Recognized HAPS as HIBS and identified spectrum below 2.7~GHz and around 47/48~GHz for compatible services \cite{haps_ondemand}. \\ \hline
\textbf{HAPS Alliance} & Publishes guidance on architectures, spectrum use, and certification pathways; engages regulators to harmonize policies \cite{hapsalliance}. \\ \hline
\textbf{EASA “Higher Airspace” Roadmap} & Outlines licensing, airworthiness, traffic management, and detect-and-avoid frameworks for stratospheric operations \cite{easa}. \\
\hline
\end{tabular}
\caption{Major Standardization and Regulatory Initiatives for HAPS Integration}
\label{tab:standards}
\end{table*}

\textbf{2) Spectrum Allocation:} HAPS may use IMT spectrum for direct access and fixed-service and mmWave bands for backhaul; therefore, coordination and interference protection with terrestrial and satellite incumbents are essential. ITU identifications (e.g., 47.2–47.5/47.9–48.2~GHz and portions of 2~GHz) enable hybrid access/backhaul options, with technical limitations to protect co-channel users \cite{haps_ondemand}.

\textbf{3) Network Integration:} In B5G/6G, HAPS are positioned as a mid-layer to enhance resilience and capacity between ground and space segments. 3GPP work items describe mobility, integrated access-backhaul, and continuity mechanisms so devices can transition across TBS–HAPS–LEO as needed \cite{haps_ondemand}.

\textbf{4) Aviation Regulations – Licensing and Airspace Management:} HAPS occupy a "new frontier" of airspace. In most jurisdictions, stratospheric operations are treated like aircraft above 60{,}000~ft (Class~E airspace in the U.S.) rather than satellites. The FAA and NASA have defined an \textit{Upper Class~E} category for flights above FL600 and are working on an \textit{Upper Class~E Traffic Management (ETM)} concept\cite{class_e_management,nasa_class_e}. Likewise, ICAO and national CAA/AST agencies are studying HAPS rules. Current airspace systems lack provisions for long-endurance uncrewed flight at 20--25~km, so regulators anticipate new procedures. For instance, NASA points out that the necessary regulations for commercial operations in the stratosphere are still under review, with companies often operating under waivers\cite{nasa_class_e}. HAPS operate in higher airspace in extreme conditions (very low temperature, high UV/radiation), where aviation and aerospace regimes intersect, and most jurisdictions are still formalizing rules. HAPS carry no crew, so certification must focus on ground/overflight safety \cite{haps_reg_class_e}. The Alliance has urged ICAO and JARUS to develop dedicated HAPS airworthiness guidelines, focusing on preventing crashes or debris impact on people below \cite{haps_reg_class_e,nasa_class_e}. Research highlights gaps in authorization pathways and preparedness for stratospheric operations.

\textbf{5) National Legislation and Data Sovereignty:} Because HAPS operate within sovereign airspace, they allow localized services and data handling aligned with domestic laws. This supports national objectives around data protection and secure communications without relying on cross-border satellite overflight \cite{eyes_strat}. Statutes (e.g., FAA Reauthorization Act 2024) and updates to UAS rules (beyond Part~107) are expected to define new categories and performance-based pathways tailored to stratospheric platforms \cite{faa_reauthorization}.

\textbf{6) Industry and Academic Contributions:} Consortia and research communities are shaping technical baselines and policy inputs through trials, whitepapers, and standards contributions—addressing interference coordination, mobility, atmospheric effects, and certification practices \cite{hapsalliance,easa}.

\section{Gaps, Challenges, and Future Directions for Triple-C NTNs}
The lack of unified three-tier TBS--HAPS--LEO architecture implies existing gaps that need to be addressed. 3GPP TR~38.811 and TR~36.763 focus on studies to support NTNs, and existing surveys largely describe the integration of space-air-ground without describing or explaining how to implement the mode-aware state machine, inter-layer interfaces or KPIs \cite{chen_saginsurvey_challenges, routing_challenge_aktas, interference_modeling_challenge } that are directly relevant to 3Cs. Existing channel, interference, and mobility models are unable to jointly capture the near-space propagation, mmWave/THz effects, ISLs, and dense terrestrial reuse \cite{routing_challenge_aktas, interference_modeling_challenge,cao_survey} phenomena due to them being platform centric instead of integration-centric. LEO access-network and 6G–satellite routing studies further show that random access, handover, \ac{HARQ}, and routing are already strained in single- or dual-tier NTNs \cite{chen_saginsurvey_challenges,cao_survey,leochallenge-zhenyu,handover-sreeram,6gntnchallenge-webpage}, and are not explicitly defined yet for a three-tier, mode-explicit TBS--HAPS--LEO stack which has multi-operator incentives and environmental constraints.

\subsection{Technical Challenges in the Triple-C Paradigm}
\begin{figure}[t]
    \centering
    \includegraphics[width=1\linewidth]{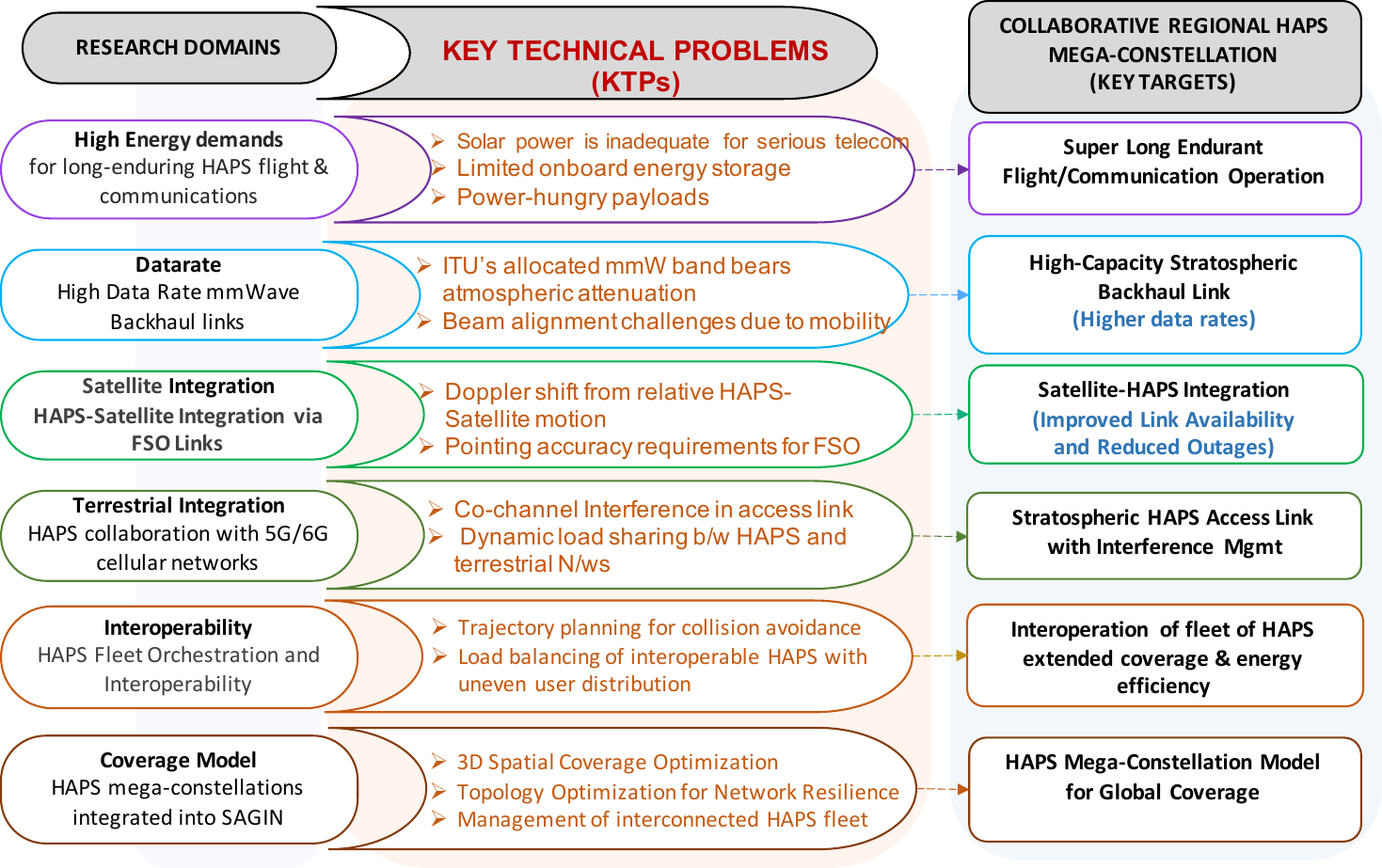}
    \caption{Cross-domain challenges in the realization and integration of HAPS.}
    \label{fig:OpenResearch}
\end{figure}

The Triple-C paradigm exposes that HAPS-assisted NTNs are not limited by a single bottleneck. These challenges jointly determine whether the integrated network can realistically meet the requirement. Fig.~\ref{fig:OpenResearch} summarizes the key technical problems and system-level targets that must be addressed for scalable Triple-C operation. Building on this overview, we now examine how these challenges manifest in each Triple-C mode.

\textbf{1) Collaboration:}
While Collaborative Triple-C modes presuppose joint beamforming, routing, and backhaul sharing across TBS, HAPS, and LEO, near-space surveys emphasize how HAPS channels differ qualitatively from conventional terrestrial and satellite links and mention the non-applicability to integrated networks; at mmWave/THz, tropospheric scintillation and low-elevation rain fading can introduce tens of dB of loss on HAPS and LEO paths \cite{routing_challenge_aktas,cao_survey,interference_modeling_challenge}. On the other hand, dense ISL operation in Q/V and beyond is still poorly characterized, with no unified model capturing cross-link and cross-tier interference \cite{interference_modeling_challenge}. Consequently, collaborative beamforming ad RRM across tiers are largely heuristic. Moreover, while 3GPP has defined baseline mmWave satellite propagation characteristics, key high-frequency effects—such as Doppler, fading, multipath structure, and joint space–time correlation—remain insufficiently characterized and still require detailed investigation\cite{giordani2021}.

LEO-\ac{SAN} works also show that even in a single satellite tier, conventional RACH and HO suffer due to low flexibility and high latency under massive access and fast-moving beams \cite{cao_survey,leochallenge-zhenyu}. For triple-Cs, a UE may see  TBS, HAPS, and multiple LEOs with heterogeneous RTTs and Doppler, but there is no MAC design that coordinates RA across tiers, maintains link-specific \ac{HARQ} processes, and re-anchors user sessions across ground and non-terrestrial gateways without context loss \cite{leochallenge-zhenyu}. SAGIN and cross-domain SDN proposals suggest splitting control, but routing remains siloed: there is no standard mechanism to jointly optimize paths that mix TBS, HAPS relays, and LEO ISLs under joint latency, reliability, and cost constraints \cite{leochallenge-zhenyu,6gntnchallenge-webpage,routing_challenge_aktas}. NTN roadmaps on slicing, AI/ML, and ORAN treat NTNs as a generic platform but do not address how to expose Triple-C modes to slices or how to place intelligence given HAPS energy and payload limits \cite{nguyen-emerging}. The study of network coding and computation offloading is mainly done in a per-tier or per-operator basis, with virtually no work on joint operation which leaves a clear gap for collaboration \cite{interference_modeling_challenge}.

\textbf{2) Competition:}
Coexistence studies of TBS, HAPS and LEO mainly focus on satellite–terrestrial interactions, while ISL-focused works show that inter-satellite interference in dense constellations remains poorly understood \cite{interference_modeling_challenge}. Due to the lack of model that combines all three into a single competitive interference picture, it has resulted  in the lack of robust power-control and spectrum-sharing mechanisms. Existing surveys rely on conventional KPIs but do not quantify metrics inter-layer fairness which is applicable in scenarios when LEO capacity grabbing starves HAPS/TBS of high-value users\cite{6gntnchallenge-webpage}. This lack of competition-aware metrics means there is a need to design schedulers, tariffs, or auctions that avoid unstable or unfair outcomes.

\textbf{3) Complementarity:}
One of the frequently cited goals of SAGIN is complementary operation, but concrete mechanisms remain vague. LEO handover studies report "handover storms" when beams and gateways shift \cite{leochallenge-zhenyu,handover-sreeram}; Triple-C complementarity would require explicit policies specifying which users move to HAPS vs. TBS when LEO footprints depart, under which QoS/energy constraints, and how context (buffers, \ac{HARQ} state, security) is transferred across tiers. There have been some works that route around jamming by exploiting competition and cooperation among nodes \cite{anti_jamming}, but there is no relevant mapping from these ideas to mode transitions (e.g., disabling competition for a compromised tier) or resilience metrics that quantify "cooperation gain"\cite{chen_saginsurvey_challenges}. ISL analyses further show that gateway placement, ISL scheduling, and constellation topology can produce tens-of-hop variations across starting satellites, leading to highly variable end-to-end delay and reliability for any HAPS/terrestrial segment anchored on that constellation \cite{leo_routing}; current 3D NTN simulators rarely include such multi-timescale ISL dynamics.

\subsection{Non-Technical Challenges in the Triple-C Paradigm}
Beyond physical-layer limits, Triple-C operation is constrained by how different stakeholders cooperate, how multi-layer resources are coordinated, and how legal/planetary boundaries are respected. We therefore highlight three non-technical dimensions: federation, orchestration, and geographical/sustainability constraints.

\textbf{1) Federation:}
Terrestrial MNOs, HAPS operators, and LEO constellation providers are typically distinct entities with different CapEx/OpEx profiles. Techno-economic studies suggest that HAPS are attractive for regional coverage and LEO for global reach\cite{haps-needed,wide_coverage_LEO}, but there is no established framework for dynamic roaming, capacity leasing, or bandwidth trading across these layers. Existing roaming is mostly static and terrestrial-only; there is no real-time ``bandwidth market'' where an MNO can automatically purchase HAPS/LEO capacity during peaks or a HAPS operator can monetize complementary disaster coverage. This lack of federation and revenue-sharing models undermines incentives to deploy Triple-C capabilities.

\textbf{2) Orchestration:}
Cross-domain SDN studies advocate separate satellite, aerial, and terrestrial controllers plus an upper-layer orchestrator \cite{routing_challenge_aktas,6gntnchallenge-webpage}, yet provide no multi-operator deployment model. \cite{nguyen-emerging} survey NTN slicing, AI/ML, and ORAN but do not discuss how these mechanisms would operate when TBS, HAPS, and LEO belong to different stakeholders with distinct incentives. Multi-agent RL and game-theoretic orchestration have been proposed for SAGIN anti-jamming and resource allocation \cite{anti_jamming}, but typically assume either fully cooperative or fully competitive agents and do not address non-stationary environments with orbital dynamics, weather, and traffic jointly evolving. AI safety and explainability for Triple-C are virtually untouched\cite{llm-ai}, despite their importance in public-safety and defence contexts, where operators may require interpretable mode transitions.

\textbf{3) Geographical boundaries and sustainability:}
Triple-C modes also raise unresolved questions on serving areas, cross-border footprints, and environmental limits. LEO beams span multiple countries, HAPS can loiter over borders, and TBS coverage is tied to national licensing. Regulatory analyses note that spectrum allocation for NTN bands (Ka/Q/V/THz) and HAPS fixed-service bands remains fragmented, with World Radiocommunication Conferences' (WRCs) processes struggling to reconcile satellite broadband, terrestrial 5G/6G, and emerging NTN uses \cite{6gntnchallenge-webpage}. Data-sovereignty and dual-use concerns add friction, as Triple-C mode transitions could route traffic through foreign gateways or operators. Additionally, there is a need to incorporate space footprint or astronomy safety alongside QoS.

\subsection{Future Outlook and Research Directions}
As Triple-C NTNs move from concept to deployment, several tightly coupled research directions emerge on how to architect, operate, and power multi-tier networks. We highlight three priorities: architectural design, radio-resource and mobility control, and energy- and sustainability-aware operation.

\textbf{1) Architectural Design:}
Near-term work must close the architectural and modeling gaps by defining a three-tier TBS--HAPS--LEO reference architecture with an explicit Triple-C state machine, inter-layer interfaces, and KPIs that quantify cooperation gain, complementarity, competition fairness, and resilience. This should be supported by joint 3D propagation and interference models for sub-6~GHz, mmWave/THz, and optical regimes that capture near-space effects, ISLs, and dense terrestrial reuse \cite{interference_modeling_challenge,routing_challenge_aktas,cao_survey}. SAGIN and LEO-\ac{SAN} studies already call for sustainable, AI-driven 3D architectures \cite{chen_saginsurvey_challenges,leochallenge-zhenyu,6gntnchallenge-webpage}; routing surveys advocate ML-enhanced, QoS-aware routing over integrated satellite–terrestrial paths \cite{routing_challenge_aktas}. These directions must be extended to treat HAPS as a first-class, energy-constrained stratospheric layer, not a generic NTN node \cite{nguyen-emerging}.

Standalone HAPS operation will continue to face challenges such as long-duration station-keeping against stratospheric winds \cite{fo_stationkeeping_2024}, mass and power limits for SDR/SAR payloads, and stringent feeder/service-link requirements \cite{itu_hibs}. In collaborative modes, Triple-C research should refine joint Doppler mitigation, timing alignment, and load balancing, and design schedulers for multi-HAPS constellations that avoid packet reordering and instability. In competitive modes, evolving spectrum policies, NTN slicing, and security frameworks \cite{fo_ntn_rrm_challenges_2025,fo_slice_security_2019,fo_slice_ntn_2024} will shape how operators position TBS, HAPS, and LEO without destabilizing the ecosystem.

\textbf{2) Radio resource management, mobility, and traffic scheduling:}
RRM research is beginning to adopt learning-based and topology-aware approaches, including digital-twin-assisted offloading and DRL-based schedulers for multi-tier LEO networks with hybrid RF/FSO backhaul \cite{digitaltwin-offloading}. For Triple-C, these must evolve into multi-timescale schedulers that couple fast LEO dynamics, slower HAPS drift, and static TBS grids; multi-tier RA/HO procedures that prevent RA congestion and HO storms when many users change serving tiers \cite{cao_survey,leochallenge-zhenyu,handover-sreeram}; and routing/scheduling schemes that exploit ISL structure while using HAPS for regional aggregation \cite{leo_routing}. Programmable SDN/NFV control and slice-aware resource allocation will likely underpin such mechanisms, but robust multi-domain APIs and intent models are still missing \cite{leochallenge-zhenyu,6gntnchallenge-webpage}. Network coding and cooperative \ac{HARQ} across tiers remain largely unexplored tools for realizing mode transitions at the data-plane level.

\textbf{3) Energy considerations and sustainability:}
Energy and sustainability constraints will strongly influence feasible Triple-C designs. For HAPS, highly efficient solar arrays, lighter structures, and improved storage are key to freeing payload power for advanced radios and on-board AI; for LEO mega-constellations, launch, replacement, and duty-cycled high-power operation dominate the cost and environmental footprint. Recent long-range optical power-beaming demonstrations (e.g., DARPA's POWER program)\cite{darpa_laser} suggest that beamed energy could eventually augment on-board generation, enabling more compute-heavy regenerative payloads and AI-driven orchestration on HAPS platforms. Accordingly, any Triple-C NTN framework should treat debris, light pollution, and energy footprint as first-class design metrics, rather than side constraints. A mature Triple-C framework should therefore treat ``energy and environmental budget'' as first-class constraints in mode selection and resource allocation, favouring solar-powered HAPS or terrestrial segments when they suffice, and reserving LEO backbones for cases where their unique latency or coverage is indispensable.

\section{Conclusion}
% \textcolor{blue}{
% \begin{itemize}
% \item 	Summary of key takeaways.
% \item 	Perspective on where HAPS fits in the long-term NTN vision.
% \item 	Call to action for collaborative research and policy alignment.
% \end{itemize}}

This paper presents a comprehensive and forward-looking perspective on the evolving relationship among terrestrial,aerial, and space networks, reframing their interaction through the lens of Cooperation, Complementarity, and Competition (the triple-C framework) to deliver a multi-layered, resilient, and adaptive communication ecosystem. 
Each mode contributes uniquely to the architecture: cooperation maximizes resilience, redundancy, and elimination of blind spots; complementarity ensures scenario-driven (i.e., based on spatial/ service demands) optimum utilization of resources; while competition accelerates healthy pursuits for being the best service provider. Nonetheless, despite their huge potential to offer breakthrough dividends, the realization of this 3Cs model is an uphill task that faces non-trivial technical and regulatory challenges. These mainly include, but are not limited to, 
 1) NTN-TNs standardization to achieve collaboration; 2) Delineation of dynamic boundaries from the prism of geographical and service zones to attain complementarity, and\\ 3) KPI based AI-driven service provider selection to play the competition.
 
 In the NTN constitution, HAPS fills a critical operational and economic void between ground and space networks pertaining to their quasi-stationarity, optical visibility, flexible deployment, reusability, and regional focus.  These unique capabilities enable it to mitigate well-known limitations of LEO systems, including dense handover cycles, Doppler dynamics, spectrum congestion, and lack of localized service guarantees. At the same time, LEO satellites extend the reach, robustness, and global continuity that HAPS alone cannot provide. The proposed Triple-C model offers a structured pathway to determine when HAPS should compete with, complement, or cooperate with LEO, depending on application, geography, resilience needs, and market constraints. Performance analysis and emulation results quantify the strengths and limitations of each layer, confirming that hybrid architectures consistently outperform siloed deployments, especially in resilience, outage/latency minimization, targeted coverage, and disaster recovery. The work also emphasized the broader considerations including NTN standardization, regulatory, economic, spectrum management, and global aviation and safety implications. 
 
Looking ahead, the integration of TBS, HAPS, and LEO satellites will be central to the 6G vision of resilient, intelligent, and pervasively available connectivity. To reach this goal, we call upon researchers, industry partners, standardization bodies, and national regulators to jointly address the highlighted technical and non-technical gaps, accelerate real-world trials, dispel misconceptions about HAPS capabilities, and embrace cross-layer co-design. Ultimately, the future of NTN is not a competition between platforms but a carefully orchestrated synergy of terrestrial, stratospheric, and orbital assets. HAPS will play a decisive role in this continuum by bridging layers, enhancing resilience, reducing cost, and enabling a new class of applications that neither satellites nor terrestrial systems can deliver alone.

%Finally, if the host of actors (including academia, relevant government bodies, and industry) collaborate while putting a premium on state-of-the-art deployable and observable research and development (R\&D), desired multi-domain effects can be reaped in medium to long-term, well beyond the currently imaginable frontiers.

\section{Acknowledgement}
We would like to express our gratitude to AI(ChatGPT) for its invaluable assistance in refining the grammar, enhancing clarity, and rephrasing sentences, particularly in the introduction section. While ChatGPT has contributed to enhancing the linguistic quality of our work, it's important to note that the core concepts, construction of the 3C paradigm, supporting arguments, and performance analysis were achieved independently, without the assistance of AI. 
 \bibliographystyle{IEEEtran}

\bibliography{IEEEabrv,HAPS_PP_refs}
%HAPS_LEO_refs1,HAPS_LEO_refsEROS,manualrefs_EROS,TripleC_Refs.bib}

\end{document}